\pdfoutput=1
\documentclass[12pt]{iopart}
\usepackage[numbers]{natbib}
\usepackage[utf8]{inputenc}
\usepackage{adjustbox}
\usepackage[breaklinks,colorlinks,linkcolor=blue,citecolor=blue]{hyperref}
\usepackage{tabularx}
\usepackage{pdflscape}
\usepackage{booktabs}
\usepackage{makecell}
\usepackage{amssymb}

\usepackage{comment}
\usepackage{listings}
\usepackage[dvipsnames,table,xcdraw]{xcolor}
\usepackage{caption}
\usepackage{subcaption}

\usepackage[acronym]{glossaries}
\glsdisablehyper
\makeglossaries

\newacronym{moabb}{\textsc{MOABB}}{Mother Of All \acrshort{bci} Benchmark}
\newacronym{bci}{BCI}{Brain-Computer Interface}
\newacronym{hmi}{HMI}{Human-Machine Interface}
\newacronym{eeg}{EEG}{Electroencephalography}
\newacronym{mi}{MI}{Motor Imagery}
\newacronym{erp}{ERP}{Event Related Potential}
\newacronym{ssvep}{SSVEP}{Steady State Visually Evoked Potential}
\newacronym{soa}{SOA}{Stimulus Onset Asynchrony}
\newacronym{isi}{ISI}{Inter Stimulus Interval}
\newacronym{cvep}{c-VEP}{Code-modulated Visually Evoked Potentials}
\newacronym{vep}{VEP}{Visually Evoked Potential}
\newacronym{tvep}{tVEP}{Transient Visually Evoked Potential}
\newacronym{fvep}{fVEP}{Frequency-modulated Visually Evoked Potential}
\newacronym{bbvep}{BBVEP}{Broad-band Visually Evoked Potential}
\newacronym{rvep}{rVEP}{Random Visually Evoked Potential}
\newacronym{erd}{ERD}{Event-Related De-synchronization}
\newacronym{ers}{ERS}{Event-Related Synchronization}
\newacronym{psd}{PSD}{Power Spectral Density}
\newacronym{auc}{AUC}{Area Under the Curve}
\newacronym{roc}{ROC}{Receiver
Operating Characteristic}
\newacronym{roc-auc}{ROC-AUC}{Area Under the \acrlong{roc} Curve}
\newacronym{cca}{CCA}{Canonical Correlation Analysis}
\newacronym{csp}{CSP}{Common Spatial Pattern}
\newacronym{trcsp}{TRCSP}{Tikhonov Regularized \acrshort{csp}}
\newacronym{fbcsp}{FBCSP}{Filter-Bank \acrshort{csp}}
\newacronym{mdm}{MDM}{Minimum Distance to Mean}
\newacronym{fgmdm}{FgMDM}{??? \acrshort{mdm}} % what is it an abreviation for?
\newacronym{lda}{LDA}{ Linear Discriminant Analysis}
\newacronym{tslr}{TSLR}{Tangent Space Logistic Regression}
\newacronym{tssvm}{TSSVM}{Tangent Space \acrshort{svm}}
\newacronym{svm}{SVM}{Support Vector Machine}
\newacronym{spd}{SPD}{Symmetric Positive Definite}
\newacronym{acm}{ACM}{Augmented Covariance Method}
\newacronym{blda}{BLDA}{Bayesian \acrshort{lda}}
\newacronym{dl}{DL}{Deep Learning}
\newacronym{ml}{ML}{Machine Learning}
\newacronym{mlp}{MLP}{Multi Layer Perceptron}
\newacronym{tc}{TC}{Temporal Convolution}
\newacronym{tcn}{TCN}{Temporal Convolutional Network}
\newacronym{dr}{DR}{Dimension Reduction}
\newacronym{snr}{SNR}{Signal-to-Noise Ratio}
\newacronym{svd}{SVD}{Singular Value Decomposition}
\newacronym{smd}{SMD}{Standardized Mean Difference}
\newacronym{itr}{ITR}{Information Transfer Rate}
\newacronym{bids}{BIDS}{Brain Imaging Data Structure}

\newcommand{\NbSubj}{\ensuremath{N}}
\newcommand{\NbDataset}{\ensuremath{D}}

\lstdefinestyle{moabbstyle}{
  basicstyle=\ttfamily\footnotesize,
  frame=lines,
  framesep=2mm,
}
\begin{document}

\title[The largest EEG-based BCI benchmark for open and reproducible science]{The largest EEG-based BCI reproducibility study for open science: the MOABB benchmark}

\author{Sylvain Chevallier, Igor Carrara, Bruno Aristimunha, Pierre Guetschel, Sara Sedlar, Bruna Lopes, Sebastien Velut, Salim Khazem, Thomas Moreau}
%\author{Anonymous authors}
%\author{Camille Noûs}

\address{Inria TAU, LISN-CNRS, Université Paris-Saclay, 91405, Orsay, France\\
Université Côte d’Azur, Inria Cronos Team, Sophia Antipolis, France\\
Donders Institute, Radboud University, Nijmegen, Netherlands\\
University of São Paulo, Sao Paulo, Brazil\\
Federal University of ABC, Santo Andre, Brazil \\
GeorgiaTech-CNRS IRL 2958, Centralesupelec, Metz, France\\
Inria Mind team, Université Paris-Saclay, CEA, Palaiseau, 91120, France}
%\address{Anonymous adress}

\ead{sylvain.chevallier@universite-paris-saclay.fr}
%\ead{anonymous@mail.com}
%\vspace{10pt}
%\begin{indented}
%\item[]March 2024
%\end{indented}

\begin{abstract}

\emph{Objective}. 
This study conduct an extensive Brain-computer interfaces (BCI) reproducibility analysis on open electroencephalography datasets, aiming to assess existing solutions and establish open and reproducible benchmarks for effective comparison within the field. The need for such benchmark lies in the rapid industrial progress that has given rise to undisclosed proprietary solutions. Furthermore, the scientific literature is dense, often featuring challenging-to-reproduce evaluations, making comparisons between existing approaches arduous.
\emph{Approach}. 
Within an open framework, 30 machine learning pipelines (separated into raw signal: 11, Riemannian: 13, deep learning: 6) are meticulously re-implemented and evaluated across 36 publicly available datasets, including motor imagery (14), P300 (15), and SSVEP (7). The analysis incorporates statistical meta-analysis techniques for results assessment, encompassing execution time and environmental impact considerations.
\emph{Main results}. 
The study yields principled and robust results applicable to various BCI paradigms, emphasizing motor imagery, P300, and SSVEP. Notably, Riemannian approaches utilizing spatial covariance matrices exhibit superior performance, underscoring the necessity for significant data volumes to achieve competitive outcomes with deep learning techniques. The comprehensive results are openly accessible, paving the way for future research to further enhance reproducibility in the BCI domain.
\emph{Significance}.
The significance of this study lies in its contribution to establishing a rigorous and transparent benchmark for BCI research, offering insights into optimal methodologies and highlighting the importance of reproducibility in driving advancements within the field.

\end{abstract}

\noindent{\it Keywords}: Brain-computer interface, EEG, Reproducibility, Riemannian classifier, Deep learning

\submitto{\JNE}
%\maketitle
\ioptwocol

\section{Introduction}
\label{sec:intro}

The field of \gls{bci} aims at developing methodologies to allow interactions with devices, like prostheses or computer environments, from decoding brain signals.
It is a very good candidate technology to assist people with motor disability, as it requires very limited motor capabilities of the subject.
To go from brain signals to decision-making, 
\gls{bci} systems are defined by several choices: a paradigm, that specifies the cognitive tasks to perform to control the interface, an acquisition device, to record the brain activity, and an algorithmic pipeline, that processes the acquired data and predicts which action the subject intends to perform.
As such, \gls{bci} is at the forefront of interdisciplinary research, integrating expertise from diverse fields such as electronics, neuroscience, human-machine interaction (HMI), signal processing, and machine learning.

\subsection{Open data}

To fasten the emergence of \gls{bci} system, the field has organized to decouple the lengthy paradigm design and data acquisition processes from the development of algorithmic pipelines to process them.
The creation and publication of many accessible and openly available datasets allow for offline development and evaluation of novel processing pipelines.
They also improve experiment replication and ensure the replicability of published results.

Due to the constraints on real-time acquisitions in open environments, \gls{eeg} has become the leading device to develop \gls{bci} systems with its high-frequency acquisition and limited constraints on its deployment.
In \gls{eeg}-based \gls{bci}, many real-world competitions and open datasets exist on a world scale to design and evaluate the best \gls{bci} systems.
\gls{bci} datasets come with various paradigms, depending on the decoding task to perform from brain signals. 
\gls{mi} is a common paradigm choice to control a \gls{bci}, with different imagined movements, but other paradigms are also very efficient like \gls{erp} generated by oddball stimulus or \gls{ssvep} produced by repetitive stimulations.
Many open datasets are thus openly available for \gls{eeg}-based \gls{bci}.

% Poor format
Yet, open data is not solely about availability.
It also requires the use of open formats that enable easy reading and exploitation of metadata.
Indeed, there exists a wide variety of \gls{eeg} acquisition devices with various hardware specifications, as well as a plethora of paradigm specifications, with their experimental design, shared annotations, and code for interpreting the cued events.
Retrieving this information is critical to developing valid systems that can correctly process the retrieved data.
Drawing inspiration from brain imaging techniques like fMRI or MEG that rely on the \gls{bids}, an extension of this format has been proposed for \gls{eeg}-based research~\cite{pernet2019eeg}. Despite this proposed format, most \gls{eeg} open data available online are stored in diverse structures and formats, which hinders the automation of data collection,  as each dataset requires specific processing scripts.

% Sometimes poor data
Moreover, while numerous books on \gls{eeg} data acquisition and \gls{bci} exist, it remains challenging to find well-founded guidelines for hardware requirements and \gls{bci} design.
These guidelines are needed to allow practitioners to make informed decisions regarding experimental design and hardware selection. In the worst-case scenario, a poorly designed \gls{bci} or an inappropriate hardware choice can result in unusable data. The unique characteristics of \gls{eeg}-based \gls{bci}s could be guided by meta-analyses of existing datasets, leading to recommendations on the number and positioning of electrodes, sampling frequency, the required number of subject trials, and the influence of the number of classes. However, to the best of our knowledge, there is no comprehensive and systematic evaluation of these parameters available in the existing literature.

\subsection{BCI pipelines}

Based on these open datasets, the development of BCI classification pipelines for EEG signals has a long-standing research history and a very active community~\cite{lotte2018review}.
Pipelines initially focused on methods based on feature extraction from raw signals, such as channel variance or local variation~\cite{lotte2007review}, combined with complex classifiers.
The first breakthrough came with the introduction of \emph{spatial filtering} methods based on covariance estimation, such as XDAWN~\cite{rivet2009xdawn}, CSP~\cite{ang2008filter}, or CCA~\cite{lin2006frequency}. These methods significantly improved classification accuracy when paired with simple classifiers. By using supervised filters, the separability of different BCI classes was improved while reducing the dimensionality of the input signal. As a result, these raw-signal-based pipelines with spatial filtering became the top-ranked approaches in BCI competitions.
The second advancement in BCI classification involved a reformulation of spatial filtering based on covariance matrices using \emph{Riemannian geometry}~\cite{barachant2010riemannian,yger2016riemannian}. Leveraging the inherent manifold structure of symmetric covariance matrices, this approach provides robust classification by exploiting the invariance properties of the covariance manifold. This method has demonstrated remarkable performance across various BCI tasks, even with noisy EEG signals, making it the preferred choice and quickly outperforming other pipelines in BCI competitions~\cite{congedo2017riemannian}.
Pipelines based on Riemannian geometry are considered state-of-the-art methods in the current literature.

Recently, \emph{deep learning methods} have been explored for EEG-based BCI classification~\cite{roy2019deep}. Although deep learning has enjoyed success in computer vision tasks, it is not straightforward to adapt these techniques to handle strongly correlated time series data, such as EEG signals.
As a result, the deep learning approaches performances on EEG signals have not outperformed existing approaches~\cite{schirrmeister2017deep}. Indeed, the main workhorse for the success of deep learning lies in the availability of vast amounts of data. However, in the context of BCI and EEG data, there is a scarcity of data at subject-level, which could potentially explain the limited performance of deep neural networks in this domain. To address this issue, incorporating auxiliary tasks such as self-supervised learning~\cite{banville2021uncovering} or data augmentation~\cite{rommel2022data} are promising approaches. However, further investigation is still required to explore the efficacy of these methods.

\subsection{Evaluation and Reproducibility in BCI}

While the literature on \gls{eeg}-based \gls{bci} pipelines is very dense, their evaluation and the interpretation of the results is a major issue for \gls{bci}.
Indeed, for many studies, it is very difficult, if not impossible to compare the produced results.
This stems from the fact that various factors hinder the experimental result comparison: specific preprocessing, cherry-picking datasets, subjects, or classes on a dataset, missing pipelines in the comparison, or lack of statistical analysis.

On the one hand, this issue is compounded by the need for shared resources and comprehensive evaluations.
Thorough evaluations often exceed the scope of a single research work, and it is thus crucial to find rigorous methods to assess the performance of BCI pipelines and ensure the reproducibility of their results.
On the other hand, the scientific community has been grappling with a reproducibility crisis across various domains~\cite{Baker2016}, and the field of \gls{bci} is not exempt. Addressing this crisis in \gls{bci} research is particularly pertinent due to the domain's unique requirements and complexities. \gls{bci} studies involve complex methodologies, intricate data acquisition techniques, and multifaceted analysis pipelines, making it challenging to replicate and validate experiments. The transdisciplinary nature of BCI research necessitates specialized knowledge from multiple disciplines, further complicating the ability to reproduce results.

Meta-analysis, a statistical technique for combining the results of multiple studies, is a powerful tool for synthesizing findings and deriving insights from a large body of research~\cite{hedges2014statistical}.
However, due to non-standardized evaluation protocols, conducting meta-analyses in the field of BCI has proven to be challenging. The variability in datasets, experimental tasks, and analysis pipelines inhibits the aggregation of results, despite efforts to establish standardization protocols. Consequently, it is difficult to obtain a comprehensive understanding of the performance of BCIs pipelines across different paradigms through the existing literature.

In response to these challenges, the field has seen the emergence of the Mother of All BCI Benchmarks (MOABB). MOABB~\cite{moabb-software} was developed as an open-source platform to facilitate the benchmarking and assessment of new datasets and classifiers in major BCI paradigms. The first version of MOABB initiated a community-driven effort to define rigorous and expert methods for conducting proper benchmarking assessments. By providing standardized evaluation procedures, MOABB enables researchers to compare and evaluate the performance of BCI methods in a transparent and reproducible manner. While its adoption by the community is broadening, the systematic evaluation of existing pipelines proposed in~\cite{jayaram2018moabb} is now outdated because it is focused on a specific BCI modality, and new pipelines have emerged in the literature.

\subsection{Environmental impact}
\label{sec:envimpact}

Addressing climate change requires comprehensive actions across all facets of human societies to adhere to the Paris Climate Agreement.
Research, including the field of AI, plays a significant role in achieving this objective.
Assessing the environmental impact of machine learning techniques is a challenging task~\cite{ligozat2022unraveling}, which has been pioneered by the natural language processing community with the prominence of conversational agents~\cite{luccioni2023estimating}.
While machine learning models used in BCI are smaller in scale, assessing the performance of various models in link with their environmental impact is critical to promoting virtuous and sustainable models.

The costs associated with deploying computers or computer clusters for training machine learning models are dependent on infrastructure requirements. The environmental impact resulting from energy consumption during model training varies based on geographical criteria, as electricity production methods differ across countries. Several libraries have been developed to provide estimates of this environmental impact, measured in terms of grams of CO2 equivalent emissions~\cite{jay2023experimental}. Although these libraries have limitations, they offer a valuable measurement to enhance our understanding of the training requirements for models. This facilitates a more comprehensive comparison of pipelines within a benchmark setting.

\subsection{Contributions}

This paper aims to address the lack of reproducibility studies in the field of EEG-based BCI and to go beyond a simple benchmark of existing machine learning pipelines. Indeed, it is important to provide an updated comparison of the previous benchmark~\cite{jayaram2018moabb}, including deep learning pipelines. Also, new BCI paradigms for controlling systems could now be included in the evaluation to propose a global take on the current BCI state.

It is an impossible task to evaluate all openly available datasets and all published pipelines in the literature, as there are exotic or unreadable data formats, BCI protocols that are used for a unique dataset, and unavailable code. For this paper, we chose to restrict to three common BCI paradigms, namely \gls{mi}, \gls{erp}, and \gls{ssvep}, that are well documented in the literature. The reason for this choice is to ensure reproducible evaluation with enough datasets and subjects to achieve decent statistical meta-analysis. 
% We thus excluded interesting BCI datasets that use a protocol or a task that is unique

For the machine learning pipelines, we follow similar guidelines to consider only approaches that could be reimplemented with Python open-source libraries. Indeed, we could only reimplement part of the published pipelines, but we try to include to the best of our effort all pipelines that have been reused in several publications or that are often cited as reference. The MOABB framework is designed to allow seamless integration of novel pipelines for new contributors and facilitate as much as possible the reproducibility of a benchmark to compare a novel pipeline with the one presented here.

We thought of the MOABB open science initiative as a long-term project. The objective is to endow the community with the ability to easily compare results on new datasets or new pipelines, that will be added after the paper publication. A website reproduces the results presented here and will be updated with novel additions. 
% TODO: add web link to the MOABB doc dedicated page
The goal is also to limit the environmental impact of research works by providing a simple means for scientists to copy/paste up-to-date results in their publication to compare with a reference benchmark.

Beyond the purely quantitative benchmark, it is difficult to have a good overview of the open EEG datasets available online. They have been recorded with different hardware, using similar experimental protocols with some variability in experiment design. As data sharing is becoming a common requirement in the scientific community, future works will often result in sharing new open data. It is important to be able to quickly identify existing datasets, and common design choices, to correctly position new experiments with well-informed knowledge. This task is difficult, as open data is scattered in the literature with no common format.

To summarize, in this paper we aim to address the following open questions:
\begin{itemize}
    \item What is the most effective approach for classifying EEG? How do their computation time and energy consumption compare for training a model?
    \item What is the best deep learning method? The best Riemannian-BCI pipeline?
    \item How many trials or channels are required in a dataset to achieve correct performances?
    \item For \gls{mi}, which motor imagery tasks give the best accuracy?
    \item What are the open datasets in different paradigms and how do they compare?
\end{itemize}

To investigate those research questions, we conducted many experiments built on open-source tools and with the help of a large community driven by open science guidelines. This paper describes the results obtained from a wide experimental campaign and their in-depth analysis, resulting in the following contributions:
\begin{itemize}
    \item the largest benchmarking in \gls{eeg}-based \gls{bci} in open science relying on open source libraries
    \item a fair and replicable evaluation with expert knowledge in \gls{bci}
    \item a deep analysis of the benchmark results, with guidelines for proposing new machine learning pipelines and new datasets or experiments.
\end{itemize}

\section{Benchmark methodology}

In this section, we describe the methodology for our benchmark, including which methods are considered, how they are trained and evaluated, and how we analyze the results.

\subsection{Analysis pipelines inclusion}

A major difficulty in the \gls{bci} literature, apart from the data and code availability, is the importance of signal processing in \gls{eeg}.
Many toolboxes are available, in different software platforms, like Matlab, Python, C/C++, Java, C\#, Julia, Delphi, and many more. Some toolboxes are sold as commercial products, with undisclosed code or signal processing techniques that are hidden or obfuscated for intellectual property reason. The choices made in those toolboxes, let apart all single projects maintained by only one person, are very different regarding signal filtering or electrode referencing and yield very different outcomes. Despite the fact that complex preprocessing pipelines, and often arguably overcomplicated ones, are detrimental for \gls{eeg} interpretation and classification~\cite{delorme2023eeg}, most of the toolboxes include them and promote their application in tutorials and guidelines.

In this paper, our analysis pipeline relies on the Python language for its large adoption in the scientific computing, neuroscience and machine learning communities, supported by robust and extensively validated libraries such as numpy~\cite{harris2020array}, scipy~\cite{2020SciPy-NMeth}, MNE~\cite{GramfortEtAl2013a}, scikit-learn~\cite{scikit-learn} and pyriemann~\cite{pyriemann}.
Manipulation of \gls{eeg} recordings is facilitated through MNE, enabling the extraction of hardware events and the conversion of recordings into numpy arrays.
As a light preprocessing, the signal is processed with a 4th-order Butterworth bandpass IIR filter, applied with a forward-backward pass, using standard MNE parametrization.
The specific bandpass frequencies are subsequently provided as they depend on the chosen BCI paradigm.
Machine learning pipelines are based on scikit-learn and pyriemann estimators.

%\section{Experiments} 
%What we choose: within session, this is a common choice for offline analysis, but is very optimistic w.r.t. results obtained in online application. A possible mitigation is to use calibration/test split in the data, that is a virtuous approach~\cite{schirrmeister2017}.

\subsection{Evaluation method}

Another difficulty for anyone who wants to reproduce literature results is that reporting classification performances on \gls{eeg}-based BCI tasks is not standardized. A crucial methodological consideration pertains to the selection of the evaluation metric (whether it be \gls{auc}, precision/recall, F1-score, or accuracy), a decision that holds notable significance in the assessment of outcomes. Moreover, in many studies, the authors focus on specific subsets of subjects within established datasets, or selectively use a restricted part of the cognitive tasks conducted during experiments.
These choices make comparing findings across papers becomes inherently cumbersome.

To maintain consistent terminology concerning EEG signals, we will establish the following definitions. 
A \emph{session} refers to a series of runs where EEG electrodes remain attached to a subject's head, and the overarching experimental parameters remain constant.
The term \emph{run} denotes the period during which an experiment is conducted without interruption or pause. Within a session, several runs are performed with potential intervals between them.
An \emph{epoch} or a \emph{trial} signifies a segment within a run during which an atomic event occurs, triggered by an external stimulus for event-related potential or steady-state evoked potential or internal volition for motor imagery. These epochs are positioned in time relative to an onset, which corresponds to the start of a stimulus or a task. 
% A trial is a part of a run, that is equivalent to an epoch for motor imagery and steady-state evoked potentials. In the case of event-related potential, period of a run during which the subject tries to send one command to the computer. In the cases of \gls{mi} and \gls{ssvep}, a trial is equivalent to an epoch. In the case of \gls{erp}, a trial is composed of multiple consecutive epochs, and the binary labels of each epoch (target or non-target) must be interprested together to decode which command was target by the subject.

% Comment on Figure:
%- Alpha for the color of training and test
%- Show the Nested approach
%- Is better to keep only WS
%\begin{figure}[t]
%    \centering
%    \includegraphics[scale=.73]{figures/evaluations/within-session.pdf}
%    \caption{\textbf{OLD} Within-session evaluation}
%    \label{fig:within-session}
%\end{figure}

\begin{figure}[t]
    \centering
    \includegraphics[scale=.73]{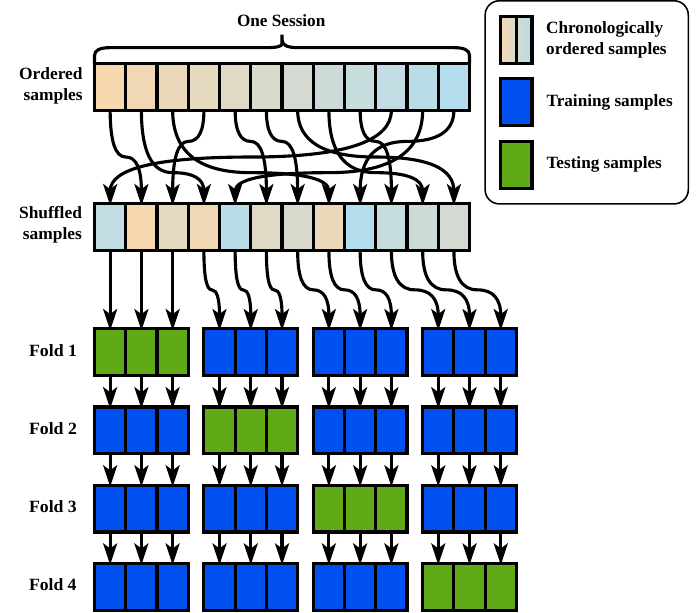}
    \caption {Within-session evaluation, small rectangles indicate a sample or \gls{eeg} trial, pastel colors on the two top lines shows the chronological order, bright color on the last three lines indicates training and testing samples/trials.}
    \label{fig:within-session}
\end{figure}

In the BCI framework, different evaluation methodologies exist for partitioning the data into training and test datasets, each tailored to address specific challenges. We differentiate between \emph{within-session} as shown in~\autoref{fig:within-session}, \emph{cross-session}, and \emph{cross-subject} evaluations, respectively illustrated in the annexes in~\autoref{fig:cross-session} and~\autoref{fig:cross-subject}.

In the context of within-session evaluation, the primary concern lies in identifying algorithms that can effectively mitigate overfitting within a single session.
In line with common practices in machine learning for cross-validation, all trials from a session are shuffled before being split in k-folds, to evaluate the generalization performance on unseen epochs. Consequently, the pipelines are trained on trials sampled throughout the session duration, which helps mitigate the impact of intra-session variability in an individual's EEG. While this allows a more statistically accurate benchmark of a pipeline, it provides an upper bound for the evaluation metrics when compared to online evaluation. Despite the difference with experimental applications of BCI system, this training methodology is commonly employed in the existing literature~\cite{barachant2011multiclass,tangermann2012review,nakanishi2015comparison,lotte2018review}, influencing our decision to adopt it for this reproducibility study.

In contrast, the cross-session (resp. cross-subject) evaluation employs a leave-one-out cross-validation technique, where only one session (resp. subject) is designated as the testing dataset.
The results from cross-session or cross-subject evaluation methods put more emphasis on transfer learning to cope with subjects' variability.
However, the questions raised by transfer learning in a BCI context~\cite{transferEEG,wei2022beetl} are manyfold -- dealing with subject alignment, training models for each subject or a single one, at a session, subject or dataset level -- and are outside the scope of this article. 
For this reason, this paper focuses on the most common evaluation strategy in the literature, which is within-session evaluation.

Nonetheless, one should acknowledge that within-session evaluation has its limitations. It addresses only partly the complications stemming from variations in sessions and subjects. Such variations may arise from internal factors, like minor electrode displacements between sessions, different calibrations of EEG hardware, or external factors, like the dynamic nature of EEG measurements in an individual based on the cognitive states~\cite{roy2022neuroergonomics}, such as alertness, drowsiness or fatigue.

Within-session evaluation of a pipeline is conducted for each subject and each session by splitting the epochs into training and test epoch sets using five stratified K-fold splits. In each fold, the testing set contains $~20 \%$ of the session epochs while maintaining the class distribution. The final evaluation score corresponds to the average over the five splits. Depending on whether the classification problem is binary or involves more than two classes, evaluation scores correspond to the \gls{roc-auc} or classification accuracy, respectively.

\subsection{Grid search} 

Another important aspect in \glspl{bci} is how the hyperparameters are selected.
They can significantly affect the performance of the algorithm and, consequently, the accuracy of the system's predictions. It is therefore crucial to employ a method that facilitates the search for optimal parameter values. Grid search stands out as a popular method for hyperparameter tuning in machine learning algorithms. In particular, it is essential to search for the correct hyper-parameters for each scenario, considering variations in evaluation procedures, datasets, subjects, and sessions. %different evaluation procedures, different datasets, different subjects, different sessions.

Using nested cross-validation for hyperparameters selection, we mitigate the risk of overfitting, as recommended in existing literature~\cite{cawley2010over}. This approach is implemented by using an inner 3-fold cross-validation.
Additionally, we have devised tailored grid search functions for each evaluation procedure, as detailed in~\autoref{app:gridsearch}.
This standardized framework streamlines hyperparameter tuning, enabling seamless performance comparison across diverse machine-learning models and promoting experiment reproducibility across varied datasets.

\subsection{Statistical analysis} 

% Within-session evaluation of a pipeline is conducted for each subject and each session by splitting the epochs into training and test epoch sets using five stratified K-fold splits. In each fold, the testing set contains $~20 \%$ of the session epochs while maintaining the class distribution. The final evaluation score corresponds to the average over the five splits. Depending on whether the classification problem is binary or involves more than two classes, evaluation scores correspond to the \gls{roc-auc} or classification accuracy, respectively. 

The statistical comparison of two different pipelines can be conducted either at a \textit{dataset-wise} level or across \textit{multiple datasets}.

The \textit{dataset-wise} comparison involves assessing the statistical differences between two pipelines within the same dataset. This is done using effect sizes and p-values. To optimize computational efficiency, the number of subjects $\NbSubj$ in the dataset determines the method of estimating p-values. For datasets with $\NbSubj < 13$, one-tailed permutation-based paired $t$-tests are used with all possible permutations. For datasets with $13 \leqslant \NbSubj \leqslant 20$, 10000 random permutations are employed. For datasets with $\NbSubj > 20$, the Wilcoxon signed-rank test~\citep{wilcoxon1992individual} is used. 
The effect size between two pipelines is measured via \gls{smd}~\citep{hedges2014statistical} estimated over the subjects.

For \textit{multiple datasets}, the comparison of statistical differences between two pipelines over \NbDataset{} is done by combining effect sizes $\left\{s_i\right\}_{i=1}^\NbDataset$ and p-values $\left\{p_i\right\}_{i=1}^\NbDataset$ with Stouffer’s Z-score method~\citep{rmj1949american} that take in account the different sizes of the datasets. Combined p-values are obtained by estimating $Z = \sum^\NbDataset_{i=1} w_i Z_i$, where $Z_i = \Phi^{-1}(1-p_i)$ with $\Phi$ the standard normal cumulative distribution function. This weighted Z-score relies on weights $w_i = \sqrt{\frac{\NbSubj_i}{\sum^\NbDataset_{i=1} w^2_i}}$ that are proportional to $\NbSubj_i$ the number of subjects in the dataset $i$.
% the inverse survival function for the datasets' p-values, followed by their weighted average, where the weights correspond to $\sqrt{\NbSubj}$ the square roots of the number of subjects normalized by the weights' second norm, and at the end the combined p-value is equal to the survival function for the obtained the weighted average.
Combined measures of the effect sizes are obtained by $S = \sum^\NbDataset_{i=1} w_i s_i$ weighted average of \gls{smd}.
%, where the weights correspond to the square roots of the number of subjects normalized by the weights' first norm.

\subsection{Code Carbon} 

The assessment of the environmental impact of research is gathering momentum, yet the methodology remains a topic of heavy debate, particularly in computer science and machine learning~\cite{ligozat2022unraveling}.

Within these fields, a significant environmental impact stems from the production processes of devices (such as acquisition devices, computers, GPUs, and clusters) and the energy consumption during their usage. Evaluating the impact of device production poses challenges, given the limited information shared by manufacturers, potential inaccuracies in estimations, and the frequent sharing and reusing of research equipment across multiple projects. This issue is notably prevalent for CPU and GPU clusters, where numerous experiments run in parallel~\cite{luccioni2023estimating}.

Energy consumption during usage is commonly acknowledged as a key indicator of the environmental impact, although it represents only a fraction of the overall impact. This measurement is typically expressed as grams of CO2 equivalent (gCO2) emissions released into the atmosphere and heavily relies on factors such as the power grid setup and energy production localization. National power companies typically offer estimates of the carbon footprint of consumed energy, articulated in gCO2 per Watt-hour. This value can significantly vary between countries based on energy production sources; for instance, countries reliant on coal or fuel power plants tend to have a higher carbon footprint compared to those utilizing hydroelectric power or solar panels.

For optimal measurement of energy consumption and carbon footprint, power meters are ideal, albeit encompassing the entire computer's energy consumption. To gain more precise insights, especially for evaluating specific programs or machine learning pipelines, software-based power meters prove useful~\cite{jay2023experimental}. Various tools exist for this purpose, many of which are open source. In our study, we opted to employ Code Carbon~\cite{codecarbon}, a Python-based tool that seamlessly integrates into experiments conducted on individual computers or clusters.

When estimating the carbon footprint of a machine learning pipeline, it is critical to analyze both the training and inference phases. During k-fold validation, inference constitutes a substantial portion of energy consumption. Consequently, the estimated carbon footprint offers valuable insights into the trade-offs between different algorithms, with considerations for CPU consumption, execution time, and the parallelization of algorithms. Notably, well-parallelized algorithms can achieve lower execution times when distributed across a large number of CPUs. The carbon footprint serves as an insightful metric for assessing the computational complexity of a pipeline, particularly when considering its adaptability to embedded or dedicated architectures, common in commercial neurotechnology products.

\section{Datasets} 

There are different types of \gls{bci} applications, referred to as \emph{paradigms} throughout this article, each relying on distinctive neurological patterns to facilitate communication between the brain and the BCI system. 
Depending on the selected paradigm, the raw EEG recordings are transformed into trials for machine learning pipelines.
These transformations include bandpass filtering, signal cropping based on stimulus events, and potential resampling to adjust the sampling frequency if needed by the machine learning pipelines. 
Employing standardized preprocessing procedures enables a fair comparison among different algorithms. 
%In \gls{moabb}, we have paradigm classes corresponding to the standard \gls{bci} paradigms.
% These classes encapsulate the process of transforming continuous data (i.e. raw recordings) into trials for machine learning pipelines.
% This is especially critical in \gls{eeg} and biosignal processing, where datasets frequently contain diverse events within the continuous data. 
%The \gls{moabb} library also allows picking only certain channels at this stage, but we kept them all in our experiments.
%Organising the pre-processing by paradigm allows for having reasonable parameters for each of them.

\subsection{Motor Imagery} 
\label{sec:mi}
The \gls{mi} paradigm involves a cognitive process where an individual mentally simulates the execution of a motor action without actually performing it. This paradigm is widely used in neuroscience to delve into the neural mechanisms governing motor control and learning.  Moreover, it finds applications in neurorehabilitation to assist in restoring motor functions for individuals with neurological disorders or injuries. 
Various tasks are associated with this paradigm in \gls{moabb}, with common examples including left-hand, right-hand, and feet imagery. The choice of evaluation metrics for classification performance varies based on the number of tasks involved in the classification process: \gls{roc-auc} is used for two-task classifications, whereas accuracy metrics are employed for multiclass scenarios.
With the \gls{mi} paradigm, signal processing includes bandpass filtering within the $[8-32]$\,Hz frequency range~\cite{nam2018brain}. 

The \autoref{table:MI} provides a comprehensive overview of \gls{mi} datasets, with class name abbreviations including RH (Right Hand), LH (Left Hand), F (Feet), H (Hands), T (Tongue), R (Resting State), LHRF (Left Hand Right Foot), and RHLF (Right Hand Left Foot). The column 'No. trials' denotes the number of trials per class, session, and run. For instance, BNCI2014\_001 comprises 12 trials per class across 4 classes, 6 runs, and 2 sessions, resulting in a total of $12\times4\times6\times2=576$ trials in the dataset. The only exception is for the PhysionetMI dataset, where in the first 3 runs, there are RH, LH, and R events; in the last 3 runs, there are H, F, and R events.

\begin{table*}[htb!!] 
  \caption{Overview of the \acrlong{mi} \gls{eeg} datasets available in \gls{moabb}.}
  \label{table:MI}
  \centering
  \resizebox{\linewidth}{!}{
    \begin{tabular}{|p{4.6cm}|c|c|c|c|c|c|c|c|p{1.5cm}|}
        \hline
        \rowcolor{SeaGreen!50}\multicolumn{10}{|c|}{\textbf{Motor imagery}}\\
        \hline
        \rowcolor{SeaGreen!10} \textbf{Dataset} & \makecell{No.\\subj.} & \makecell{No.\\ch.} & \makecell{No.\\classes} & \Gape[0pt][2pt]{\makecell{No. trials\\/session/class}} & \makecell{Trial\\len.(s)} & \makecell{S.freq.\\ (Hz)} & \Gape[0pt][2pt]{\makecell{No.\\sessions}} & \makecell{No.\\runs} & \makecell{Classes} \\
        \hline
        \rowcolor{SeaGreen!20} \textbf{AlexMI}~\scriptsize{\citep{barachant2012commande}} & 
        8 & 16 & 2(3) & 20 $\pm$ 0 & 3 & 512 & 1 & 1 & \scriptsize{RH, F, (R)}\\
        \hline
        \rowcolor{SeaGreen!10} \textbf{BNCI2014\_001}~\scriptsize{\citep{tangermann2012review}}& 9 & 22 & 3(4) & 72 $\pm$ 0 & 4 & 250 & 2 & 6 & \scriptsize{RH, LH, F, (T)}\\
        \hline
        \rowcolor{SeaGreen!20} \textbf{BNCI2014\_002}~\scriptsize{\citep{steyrl2016random}}& 14 & 15 & 2 & 80 $\pm$ 0 & 5 & 512 & 1 & 8 & \scriptsize{RH, F}\\
        \hline
        \rowcolor{SeaGreen!10} \textbf{BNCI2014\_004}~\scriptsize{\citep{leeb2007brain}} & 9 & 3 & 2 & 72.4 $\pm$ 9.5 & 4.5 & 250 & 5 & 1 & \scriptsize{RH, LH} \\
        \hline
        \rowcolor{SeaGreen!20} \textbf{BNCI2015\_001}~\scriptsize{\citep{faller2012autocalibration}}& 12 & 13 & 2 & 100 $\pm$ 0 & 5 & 512 & \makecell{3 subj. 8-11\\2 others} & 1 & \scriptsize{RH, F}\\
        \hline
        \rowcolor{SeaGreen!10} \textbf{BNCI2015\_004}~\scriptsize{\citep{scherer2015individually}}& 9 & 30 & 2(5) & 39.4 $\pm$ 1.6 & 7 & 256 & 2 & 1 & \scriptsize{RH, F}\\
        \hline
        \rowcolor{SeaGreen!20} \textbf{Cho2017}~\scriptsize{\citep{cho2017eeg}}& 
        52 & 64 & 2 & 101.2 $\pm$ 4.7 & 3 & 512 & 1 &  1 & \scriptsize{RH, LH}\\
        \hline
        \rowcolor{SeaGreen!10} \textbf{Lee2019\_MI}~\scriptsize{\citep{lee2019eeg}}& 
        54 & 62 & 2 & 50 & 4 & 1000 & 2 & 1 & \scriptsize{RH, LH} \\
        \hline
        \rowcolor{SeaGreen!20} \textbf{GrosseWentrup2009}~\scriptsize{\citep{grosse2009beamforming}}& 10 & 128 & 2 & 150 $\pm$ 0 & 7 & 500 & 1 & 1 & \scriptsize{RH, LH} \\
        \hline
        \rowcolor{SeaGreen!20} \textbf{PhysionetMI}~\scriptsize{\citep{schalk2004bci2000}}& 109 & 64 & 4(5) & 22.6 $\pm$ 1.3 & 3 & 160 & 1 & 6*** & \scriptsize{RH, LH, H, F, (R)} \\
        \hline
        \rowcolor{SeaGreen!10} \textbf{Schirrmeister2017}~\scriptsize{\citep{schirrmeister2017deep}}& 14 & 128 & 3(4) & 240.8 $\pm$ 37.7 & 4 & 500 & 1 & 2 & \scriptsize{RH, LH, F, (R)} \\
        \hline
        \rowcolor{SeaGreen!20} \textbf{Shin2017A}~\scriptsize{\citep{shin2016open}} & 29 & 30 & 2 & 10 $\pm$ 0 & 10 & 200 & 3 & 1 & \scriptsize{RH, LH}\\
        \hline
        \rowcolor{SeaGreen!20} \textbf{Weibo2014}~\scriptsize{\citep{yi2014evaluation}}& 10 & 60 & 4(7) & 79 $\pm$ 3 & 4 & 200 & 1 & 1 & \scriptsize{RH, LH, H, F, (LHRF), (RHLF), (R)}\\
        \hline
        \rowcolor{SeaGreen!10} \textbf{Zhou2016}~\scriptsize{\citep{zhou2016fully}} & 4 & 14 & 3 & 50 $\pm$ 3.5 & 5 & 250 & 3 & 2 & \scriptsize{RH, LF, F}\\
        \hline
\end{tabular}}
\end{table*}

\subsection{P300/ERP}
The P300 paradigm, which is a specific \gls{erp} paradigm, serves as a framework for categorizing psychophysical experiments \cite{luck_2014}. It provides a visual representation of a distinct component characterized by a prominent positive deflection occurring approximately 300 ms after stimulus onset. Typically, this component is elicited in cognitive tasks involving unpredictable and infrequent changes in stimuli.
This is improperly called P300 \gls{erp} in the \gls{bci} community (see P300 \gls{bci} or P300 speller), whereas the cognitive components are more diverse than just the P300 wave.
See~\cite{luck_2014} for a detailed discussion about this point.

In the present study, our primary focus is on P3b, a specific subtype that examines stimulus changes relevant to the task at hand. This particular brain wave response manifests when the cognitive tasks involve stimuli that are predictable to some extent but are still imbued with an element of unpredictability. 
As is commonly the case in the literature, we classify events as targets or non-targets, \emph{i.e.}, at the epoch level, resulting in a binary classification task. We evaluate its performance with the \gls{roc-auc} metric, which handles the inherent imbalance in the problem at hand.
With the \gls{erp} paradigm, the signal is bandpass filtered to the 1-24\,Hz frequency band~\cite{luck_2014}.

The \autoref{table:P300} shows an overview of all the \gls{erp} datasets. The column ``No. epochs NT/T'' indicates the number of \textit{NonTarget} and \textit{Target} epochs per session and run.

\begin{table*}[t]
    \caption{Overview of the \gls{erp} \gls{eeg} datasets available in \gls{moabb}.}
    \label{table:P300}
    \centering
  \resizebox{\linewidth}{!}{
    \begin{tabular}{|p{4.2cm}|c|c|c|c|c|c|c|c|}
        \hline
        \rowcolor{Salmon!50}\multicolumn{9}{|c|}{\textbf{P300 / ERP}}\\
        \hline
        \rowcolor{Salmon!10} \textbf{Dataset} & \makecell{No.\\subj.} & \makecell{No.\\ch.} & \Gape[0pt][2pt]{\makecell{No. epochs NT/T\\/session}} & \makecell{Epoch\\len.(s)} & \makecell{S.freq.\\ (Hz)} & \Gape[0pt][2pt]{\makecell{No.\\sessions}} & \makecell{No.\\runs} &\makecell{Keyboard}\\
        \hline
        \rowcolor{Salmon!20} \textbf{BI2012}~\scriptsize{\citep{van2019building}}& 25 & 16 & 638.2 $\pm$ 1.9/127.6 $\pm$ 0.7 & 1 & 128 & 1  & 1 &36 aliens\\
        \hline
        \rowcolor{Salmon!10} \textbf{BI2013a}~\scriptsize{\citep{vaineau2019brain,barachant2014plug,congedo2011brain}}& 24 & 16 & 400.3 $\pm$ 2.3/80.1 $\pm$ 0.5 & 1 & 512 & \makecell{8 subj. 1-7\\1 subj. 8-24} & 1 & 36 aliens\\
        \hline
        \rowcolor{Salmon!20} \textbf{BI2014a}~\scriptsize{\citep{korczowski2019brain_a}}& 64 & 16 & 794.5 $\pm$ 276.7/158.9 $\pm$ 55.3 & 1 & 512 & 1 & 1 &36 aliens\\
        \hline
        \rowcolor{Salmon!10} \textbf{BI2014b}~\scriptsize{\citep{korczowski2019brain_b}}& 37 & 32 & 201.3 $\pm$ 61.5 /40.3 $\pm$ 12.3 & 1 & 512 & 1 & 1 &36 aliens\\
        \hline
        \rowcolor{Salmon!20} \textbf{BI2015a}~\scriptsize{\citep{korczowski2019brain_2015_a}}& 43 & 32 & 461.8 $\pm$ 220.9/92.3 $\pm$ 44.1 & 1 & 512 & 3 & 1 & 36 aliens\\
        \hline
        \rowcolor{Salmon!10} \textbf{BI2015b}~\scriptsize{\citep{korczowski2019brain_2015_b}}& 44 & 32 & 2158.7 $\pm$ 6.3/479.9 $\pm$ 0.3 & 1 & 512 & 1 & 4 & 36 aliens\\
        \hline
        \rowcolor{Salmon!20} \textbf{BNCI2014\_008}~\scriptsize{\citep{riccio2013attention,farwell1988talking}}& 8 & 8 & 3500 $\pm$ 0/700 $\pm$ 0 & 1 & 256 & 1 & 1 & 36 char.\\
        \hline
        \rowcolor{Salmon!10} \textbf{BNCI2014\_009}~\scriptsize{\citep{arico2014influence}}& 10 & 16 & 480 $\pm$ 0/96 $\pm$ 0 & 0.8 & 256 & 3 & 1 & 36 char.\\
        \hline
        \rowcolor{Salmon!20} \textbf{BNCI2015\_003}~\scriptsize{\citep{guger2009many}}& 10 & 8 & 2250 $\pm$ 1500 /270 $\pm$ 60  & 0.8 & 256 & 1 & 2 & 36 char.\\
        \hline
        \rowcolor{Salmon!10} \textbf{EPFLP300}~\scriptsize{\citep{hoffmann2008efficient}}& 8 & 32 & 685.2 $\pm$ 16.9/137.2 $\pm$ 3.5 & 1 & 2048 & 4 & 6 & 6 images\\
        \hline
        \rowcolor{Salmon!20} \textbf{Huebner2017}~\scriptsize{\citep{hubner2017learning}}& 13 & 31 & 3275.3 $\pm$ 2.1 /1007.8 $\pm$ 0.6 & 0.9 & 1000 & \Gape[0pt][2pt]{\makecell{2 subj. 6\\3 others}} & 9 & 42 char.\\
        \hline
        \rowcolor{Salmon!10} \textbf{Huebner2018}~\scriptsize{\citep{huebner2018unsupervised}}& 12 & 31 & 3638.4 $\pm$ 7.7 /1119.6 $\pm$ 2.5 & 0.9 & 1000 & 3 & 10 & 42 char.\\
        \hline
        \rowcolor{Salmon!20} \textbf{Lee2019\_ERP}~\scriptsize{\citep{lee2019eeg}}& 54 & 62 & 3450/690 & 1 & 1000 & 2 & 1 & 36 char.\\
        \hline
        \rowcolor{Salmon!10} \textbf{Sosulski2019}~\scriptsize{\citep{sosulski2019spatial,sosulski2021online}}& 13 & 31 & 75 $\pm$ 0 /15 $\pm$ 0 & 1.2 & 1000 & \Gape[0pt][2pt]{\makecell{4 subj. 1\\3 others}} & 20 & 2 tones\\
        \hline
        \rowcolor{Salmon!20} \textbf{Cattan2019\_VR}~\scriptsize{\citep{cattan2019dataset}}& 21 & 16 & 600 $\pm$ 0/120 $\pm$ 0 & 1 & 512 & 2 & 60 & 36 crosses\\
        \hline
    \end{tabular}}
    \end{table*}

\subsection{SSVEP}

Steady State Visually Evoked Potentials are generated when presenting repetitive sensory stimuli to the subject. While tactile and auditory stimulations are seldom used, visual stimulation is very common, both for control~\cite{chevallier2018brain} or cognitive probes~\cite{wu2007influence}. The frequency of the stimulus repetition induces a brain oscillation in the associated sensory area. The amplitude of the generated oscillation follows the 1/f law, meaning that stimulation in low frequency (5-7  Hz) induces responses of higher amplitude than higher frequency (20-25 Hz). Stimulation above 40 Hz could be difficult to detect due to the weak generated oscillations. In \gls{bci} applications, \glspl{ssvep} have been used for building spellers~\cite{nakanishi2014enhancing} and for button-pressing~\cite{kalunga2016online}, but those applications are limited by the number of available frequencies of stimulation. It is possible, using systems with precisely synchronized stimulation and recording, to encode information both in frequency and phase, therefore multiplying the choices possible~\cite{nakanishi2017enhancing}.
Similarly to the \gls{mi} paradigm, we evaluate the classifiers using the \gls{roc-auc} metric if only two classes are used and the accuracy metric if there are more.
With the \gls{ssvep} paradigm, the signal is bandpass filtered to the 7-45\,Hz frequency band~\cite{chevallier2018riemannian}.
The \autoref{table:SSVEP} encompasses all the \gls{ssvep} datasets considered in this study.

\begin{table*}[t]
    \caption{Overview of the \gls{ssvep} \gls{eeg} datasets available in \gls{moabb}.
    \label{table:SSVEP}}
    \centering
    \resizebox{\linewidth}{!}{
    \begin{tabular}{|p{4cm}|c|c|c|c|c|c|c|c|p{3.0cm}|}
        \hline
        \rowcolor{YellowOrange!50}\multicolumn{10}{|c|}{\textbf{SSVEP}}\\
        \hline
        \rowcolor{YellowOrange!10} \textbf{Dataset} & \makecell{No.\\subj.} & \makecell{No.\\channels} & \makecell{No.\\classes} & \Gape[0pt][2pt]{\makecell{No. trials\\/session/class}} & \makecell{Trial\\len.(s)} & \makecell{S.freq.\\ (Hz)} & \makecell{No.\\sess.} & \Gape[0pt][2pt]{\makecell{No. runs}} & \makecell{Classes} \\
        \hline
        \rowcolor{YellowOrange!20} \textbf{Lee2019\_SSVEP} \scriptsize{\citep{lee2019eeg}}& 54 & 62 & 4 & 25 & 1 & 1000 & 2 & 1 & 4 (5.45-12) \\
        \hline
        \rowcolor{YellowOrange!10} \textbf{MAMEM1} \scriptsize{\citep{oikonomou2016comparative}}& 10 & 256 & 5 & \Gape[0pt][2pt]{\makecell{16.8 $\pm$ 3.5 classes 8.57,10.0\\
        21.0 $\pm$ 4.4 classes 6.66,7.5,12.0}} & 3 & 250 & 1 & \makecell{3 subj. 1,3,8; 4 subj. 4,6\\
                    5 others} & 5 (6.66-12.00)\\
        \hline
        \rowcolor{YellowOrange!20} \textbf{MAMEM2} \scriptsize{\citep{oikonomou2016comparative}}& 10 & 256 & 5 & \Gape[0pt][2pt]{\makecell{20 class 12.0; 30 class 8.57\\25 others}} & 3 & 250 & 1 & 5 & 5 (6.66-12.00)\\
        \hline
        \rowcolor{YellowOrange!10} \textbf{MAMEM3} \scriptsize{\citep{oikonomou2016comparative}}& 10 & 14 & 4 & \makecell{20.0 $\pm$ 0.0 class 6.66; 25.0 $\pm$ 0.0 class 8.57\\
        30.0 $\pm$ 0.0 class 10.0; 25.0 $\pm$ 0.0 class 12.0} & 3 & 128 & 1 & 10 & 4 (6.66-12.00)\\
        \hline
        \rowcolor{YellowOrange!20} \textbf{Nakanishi2015} \scriptsize{\citep{nakanishi2015comparison}}&9 & 8 & 12 & 15.0 $\pm$ 0.0& 4.15 & 256 & 1 & 1 & 12 (9.25-14.75)\\
        \hline
        \rowcolor{YellowOrange!10} \textbf{Kalunga2016} \scriptsize{\citep{kalunga2016online}}& 12 & 8 & 4 & 20.0 $\pm$ 7.7 & 2 & 256 & 1 & \Gape[0pt][2pt]{\makecell{5 subj. 12; 4 subj.10\\3 subj. 7; 2 others}} & 4 (13,17,21,rest)\\
        \hline
        \rowcolor{YellowOrange!20} \textbf{Wang2016} \scriptsize{\citep{wang2016benchmark}}& 34 & 62 & 40 & 6.0 $\pm$ 0.0 & 5 & 250 & 1 & 1 & 40 (8-15.8)\\
        \hline
    \end{tabular}}
\end{table*}

\begin{figure*}
    \centering
    \includegraphics[width=1.0\linewidth]{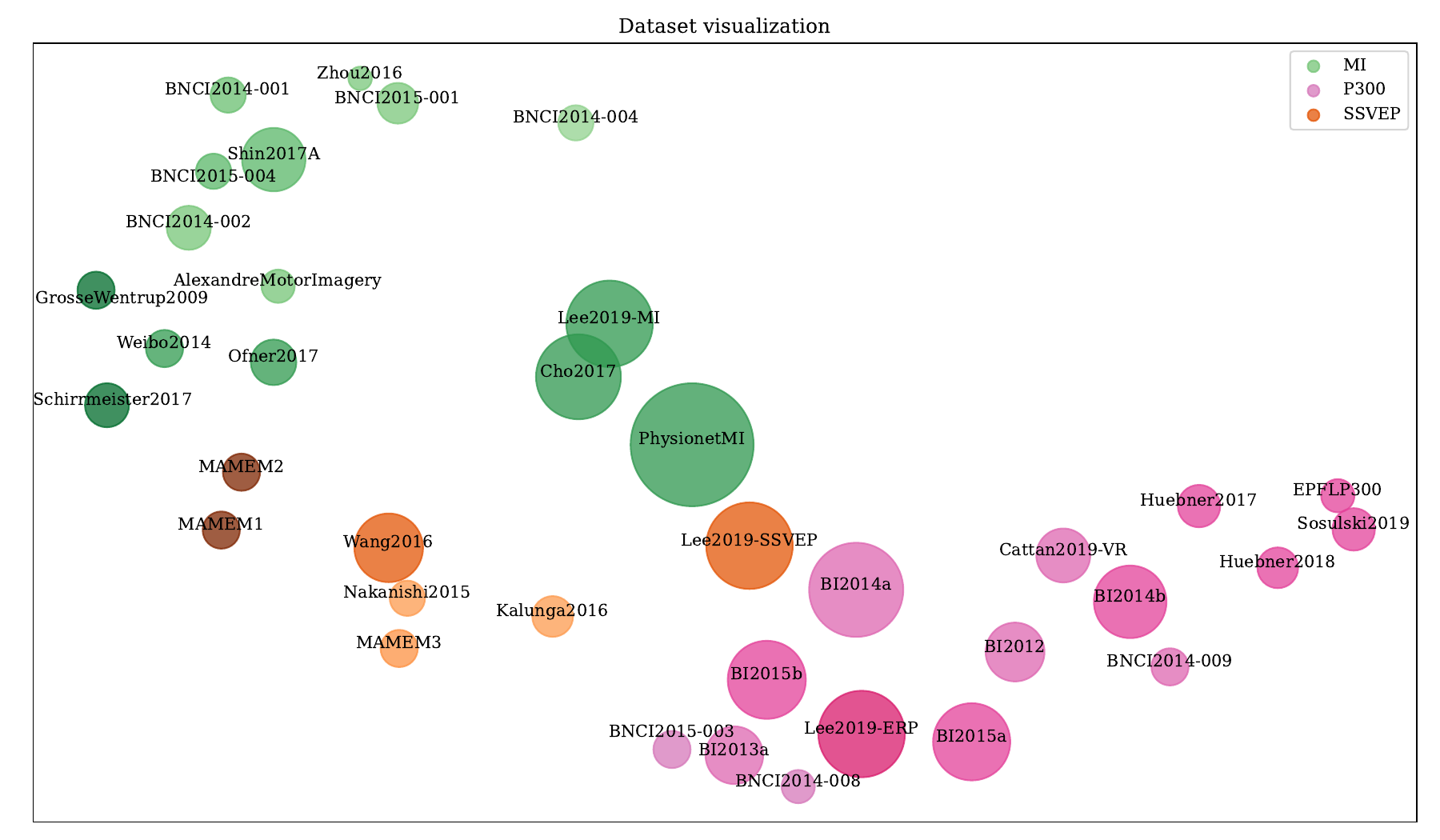}
    \caption{Visualization of the MOABB datasets, with \acrlong{mi} in green, \acrlong{erp} in pink/purple and \acrlong{ssvep} in yellow/brown. The size of the circle is proportional to the number of subjects and the contrast depends on the number of electrodes.}
    \label{fig:datasets}
\end{figure*}

The~\autoref{fig:datasets} displays an embedding of all the datasets in 2 dimensions, using UMAP for dimensionality reduction on all feature information regarding the datasets, as listed in Tables~\ref{table:MI},~\ref{table:P300} and~\ref{table:SSVEP}. 
The color indicates the paradigms (blue for \acrlong{mi}, green for \acrlong{erp}, and red for \acrlong{ssvep}), and the name of each dataset is written on top of the circle.
The color intensity is related to the number of electrodes, datasets with a low number of electrodes are in lighter color, and the size of the circles is proportional to the number of subjects.
The UMAP embedding preserves the local topology. This highlights that, despite different paradigms, datasets with many electrodes and many subjects are in a central position. Datasets with fewer subjects and fewer electrodes are closer to the border of the figure.
The BNCI2014\_001 dataset, commonly called BCI Competition IV dataset 2a, is the most widely used in BCI literature.
There are closely related datasets with roughly the same number of subjects, electrodes, and trials (BNCI2015\_004, BNCI2015\_001), with more subjects (Shin2017A) or with fewer subjects (Zhou2016).
This group of datasets is useful for the fast evaluation of new pipelines and to see how well results generalize with the same kind of dataset.
It is also possible to ensure a good coverage of the dataset features, using only a few datasets. With \acrshort{mi}, a selection of BNCI2014\_001, BNCI2014\_004, Schirrmeister2017, and PhysioNetMI might be sufficient to evaluate an approach on datasets with very different configurations.

\section{Pipelines}

As discussed in~\autoref{sec:intro}, most classification algorithms in \gls{bci} research for \gls{eeg} signals fall into one of three main categories: those based on raw signals (referred to as ``Raw'' hereafter), algorithms relying on covariance matrices seen as elements of a Riemannian manifold (denoted ``Riemannian'') and the \gls{dl} approaches.

The Raw signal methods typically employ supervised spatial filters to simultaneously enhance the component related to the cognitive task while reducing the dimensionality of the \gls{eeg} data. In contrast, Riemannian pipelines consider the signal through its estimated covariance matrices, leveraging the natural metric acting on the curved geometry of SPD matrices, which remains invariant by congruence transformations~\cite{yger2016riemannian, congedo2017riemannian}. Those approaches are thus mostly invariant to any spatial transformations applied on the signal, making them highly effective in \gls{bci} applications. 

Lastly, deep neural networks learn spatial and temporal filters directly from raw \gls{eeg} data. Although there is a wealth of literature on deep learning models, there are few models available or evaluated with reproducible frameworks. Initiatives like Braindecode~\cite{schirrmeister2017deep} or, to a lesser extent for \gls{bci}, torchEEG~\cite{torcheeg}, offer an open-source implementation of the most efficient models. %
This study considers a set of raw, Riemannian, and deep learning models, which are detailed in~\autoref{table:Pipeline}, along with the hyperparameters used for the grid-search approach, listed in the annexes within~\autoref{table:gridsearch_parameter}.

% TODO: add references for missing pipelines in table
\begin{table*}[!ht]
\begin{center}
\caption{Pipelines considered in this study, the color indicates the paradigm. Green is for motor imagery, pink for P300 and orange for SSVEP. 
%For more detail about the different pipelines please refer to the~\ref{app:pipelines}.
} 
\label{table:Pipeline} 
\footnotesize
\begin{adjustbox}{width=\textwidth}
\begin{tabular}{llr|llr}
\hline
\textbf{Pipeline Name} & \textbf{Category} & \textbf{References} & \textbf{Pipeline Name} & \textbf{Category} & \textbf{References} \\
\hline 
\rowcolor{SeaGreen!50} LogVar + LDA & Raw & & TS + LR & Riemannian & \cite{barachant2010riemannian} \\
\rowcolor{SeaGreen!50} LogVar + SVM & Raw & & TS + SVM & Riemannian & \cite{barachant2010riemannian} \\
\rowcolor{SeaGreen!50} CSP + LDA & Raw & \cite{koles1990spatial,blankertz2007optimizing} & ACM + TS + SVM & Riemannian & \cite{carrara2023classification} \\
\rowcolor{SeaGreen!50} CSP + SVM & Raw & \cite{koles1990spatial,blankertz2007optimizing} & ShallowConvNet & Deep Learning & \cite{schirrmeister2017deep} \\
\rowcolor{SeaGreen!50} TRCSP + LDA & Raw & \cite{lotte2010regularizing} & DeepConvNet & Deep Learning & \cite{schirrmeister2017deep} \\
\rowcolor{SeaGreen!50} DLCSPauto + shLDA & Raw & \cite{koles1990spatial,blankertz2007optimizing}, & EEGNet 8 2 & Deep Learning & \cite{lawhern2018eegnet} \\
\rowcolor{SeaGreen!50} FBCSP+SVM & Raw & \cite{ang2008filter} & EEGTCNet & Deep Learning & \cite{ingolfsson2020eegtcnet} \\
\rowcolor{SeaGreen!50} FgMDM & Riemannian & \cite{barachant2010riemannian} & EEGITNet & Deep Learning & \cite{salami2022eeg} \\
\rowcolor{SeaGreen!50} MDM & Riemannian & \cite{barachant2011multiclass} & EEGNeX 8 32 & Deep Learning & \cite{chen2022toward} \\
\rowcolor{SeaGreen!50} TS + EL & Riemannian & \cite{corsi2022functional} & & &\\
\hline
\rowcolor{Salmon!20} XDAWN + LDA & Raw & \cite{rivet2009xdawn} & XDAWNCov + TS + SVM & Riemannian & \cite{chevallier2018brain}\\
\rowcolor{Salmon!20} XDAWNCov + MDM & Riemannian & \cite{barachant2014meg} & ERPCov + MDM & Riemannian & \cite{barachant2014plug}\\
\rowcolor{Salmon!20} ERPCov$(svd_n=4)$ + MDM & Riemannian & \cite{barachant2014plug} & & &\\
\hline
\rowcolor{YellowOrange!10} TRCA & Raw & \cite{nakanishi2017enhancing} & SSVEP MDM & Riemannian  & \cite{chevallier2018riemannian} \\
\rowcolor{YellowOrange!10} CCA & Raw & \cite{lin2006frequency} & SSVEP TS + LR & Riemannian & \cite{chevallier2018riemannian} \\
\rowcolor{YellowOrange!10} MsetCCA & Raw & \cite{zhang2014frequency} & SSVEP TS + SVM & Riemannian & \cite{chevallier2018riemannian} \\
\hline
\end{tabular}
\end{adjustbox}
\end{center}
\end{table*}

\subsection{Raw signal} % Sara

The category of pipelines referred to as \textit{Raw} consists of \gls{bci} classifiers employing traditional statistical analysis, as well as temporal and/or spatial filtering tools to extract features.
%LogVariance features are defined as the logarithm of the variance estimated per \gls{eeg} channel. 
Variance-based pipelines represent one of the initial BCI pipeline concepts proposed for motor imagery. Several approaches have highlighted the value of utilizing intra-channel variance for online decoding tasks. These pipelines calculate the variance of each electrode within an epoch to create a positive definite real-valued vector. To address noise or artifacts, it is common practice to logarithmically transform the observed variance~\cite{lotte2007review}. This approach essentially boils down to only considering the diagonal elements of the covariance matrices. Classifiers such as Linear Discriminant Analysis (LDA) or Support Vector Machine (SVM) can be utilized on the resulting vector to predict the label of the epoch.

\gls{csp} approach learns spatial filtering matrices in a supervised manner, minimizing the variance of the band power feature vectors within the same class, while maximizing the between-class variance~\cite{muller1999designing,blankertz2007optimizing}.
To enhance the robustness of the \gls{csp} against noise and overfitting, \gls{trcsp} approach has been proposed in~\cite{lotte2010regularizing}.
In contrast to the classical \gls{csp}, which uses band-pass filtered \gls{eeg} signals that may vary per subject, \gls{fbcsp} addresses this issue by extracting \gls{csp} features for each band-pass filter from the filter bank~\cite{ang2008filter}.
The~\autoref{app:csp} details the \gls{csp} algorithm.
%\gls{csp} is a well established method used to discriminate between different mental states based on \gls{eeg} signals. It calculates spatial filters that optimally separate signals from different conditions based on their bandpower. This approach is particularly effective in capturing \gls{erd} and \gls{ers} effects, which are indicative of specific cognitive processes \cite{Koles1990}. \gls{csp} has found wide applications in \gls{bci} systems for accurate classification of mental states \cite{blankertz2007optimizing}. 

\gls{cca} has emerged as a prominent approach for classifying \gls{ssvep} signals, initially introduced in the work by~\cite{lin2006frequency}. 
Subsequently, \gls{cca} has been successfully employed in numerous notable \gls{ssvep}-based \gls{bci} studies, such as those conducted by~\cite{bin2009online} and~\cite{nakanishi2014enhancing}. 
\gls{ssvep} signals exhibit correlation to flickering visual stimuli, with their signal phase and frequency corresponding to stimulus characteristics. Leveraging this relationship, \gls{cca} aids in extracting \gls{eeg} spatial components with the strongest correlation to \gls{ssvep} stimuli. Further details on the different pipelines can be found in~\autoref{app:cca}.

\subsection{Riemannian geometry}
\label{sec:riempip}

The introduction of Riemannian geometry into \gls{bci} processing marked a pivotal moment for the \gls{bci} community~\cite{barachant2011multiclass}.
The fundamental concept underlying this approach is to represent the signal using covariance matrices or their derivatives.
As covariance matrices are \gls{spd} matrices, they live in a Riemannian space~\cite{yger2016riemannian}. We provide here a general description of the framework needed to define algorithms based on the Riemann distance for classification tasks. For a more in-depth description of the Riemannian framework, we refer the reader to~\cite{boumal2023intromanifolds} which gives a pedagogical introduction to these concepts.

Due to the curvature of the SPD matrix space, traditional Euclidean geometry is ill-suited and introduces a swelling effect. 
Particularly, the Euclidean distance could result in a wrong characterization of the relationship between \gls{spd} matrices. Instead, Riemannian methods rely on a distance that respects the geometry of the SPD matrices space, based on geodesics, i.e., the shortest path that connects 2 elements and stays in the space of \gls{spd} matrices. A common choice of distance is the affine-invariant one~~\cite{moakher2005differential}. 
Considering the manifold of \gls{spd} matrices $\mathcal{M}_n = \{ \mathbf{P} \in \mathbb{R}^{n \times n} | \mathbf{P} = \mathbf{P}^\top  \mathrm{ and } x^\top\mathbf{P}x > 0, \forall x \in \mathbb{R}^n \}$, the affine-invariant distance for $\mathbf{P}_{1}, \mathbf{P}_{2} \in \mathcal{M}$ is defined as
\begin{equation}
\delta_{R}\left(\mathbf{P}_{1}, \mathbf{P}_{2}\right)=\left[\sum_{i=1}^{n} \log ^{2} \lambda_{i}\left(\mathbf{P}_{1}^{-1} \mathbf{P}_{2} \right)\right]^{1 / 2}
\label{eq:airm}
\end{equation}
where %$\|\cdot\|_{F}$ is the Frobenius norm, $Log()$ is the logarithm of a matrix and 
$\lambda_i (\mathbf{P})$ is the $i$-th eigenvalues of $\mathbf{P}$. % $\mathbf{P}_1^{-1}\mathbf{P}_2$. 

Similarly, the concept of the mean in Riemannian geometry must be redefined to ensure it belongs to the manifold; it is known as the Frechet mean~\cite{moakher2005differential} and is defined as:
\begin{equation}
\label{eq:frechet_mean}
\hat{\mathbf{G}} =\underset{\mathbf{P} \in P(n)}{\operatorname{argmin}} \sum_{i=1}^{m} \delta_{R}^{2}\left(\mathbf{P}, \mathbf{P}_{i}\right)
\end{equation}

As we are operating within a Riemannian manifold, traditional machine-learning classification algorithms cannot be directly applied. Instead, there are two options: either create new algorithms to classify \gls{spd} matrices on the Riemannian manifold or map the matrices to an associated Euclidean space and then apply standard classification algorithms.
In the former scenario, algorithms like Minimum Distance to Mean give robust accuracy.
In the latter scenario, it is possible to rely on the tangent space to a point of the manifold. It is possible to circulate between the manifold and the tangent space using the \(\operatorname{Log}\) and \(\operatorname{Exp}\) map functions. The \(\operatorname{Log}\) (resp. \(\operatorname{Exp}\)) maps the manifold to the tangent space (resp. the tangent space to the manifold):
\begin{align}\label{eq:TS_cov}
    \operatorname{Exp}_{\mathbf{P}}\left(\mathbf{S}_{i}\right)  =  \mathbf{P}^{1 / 2} \operatorname{Exp}\left(\mathbf{P}^{-1 / 2} \mathbf{S}_{i} \mathbf{P}^{-1 / 2}\right) \mathbf{P}^{1 / 2}\\
    \operatorname{Log}_{\mathbf{P}}\left(\mathbf{P}_{i}\right)= \mathbf{P}^{1 / 2}  \operatorname{Log}\left(\mathbf{P}^{-1 / 2} \mathbf{P}_{i} \mathbf{P}^{-1 / 2}\right) \mathbf{P}^{1 / 2}
\end{align}
Projected in the tangent space, any machine learning algorithm could be applied. One limitation is the size of the considered space, that is $\frac{n(n+1)}{2}$ for $\mathcal{M}_n$. Machine learning algorithms like Support Vector Machine (SVM), ElasticNet (EL) or Logistic Regression (LR) are among the most popular for classification in the tangent space. Details regarding the pipelines implementation are available in ~\autoref{ann:riempip}.

\subsection{Deep learning} % Igor
\label{sec:ppl-dl}

\gls{dl} methods have demonstrated in recent years considerable promise in various tasks that involve handling massive volumes of digital data and those in different fields such as computer vision~\cite{krizhevsky2017imagenet} and Natural Language Processing~\cite{vaswani2017attention,schneider2019wav2vec}. 

This also holds for \gls{bci} applications. The \gls{bci} field has been significantly impacted by the integration of \gls{dl} techniques~\cite{roy2019deep, craik2019deep}, which exhibit strong generalization capabilities, with transfer learning emerging as a key focus in \gls{bci} research. One notable advantage of DL models is their capacity to leverage vast datasets, a task typically challenging for classical \gls{ml} algorithms. Moreover, by conducting all processing and classification steps within a neural network, DL models enable optimized end-to-end learning.
% \gls{dl} algorithms have begun to be applied to a wide variety of fields in recent years [CITA] thanks to the use of GPUs, which has speeded up the learning process. 
%\gls{dl} can be particularly useful in the field of \gls{bci} as they are algorithms capable of extracting meaningful information from a huge amount of data without requiring the supervision of an expert in the field. 
%In fact, normal Machine Learning algorithms require a Feature Extraction phase that can only be optimized by experts in the field, contrarily to \gls{dl} algorithms that contain within them a fully automated feature extraction phase.
In this paper, we will focus specifically on \gls{dl} methods that have been applied to \gls{mi} paradigms.

% Explain that this kind of algorithm are only recently applied by the \gls{bci} community because it have 2 big limitation: Small Dataset, High \gls{snr}

% explain the integration of both TF and Pytorch (Skorc and scikeras)
The predominant Python libraries for \gls{dl} are \textsc{Tensorflow}~\cite{tensorflow2015} and PyTorch~\cite{pytorch}. To facilitate broader access to \gls{moabb}, we developed integration for both libraries using wrappers from \textsc{Scikeras}~\cite{scikeras} (for \textsc{Tensorflow}) and \textsc{Skorch}~\cite{skorch} (for PyTorch). We were also keen on integrating the \textsc{Braindecode}~\cite{schirrmeister2017deep} library, which incorporates several \gls{dl} algorithms for \gls{eeg} processing using \textsc{PyTorch}.
% Explain how the signal is processed 
% Different method of work, Keras work with epoch, skorch with dataset create with braindecode

Most \gls{dl} architectures for \gls{eeg} decoding operate on minimally pre-processed (bandpass filters) or raw epochs. To address the spatio-temporal complexity of \gls{eeg} signals, convolutional layers are commonly employed with separable 1D convolution along the temporal and spatial dimension (i.e., \gls{eeg} channels). By segregating the spatial and temporal convolutions, these models account for the fact that all EEG channels observe all brain sources. 
%instead of one 2D convolution, models the fact that all \gls{eeg} channels observe all brain sources. The kernel of the spatial convolution typically covers all the \gls{eeg} channels and therefore does not "move". Such a spatial convolution would be equivalent to applying the same fully-connected layer to every time sample. 
% In general the \gls{eeg} signal can be described as multidimensional signal $X \in R^{epoch, ch, T}$, where $epoch$ is the size of the dataset, $ch$ is the number of channel and $T$ is the time of the recording. So each epoch is treated as a monochromatic image to see a parallel in computer vision. On this "image" the most common used architecture is the Convolutional Neural Network (CNN), used directly on the raw signal, that is used to extract relevant information, while the classification task is usually performed by a \gls{mlp} [CITA references for CNN and MLP general].

% Explain Preprocessing
Although the goal is to minimize pre-processing steps before passing data to \gls{dl} pipelines, a few steps are usually necessary. 
Bandpass filtering is applied to ensure a fair comparison between \gls{ml} and \gls{dl} methods.
Neural network architectures reviewed here are designed for decoding \gls{eeg} signals at a certain sampling frequency. 
Resampling datasets to match DL architectures' expected frequencies is performed to avoid interfering with the relative temporal length of their kernels.
Standard rescaling of the signal is implemented before \gls{dl} pipelines in adherence to common deep learning practices~\cite{lecun2002efficient}.

% Explain No data Augmentation is used
Data augmentation, a technique generating synthetic training samples through transformations, is well-established in computer vision for producing state-of-the-art results~\cite{chen2020simple}.
However, in the realm of EEG applications, data augmentation poses unique challenges. 
% and other domains since can drastically reduce the overfitting problem and allow the use of more complex algorithms. 
Despite its potential to reduce overfitting and enable complex algorithms, it is still an active area of research.
Assessing whether a generated signal accurately captures the physiological attributes of EEG remains an open question.
Thus, this study abstained from employing data augmentation procedures in the context of EEG decoding.
For a comprehensive examination of various existing techniques for \gls{eeg}, we refer the reader to~\cite{rommel2022data}.

Consistent parameters were used across all \gls{dl} experiments, training networks for $300$ epochs with a batch size of $64$, Adam optimizer~\cite{adam} with a base learning rate of $0.001$ and cross-entropy as the loss function.  An early stopping strategy with a patience parameter of $75$ epochs was crucial to prevent overfitting. 
While the considered \gls{dl} architecture hyperparameters align with state-of-the-art standards through signal resampling, further optimization could enhance model performance via a grid-search approach tailored to individual scenarios.

\section{Experimental results}
%choice for within-session cross-validation, literature oriented but not applicable in real BCI setup
%We excluded the rest from the datasets

%What are our recommendation for pipelines (wrt nb samples, nb of electrodes, training time, robustness), datasets (good discriminative datasets, which one use for a bench, for high number of electrodes, etc), paradigms (which one is easier/harder)?

%Open datasets are mainly acquired in lab condition and are using very controled setups. This does not reflect the EEG recorded in real condition.

% BNCI2014-001 is a good dataset, with high accuracy, could we make any recommandation on this? Is it a good dataset to test new idea? How is it when compared to other? Have we a better recommandation, for a "way-to-go" dataset that could be easily use to assess new ideas (like 5 worst/5 best in Cho2017)?

%Is there a magic formula (that we could learn?), that combines number of trials, number of time samples/trial, number of electrodes to give a rough prediction of the accuracy?

The experimental results outlined in this section encapsulate the key insights gathered during this benchmark study. 
While we provide only the most important findings to the reader in this section, we also report all the raw results to ensure proper reproducibility.
Due to space limitations, all the detailed evaluations, including \gls{roc-auc} or accuracy scores, are available in the appendix. Tables~\ref{tab:All_agg_dataset}, \ref{tab:lhrh_agg_dataset}, \ref{tab:rf_agg_dataset}, \ref{tab:P300_agg_dataset} and \ref{tab:SSVEP_agg_dataset} 
% \ref{tab:rfRiemannian_agg_dataset}, \ref{tab:rfDeep_agg_dataset}, 
% \ref{tab:lhrhDeep_agg_dataset}, \ref{tab:lhrhRiemannian_agg_dataset}, \ref{tab:rfRaw_agg_dataset}, and \ref{tab:lhrhRaw_agg_dataset} 
display the average scores of each pipeline across all subjects and sessions within a specific dataset. Additionally, Figures~\ref{fig-app:LHRH-raincloud}, \ref{fig-app:RHF-raincloud}, \ref{fig-app:SSVEP-raincloud}, \ref{SSVEP:Riemannia>Raw} and \ref{fig-app:P300-raincloud} present the pipeline groups' scores for each subject and session across all datasets to aid in result interpretation. 

\subsection{Riemannan pipelines outperforms others pipelines}

%\subsection{Motor Imagery: Riemannian methodology-based pipeline outperform its counterparts in performance.}

\begin{figure*}
    \centering
    \includegraphics[width=\linewidth]{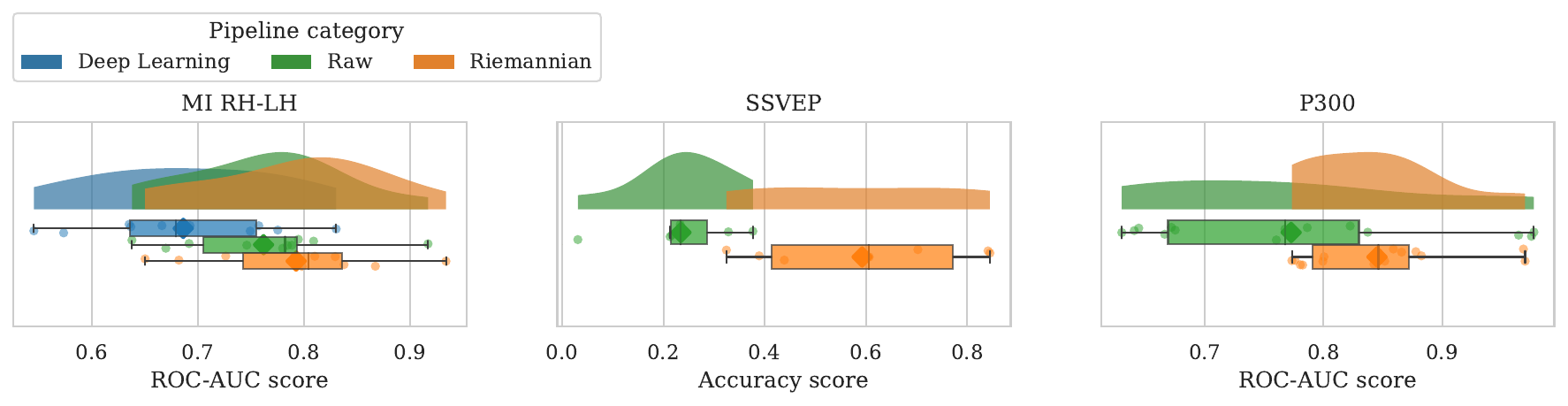}
    \caption{
    Average performance of pipelines grouped by category (Deep Learning, Riemannian, and Raw) across the \gls{mi} (right-hand vs left-hand), \gls{ssvep}, and \gls{erp} paradigms displayed as raincloud plots. Each point in the plot corresponds to the average score of one dataset across all pipelines within a specific category, encompassing all subjects and sessions.}
    \label{fig:riemann>all}
\end{figure*}

The Riemannian distance-based classification pipeline consistently outperforms results obtained through \gls{dl} and Raw pipelines across all datasets, on all paradigms. Figure~\ref{fig:riemann>all} illustrates this superiority in the context of right-hand vs left-hand classification, SSVEP and P300. The \gls{roc-auc} and accuracy results are shown based on each pipeline's performance relative to its category (Raw, Riemannian, \gls{dl}). It should be emphasized that each dot on the plots represents the average of all pipelines from a category on a dataset, each dataset encompassing 10 to 100 subjects. In the distributions, each dataset has the same weight, regardless of the number of subjects or sessions. The results presented here are thus summarizing the largest BCI study to date.

The plots demonstrate the dominance of the Riemannian approaches, showcasing its superior performance not only in overall averages but also consistently across all datasets examined. This trend is further reinforced by the noticeable peak shift in the distribution of results, providing additional confirmation of the effectiveness of the Riemannian approaches. Similar findings are observed across various classification tasks within the Motor Imagery paradigm, as could be seen in the appendix on Figures~\ref{fig-app:LHRH-raincloud} and \ref{fig-app:RHF-raincloud}.

The suboptimal performance of \gls{dl} pipelines can be attributed to two main factors. Firstly, the hyperparameters of \gls{dl} pipelines were not fine-tuned for each dataset; instead, the parameters described in the original articles were employed. 
This marks a significant divergence from the Riemannian and Raw approaches, which had their hyperparameters optimized through a nested cross-validation strategy.
We could not complete a similar hyperparameter search for \gls{dl} as the search space is too large and it involves complex changes in the architecture shape, such as the kernel size, or the activation functions that could not be automatized easily.
Secondly, we chose not to include any data augmentation steps in our research, as explained in Sect.~\ref{sec:ppl-dl}. It is noteworthy that such procedures have been demonstrated to exert a substantial impact on performance, as evidenced by previous studies~\cite{rommel2022data}. Still, it required a lengthy and dataset-specific parametrization. 
It is important to recall that the objective of this study is to assess existing and published algorithms, to evaluate their off-the-shelve performances, and not to investigate how to properly tune them.

%\subsection{Best scenario for Riemannian pipelines}
\subsection{Riemannian pipelines work well with limited number of electrodes}
%\subsection{Motor Imagery: Riemannian Pipeline are the best in a reduced number of channel situation}
Riemannian pipelines perform best in scenarios involving a reduced number of channels for \gls{mi} tasks.
Employing a limited number of electrodes in \gls{eeg} recordings offers numerous advantages for practical \gls{bci} experiments. This approach simplifies the setup, reducing both complexity and cost. Additionally, it enhances user comfort, providing greater flexibility and ease of use in diverse settings, including home-based or mobile applications. However, using fewer electrodes may compromise signal quality and spatial resolution, potentially resulting in lower classification accuracy and reduced robustness against artifacts and noise.

\begin{figure*}
    \centering
    \includegraphics[width=\linewidth]{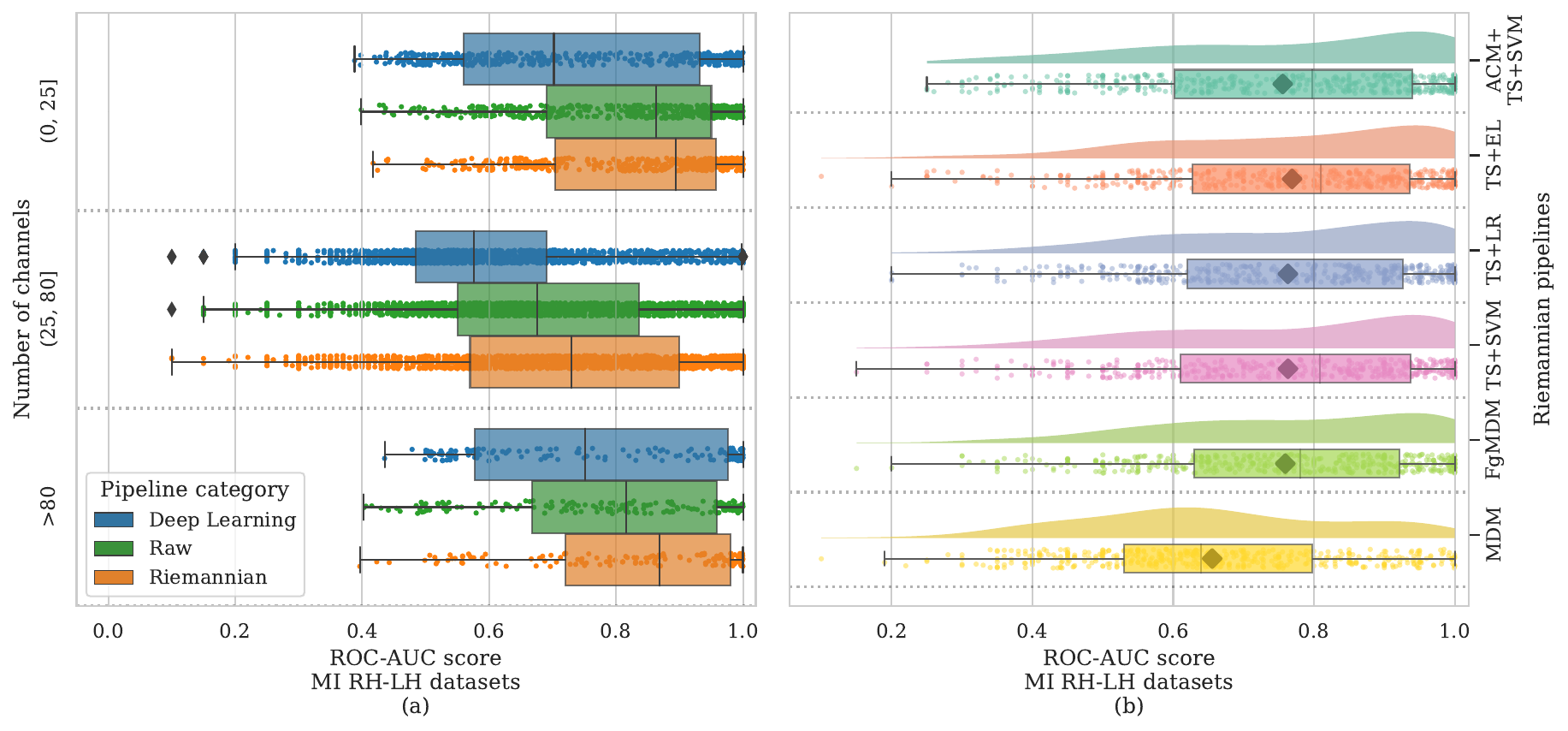}
    \caption{(a) \gls{auc} scores are averaged across all sessions, subjects, and datasets within the right-hand vs left-hand \gls{mi} paradigm for each category (Deep Learning, Riemannian, Raw), segmented by the number of channels on the y-axis. Box plots overlaid with strip plots show individual ROC-AUC scores.
    (b) Distribution of \gls{auc} scores for the Riemannian \gls{mi} pipelines is depicted for the right-hand vs left-hand classification task. The boxes and horizontal black bars denote quartile ranges.}
    \label{fig:riem-best-case}
\end{figure*}

To address these challenges, it becomes crucial to design classification algorithms capable of delivering high performance with a limited number of electrodes. As depicted in Figure~\ref{fig:riem-best-case}-(a), the Riemannian pipelines excel in performance with datasets containing $[0, 25]$ electrodes. Notably, their performances tend to decrease as the number of channels increases. Conversely, \gls{dl} pipelines require a substantial amount of information to achieve satisfactory performance, while facing the limitations described earlier, emphasizing the importance of balancing electrode count and classification effectiveness.

As observed, there is a decline in performance in settings with a moderate number of channels. This decrease in effectiveness can be attributed to several factors. Primarily, an increased number of electrodes elevates the problem's complexity, as the \gls{ml} algorithm needs to extract relevant information more efficiently. Furthermore, this category encompasses datasets with varying average scores, impacting overall performance. Conversely, scenarios with a high number of electrodes exhibit less pronounced detrimental effects. This could be a side effect; a large number of electrodes implies higher-grade equipment and specialized technicians. The enhanced data recording procedures and superior data quality associated with a larger number of channels might thus alleviate previous issues.

%\subsection{Motor Imagery: Riemannian method based on Tangent space projection outperform the one on the Riemannian surface}
Classification algorithms based on Riemannian distances can be implemented in two distinct approaches. The first method involves performing classification directly on the Riemannian manifold, using algorithms like the MDM (Minimum Distance to Mean) algorithm. The alternative approach involves projecting data onto the tangent space and conducting classification using conventional ML algorithms (SVM, LR, EL).
Comparing the two strategies, it is observed that the Riemannian method based on Tangent space projection consistently outperforms the approach centered on the Riemannian surface as shown in Figure~\ref{fig:riem-best-case}-(b).

%\begin{figure*}
%    \centering
%    \includegraphics[width=\linewidth]{figures/FigurePaper_test/Figure4_mi_lhrh_a_no_channels_b_riemannian_agg.pdf}
%    \caption{\textbf{OLD:} (a) ROC-AUC scores averaged over all the sessions of all the subjects of all the datasets of the \textit{righ hand - left hand} \gls{mi} paradigm and over all pipelines of the corresponding category (\textit{Deep learning, Riemannian, Raw}) per different group of the number of channels (y-axis). Box-plots are overlied with strip-plots, where each point represent one ROC-AUC score. 
%    (b) Distributions of ROC-AUC scores on the \textit{right hand vs left hand} classification task of the Riemannian motor imagery pipelines. The boxes and horizontal black bars indicate the quartiles.}
%    \label{fig:riem-best-case}
%\end{figure*}

%\subsection{Deep learning pipelines}
\subsection{Deep learning requires a high number of trials}

\begin{figure*}
    \centering
    \includegraphics[width=\linewidth]{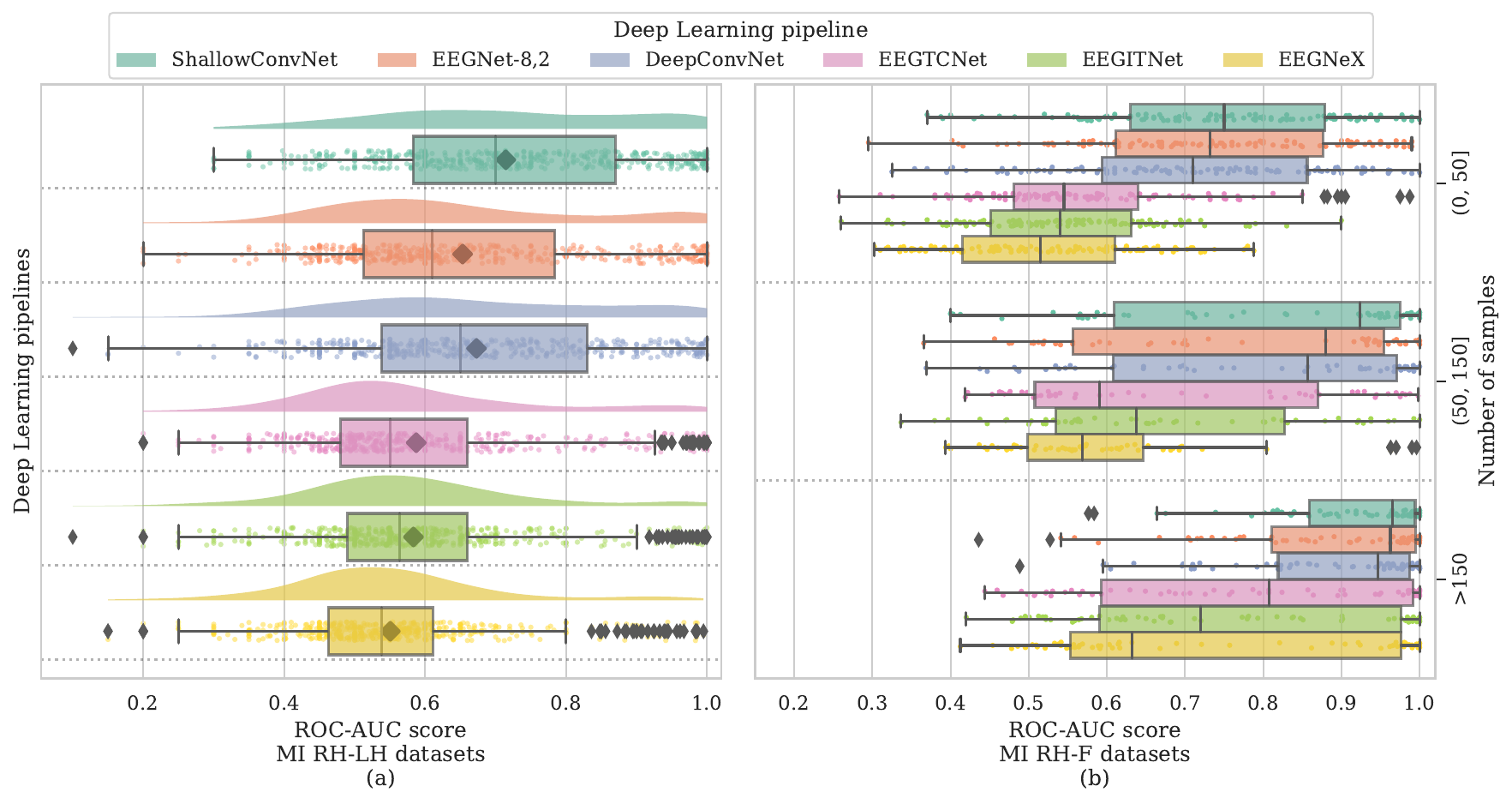}
    \caption{(a) Distributions of \gls{auc} scores averaged over all datasets for the right-hand vs left-hand classification task within the \gls{dl} pipelines.
    (b) \gls{auc} scores averaged across all sessions, subjects, and datasets within the right-hand vs feet \gls{mi} paradigm for the \gls{dl} pipelines, segmented based on the number of epochs on the y-axis.}
    \label{fig:dlacc}
\end{figure*}

%\subsection{Motor Imagery: ShallowConvNet is the best DL pipeline.}
It is essential to emphasize that the lower performance of DL pipelines, in comparison to the other pipeline categories, is closely linked to the specific DL architecture under consideration. This observation is illustrated in Figure~\ref{fig:dlacc}-(a). Amongst \gls{dl} pipelines, ShallowConvNet stands out with the highest \gls{auc}. Interestingly, a distinctive dichotomy appears within \gls{dl} models. The first group, consisting of ShallowConvNet, EEGNet-8.2, and DeepConvNet, exhibits superior performance. In contrast, the second group, gathering EEGTCNet, EEGITNet, and EEGNeX, shows lower performance levels. This divergence in performance is likely attributed to the optimization of hyperparameters in the models. It demonstrates that  \gls{dl} models that have been more extensively tested, and hence correctly parametrized, yield higher classification performance on all datasets.
% It is also interesting to note that this architecture achieves superior results compared to Raw pipelines. %as we will highlight in more detail in the section~\ref{best}.

%\subsection{Motor Imagery: To obtain satisfactory performance DL pipeline require more trial}
The number of trials employed in training algorithms carries particular importance, specifically in the context of DL pipelines. A clear trend emerges when evaluating DL algorithm performance, indicating that achieving satisfactory results typically requires more than 150 trials per class, as shown in Figure~\ref{fig:dlacc}-(b). Again we observe the same two distinct groups described above, the group comprising ShallowConvNet, EEGNet-8.2, and DeepConvNet exhibits a higher resilience to the impact of the number of trials. Notably, this group manages to achieve already satisfactory performance with as few as 50 trials per class, showcasing a relatively robust response to variations in the training dataset size, highlighting the distinct capabilities of certain architectures to yield superior performance with a more limited number of trials.

\subsection{Recommended  number of trials depends on the task complexity}

%\subsection{Motor Imagery: Number of trial needed depend on the number of class}
%The number of trial needed depend on the number of class in motor imagery. For RHF, the results are shown on Figure~\ref{MI:DL_more_trial}.
%In the case of more classes then a binary one the effect is even more marked, we need at least 200 trial to have a stable algorithm Figure~\ref{MI:More_trial_Class}. 

\begin{figure*}
    \centering
    \begin{subfigure}[t]{0.49\textwidth}
        \centering
        \includegraphics[width=\linewidth]{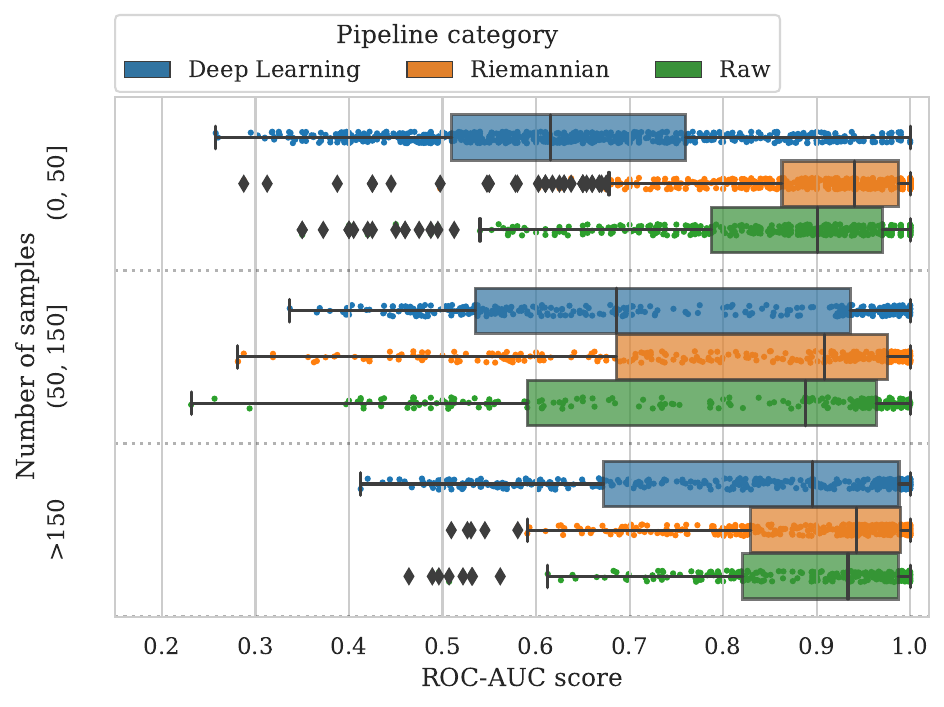}
        \caption{right-hand/feet} 
        \label{MI:DL_more_trial}
    \end{subfigure}
    \hfill
    \begin{subfigure}[t]{0.49\textwidth}
        \centering
        \includegraphics[width=\linewidth]{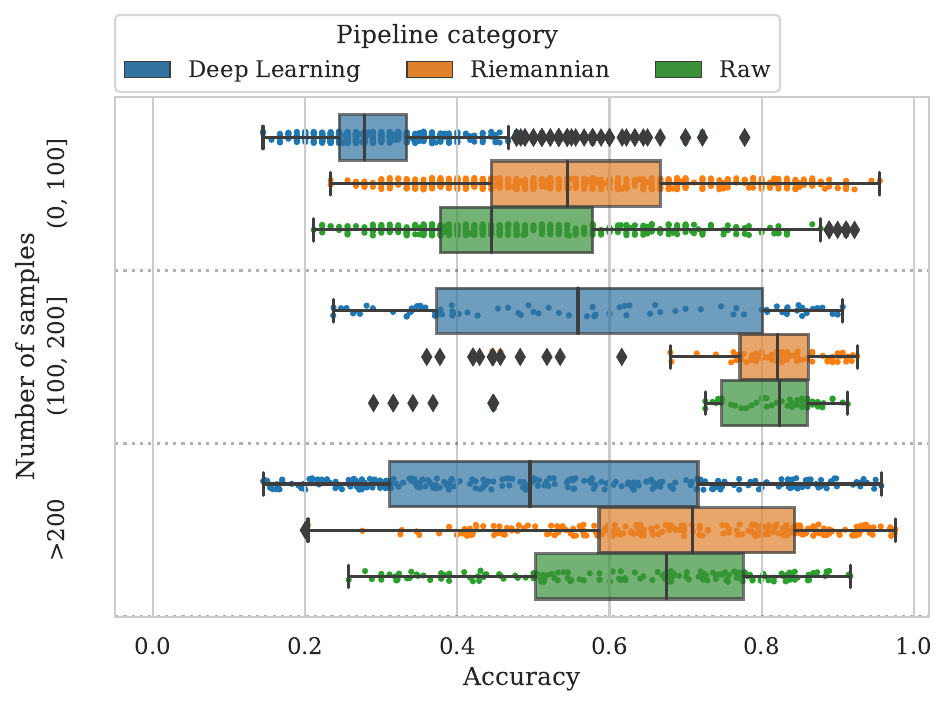}
        \caption{all classes} 
        \label{MI:More_trial_Class}
    \end{subfigure}
    \caption{(\subref{MI:DL_more_trial}) \gls{auc} scores for datasets segmented by the number of trials in the right-hand vs feet \gls{mi} paradigm for each pipeline category (\textit{Deep learning, Riemannian, Raw}).
    (\subref{MI:More_trial_Class}) Same results for \gls{mi} using all available classes and the accuracy metric. 
    }
 \end{figure*}

In scenarios where tasks exhibit a clear distinguishability, such as the \gls{mi} task involving right-hand/feet movements, achieving impressive \gls{auc}  performance with a limited number of trials is feasible, as clearly depicted in Figure~\ref{MI:DL_more_trial}. Notably, both the Raw and Riemannian pipelines demonstrate exceptional classification scores even with fewer than 50 trials. Increasing the trial count beyond 50 does not yield significant improvements in \gls{auc}  results for these pipelines. On the other hand, for \gls{dl} pipelines to reach good \gls{auc}, datasets associated with more than 150 trials are imperative to achieve satisfactory \gls{auc}  scores, highlighting the pivotal role trial quantity plays in \gls{dl} model performance. This discrepancy highlights the varying requirements and efficiencies of different pipeline approaches based on the task complexity and nature of the data.

When dealing with tasks of higher complexity, such as \gls{mi} paradigms involving 3 to 7 distinct classes (refer to Section~\ref{sec:mi} for detailed explanation), obtaining optimal accuracy becomes considerably more challenging. In such intricate tasks, a larger number of trials is imperative for pipelines to achieve the highest levels of accuracy. This phenomenon is  illustrated in Figure~\ref{MI:More_trial_Class}, where the accuracy variability across different datasets becomes apparent. The fluctuation in accuracy levels observed here is intricately tied to the unique characteristics and quality of individual datasets, portraying the nuanced dynamic between dataset quality, trial quantity, and the performance of the classification pipelines. Specifically, the noticeable decline in accuracy for datasets with over 200 trials compared to datasets with 100 to 200 trials underscores the diverse demands and responses of different pipelines to varying trial quantities in complex classification tasks.

% In tasks with strong discriminability, such as \gls{mi} using right-hand/left-hand movements, it is feasible to achieve high \gls{auc} performance with a limited number of trials, as illustrated in Figure~\ref{MI:DL_more_trial}. Both Raw and Riemannian pipelines exhibit excellent classification scores with fewer than 50 trials. Interestingly, increasing the number of trials beyond 50 does not notably enhance \gls{auc} outcomes for these pipelines. In contrast, \gls{dl} pipelines require datasets with more than 150 trials to attain satisfactory \gls{auc} scores, with poorer results seen for datasets containing fewer trials.

% In scenarios involving higher task complexity, like \gls{mi} paradigms with 3 to 7 classes (refer to Section\ref{sec:mi} for further insights), the overall accuracy tends to be lower, necessitating a greater number of trials for optimal accuracy. This trend is depicted in Figure\ref{MI:More_trial_Class}. The accuracy variability in this context is heavily influenced by the datasets and their inherent quality, which can explain why all pipelines exhibit decreased accuracy levels for datasets with more than 200 trials compared to datasets containing 100 to 200 trials.

%\subsection{SSVEP: Number of trial influence}

\subsection{Best practices in motor imagery}
%\subsection{Motor Imagery: For binary classification Right Hand vs Feet obtain the highest performance}
Motor imagery is a widely utilized paradigm in the BCI community, and a detailed analysis was conducted on the \gls{mi} results obtained in this benchmark to assist practitioners in experimental design and pipeline selection.

The highest performance levels in \gls{mi} are achieved in the binary classification distinguishing between right-hand and feet movements. 
This task notably surpasses the performance of the right-hand/left-hand classification, highlighting a significant disparity in performance between the two tasks. This performance discrepancy is visible when comparing the results of right-hand/left-hand and right-hand/feet tasks in the supplementary Figures~\ref{fig-app:LHRH-raincloud} and~\ref{fig-app:RHF-raincloud}, respectively. This trend is consistent across all pipeline categories -- Raw, Riemannian and DL -- and is prevalent across datasets, with 9 datasets for right-hand/feet tasks and 10 for right-hand/left-hand tasks. Five datasets are common to both tasks, with the \gls{auc} notably higher for the right-hand/feet task. 

This results holds significant implications for the BCI community, as it provides for the first time valuable insight into selecting the task that yields the highest accuracy in \gls{mi}. 
The findings presented here are highgly reliable, drawn from a diverse range of subjects recorded under various protocols and using different hardware setups. Furthermore, since this trend is observable even when analyzing subjects performing both tasks, any confounding effects related to recording conditions can be effectively eliminated. This insight holds particular importance for \gls{bci} practitioners when designing experimental protocols. 

\begin{figure*}
    \centering
    \includegraphics[width=\linewidth]{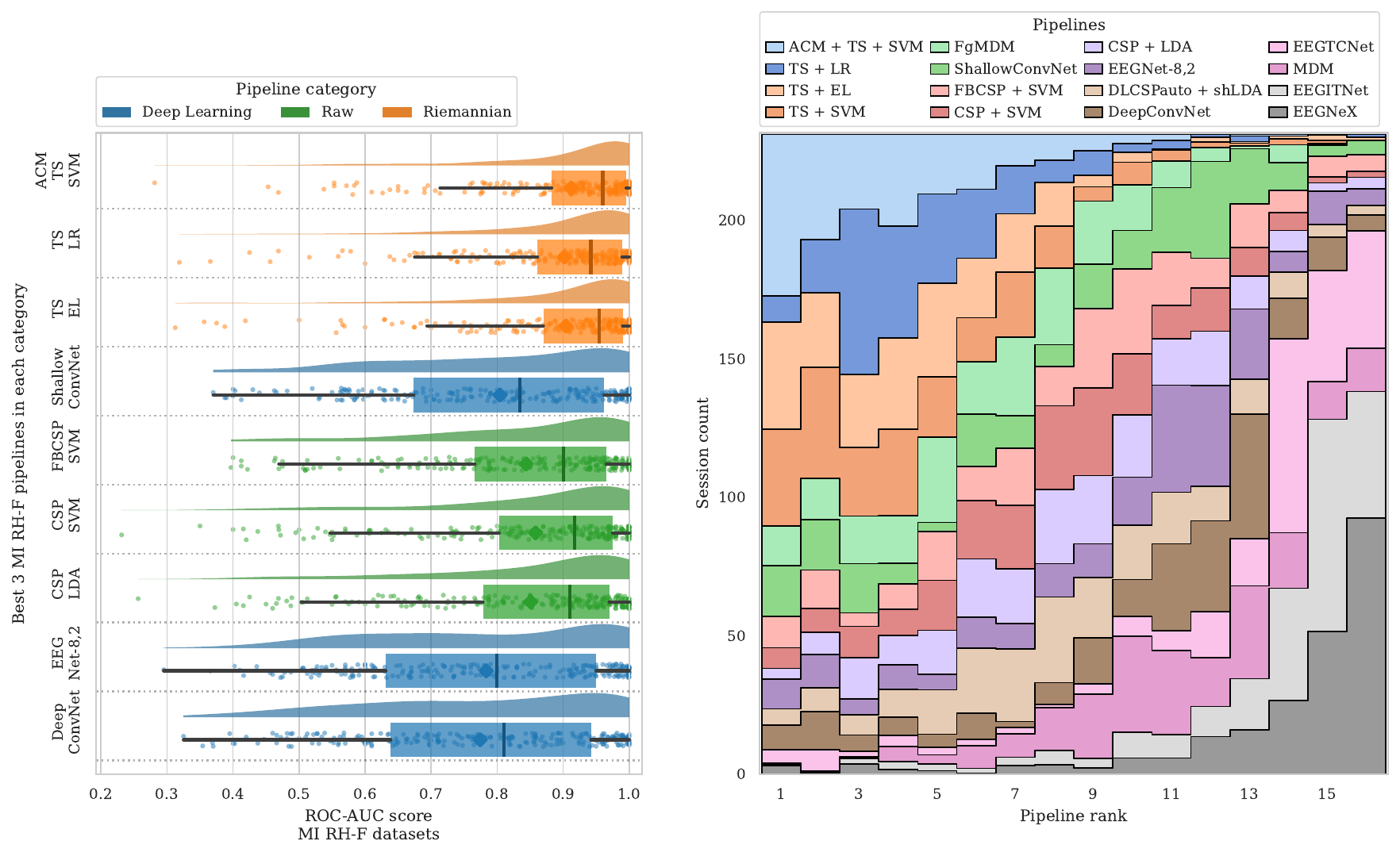}

    \hfill
    % \begin{subfigure}[t]{0.64\textwidth}
    %     \centering
    %     \includegraphics[width=\linewidth]{figures/FigurePaper/Figure1b_mi_rf_session-ranking_step.pdf}
    %     \caption{ranks} 
    %     \label{MI:BestPipelineRank}
    % \end{subfigure}
    \caption{(a) \gls{auc} scores are presented for the best three motor imagery pipelines in each category for the right-hand vs feet classification task, ordered by their average score per dataset. 
    (b) Pipeline rankings within individual sessions for the right-hand vs feet task, with each pipeline color-coded. The x-axis displays different ranks achieved by pipelines, while the y-axis indicates the number of sessions each pipeline achieves a specific rank.
    }
    \label{MI:BestPipeline}
 \end{figure*}

To delve deeper into how different pipelines compare in the right-hand/feet classification task, Figure~\ref{MI:BestPipeline}-(a) illustrates the \gls{auc} scores of the top three pipelines in each category, arranged by their average scores per dataset. 
The very good performances of Raw and Riemannian pipelines are visible on this plot. The performances of the \gls{dl} pipelines are below Raw and Riemannian pipelines for the reasons outlined previously. 

While the \gls{auc} performance measurements offer valuable insights, Figure\ref{MI:BestPipeline}-(a) reveals significant subject variability. In practical \gls{bci} applications, the choice of the pipeline is often influenced by the algorithm's ranking, with the best-performing algorithm selected for each subject. To evaluate how pipelines fare based on this criterion, pipelines were ranked in each session for all subjects, ranging from 1 (the best) to 16 (the worst) based on their relative scores. Figure~\ref{MI:BestPipeline}-(b)visualizes the frequency of sessions (y-axis) where a pipeline achieves a specific rank (x-axis), using distinct colors for each pipeline.

While the results align with the average \gls{auc} scores presented in Figure~\ref{MI:BestPipeline}-(a), they provide different qualitative insights. Firstly, the ACM+TS+SVM is outperforming other pipelines, consistently securing top positions and rarely falling below the 7th spot. Secondly, the TS+LR and TS+EL Riemannian pipelines exhibit comparable \gls{auc} scores, yet TS+EL consistently outperforms other pipelines, whereas TS+LR seldom claims the top spot but frequently ranks among the top 3 pipelines.  Lastly, \gls{dl} pipelines generally rank below the 9th position, excluding the notable exception of the ShallowConvNet pipeline, which attains top 3 rankings in several sessions, despite its lower average \gls{auc} score placing it behind CSP-based pipelines. 

%LR : 14-01, 14-04, cho, grosse, lee, physio, schirr, shin, weibo, zhou
%RF : Alex, 14-01, 14-02, 15-01, 15-04, physio, schirr, weibo, zhou
%commun (Riemann auc LR-RF) : 14-01 (87/93), physio (65/91), schirr (83/95), weibo (79/86), zhou (94/95).

%\subsection{Motor Imagery: Best 3 Pipelines}
%\label{best}
%TODO: Review
%The results of the best 3 Pipelines of each category for the \textit{right hand vs feet} are shown Figure~\ref{MI:BestPipeline}-left along the rank of several pipelines on Figure~\ref{MI:BestPipeline}-right. The best pipelines are riemanian and the best 3 riemanian pipelines are ACM+TS+SVM, TS+SVM and TS+EL. Moreover ACM+TS+SVM seems consistent because it is rank 1 compared to the other pipelines most of time. The other 2 riemanian pipelines seems less consistent as they appear less times rank number 1 or 2, in contrary to TS+LR which is a riemanian pipeline too.
% For the Deep Learning category, the best 3 pipelines are Shallow ConvNet, EEGNet8.2 and DeepConvNet that are all mostly ranked under 9 compared to the other pipelines.

%\subsection{Motor Imagery: Execution Time}
Accuracy is a primary criterion for selecting a BCI pipeline, but it's also essential to consider calibration time. Pipelines are often trained immediately following a calibration run or updated after running multiple trials, making execution time a crucial factor in pipeline selection. User interaction is typically paused during pipeline training on the training dataset, emphasizing the significance of efficient execution. This aspect is crucial not only for real-time BCI operation but also for offline evaluation and hyperparameter optimization, as it directly impacts experiment duration.

\begin{figure*}
    \centering
    \includegraphics[width=\linewidth]{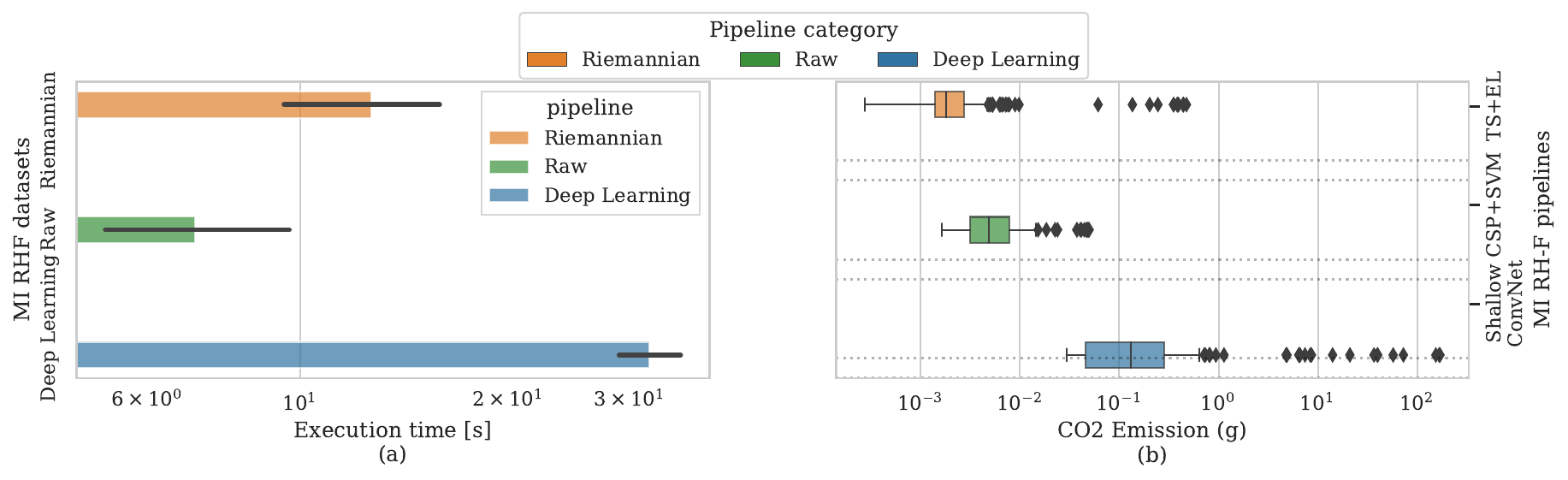}
    \caption{
    (a) Average execution times in seconds per dataset for the \gls{mi} right-hand vs feet paradigm, segmented by pipeline category (\gls{dl}, Riemannian, Raw).
    (b) Carbon emissions in gCO2 equivalent for the high-ranking pipelines (Raw CSP+SVM, Riemannian TS+EL, \gls{dl} ShallowConvNet) in the \gls{mi} right-hand vs feet task.}
    \label{fig:execo2}
 \end{figure*}

To address this issue, the average execution times for pipeline categories (Raw, Riemannian, and \gls{dl}) are presented in Figure~\ref{fig:execo2}-(a). These findings, focused on MI right-hand vs feet classification, are applicable across other tasks and paradigms as well. The measurements were conducted using the French Jean Zay CPU/GPU HPC environment, featuring Intel Cascade Lake 6248 CPUs and Nvidia V100 16 GB RAM, encompassing both training and inference phases for a single fold of cross-validation.

Raw pipelines demonstrate the shortest computational time requirements, closely followed by Riemannian pipelines. DL pipelines, on the other hand, exhibit longer training durations on average but remain within an acceptable 30-second range for experimental systems. It's noteworthy that these results were obtained using an early stopping strategy to prevent overfitting, which also contributes to reducing training time.

Environmental impact assessment is crucial in AI-related domains given the exponential growth in these areas. In this study, the direct environmental impact, measured in gCO2 equivalent generated during training and inference phases, is evaluated. The absolute values are significantly influenced by the country's energy production methods, hence the CPU/GPU server localization are important to measure the generated gCO2 equivalent. Precisely measuring algorithm energy consumption is challenging, as existing libraries differ in solutions and measurements. Using Code Carbon~\cite{codecarbon}, the environmental footprint, expressed in gCO2 equivalent emissions, for Riemannian TS+EL, Raw CSP+SVM, and ShallowConvNet pipelines is documented in Figure~\ref{fig:execo2}-(b). This provides a unified measure of required computational resources, illustrating that the Riemannian pipeline consumes less energy despite its longer training times as shown in Figure~\ref{fig:execo2}-(a).

\section{Future Directions for reproducible BCI machine learning pipelines} % Pierre August 23

%\subsection{Reproducibility Problem}
%\textcolor{red}{TODO}
%What are our recommendation for reproducibility, how to write a "good" article in BCI (on reproducbility side)?

%Thanks to its contributors, the \gls{moabb} library is constantly growing and improving. 
The path towards open and reproducible approaches in \gls{bci} improved with initiatives in open \gls{eeg} hardware~\cite{frey2016comparison,cardona2023novel}, libraries for experimental design~\cite{renard2010openvibe,peirce2022building,clisson2019timeflux} and, indeed, machine learning pipelines~\cite{moabb-software,pyriemann}. For the latter, there is still room for improvements, along two main axes.
The first one is to get closer to experimental situations and the second one is to allow fast benchmarking of the most recent \gls{bci} decoding techniques.
An initiative\footnote{\url{https://github.com/benchopt/benchmark_bci}} to tackle the second axes relies on \textsc{benchopt}~\cite{Moreau2022} and aims to provide an easy environment to evaluate novel \gls{bci} techniques along with reproducible evaluation conditions.
%Concretely, this translates into tackling the following limitations of the current \gls{moabb} library.  

%mentionner le tableau de résultat upgradé

\subsection{Pseudo-online benchmarking} 

The first limitation is that, in order to maintain compatibility with numerous datasets, the inherent chronology between epochs is disregarded, and the evaluation within each session consists of a simple 5-fold cross-validation over shuffled epochs. However, by leveraging historical knowledge, certain unsupervised classification techniques can rival supervised ones~\cite{huebner2018unsupervised, sosulski2023umm}. This evaluation presents a significant challenge as it completely overlooks the non-stationarity of the data, causing some classifiers to perform well under these conditions but fail in an online scenario. Consequently, in the near future, it is important to integrate pseudo-online benchmarking of algorithms, like along the lines proposed in~\cite{carrara2024pseudo}.
%\gls{moabb} will facilitate pseudo-online benchmarking of algorithms~\cite{carrara2024pseudo}.

In real online experiments, we anticipate a decrease in accuracy with excessively long calibration periods. User feedback could prove to be highly motivating, a factor that remains undisclosed in offline evaluations.

Regarding pseudo-online evaluation, we hypothesize that a gradual distribution shift occurs during the session. Achieving high accuracy becomes considerably more challenging as training data only capture a limited range of subject variability. Additionally, we anticipate observing learning curves that ascend and level off due to this shift; a plateau phenomenon not observed in offline analyses, where accuracy appears to increase as more data is amassed.

\subsection{CVEP paradigm} 

Most open data in \gls{bci} include \gls{mi}, \gls{erp}, and \gls{ssvep} paradigms. A recent addition to the landscape of evoked potentials, alongside the well-established \gls{erp} and \gls{ssvep} scenario, is the emerging \gls{cvep} paradigm~\cite{martinez2021brain}. Drawing an analogy to telecommunications, these three evoked paradigms can be likened to time-domain, frequency-domain, and code-domain multiple access schemes~\cite{gao2014visual}. Notably, research has demonstrated that the \gls{cvep} paradigm exhibits superior performance compared to both \gls{erp} and \gls{ssvep} paradigms~\cite{bin2009vep}, garnering increasing attention and leading to the development of high-speed \glspl{bci} measured through \gls{itr}; see, for instance,~\cite{nagel2019world, thielen2021full, sun2022120}). 
%Consequently, \gls{moabb} will incorporate support for the \gls{cvep} paradigm in the near future.

\subsection{Character-level benchmarking of \gls{erp} and \gls{cvep}}

The decoding algorithms for \gls{erp} (and soon \gls{cvep}) are mostly benchmarked at the epoch classification level, typically addressing a binary problem where the algorithms need to predict whether an epoch corresponds to a \textit{target} or a \textit{non-target} in the original application. However, this decoding approach overlooks the specific character or target being attended to in the original application, despite the availability of information regarding the stimulus sequence used for each character. Recent advancements have introduced methods that leverage application-level information to reduce the number of target hypotheses~\cite{kindermans2012p300,bianchi2021improving,barthelemy2023end}. Moreover, unsupervised classifiers have been proposed that capitalize on the structural sequence information, showcasing the potential to surpass established, even supervised algorithms~\cite{sosulski2023umm,thielen2015broad}. Unfortunately, the emphasis on binary decoding has led to incomplete availability of the necessary structural information in all existing \gls{erp} datasets. 
% In the upcoming period, \gls{moabb} will facilitate benchmarking of character-level classifiers on \gls{erp} datasets (and subsequently on \gls{cvep} datasets) while promoting the inclusion of application-level data with each new dataset.

\subsection{Cross-dataset transfer learning}

Transfer learning has consistently posed a significant challenge in the realm of \gls{bci}~\cite{jayaram2016transfer,kalunga2018transfer}. While current support includes benchmarks for cross-session and cross-subject scenarios, the absence of benchmarks across datasets remains a gap. Recent advancements in \gls{dl} models have yielded remarkably high performances in solving cross-dataset transfer learning challenges, particularly in \gls{mi} paradigms~\cite{khazem2021minimizing,guetschel23transfer,wei2022beetl, aristimunha2023evaluating}. This surge in interest toward cross-dataset transfer no longer stems solely from fundamental research but also unfolds as a promising avenue for future \glspl{bci}. 
%Consequently, one of the key objectives of \gls{moabb} is to establish a robust benchmark for cross-dataset evaluations.

% Remark: moved from introduction. Could be removed.
% Moreover, the emergence of transfer learning and domain generalization techniques has brought the possibility of predicting the brain activity of an unseen subject using data from different subjects, acquired with either the same or different BCI systems~\cite{wei2022beetl}, and trained on the same or different BCI tasks~\cite{transferEEG}. However, the methodologies for transfer learning in EEG-based BCIs still require further development to establish sound and well-defined practices, as the results in this area remain unclear.

% Concluding this section, it is worth emphasizing that \gls{moabb} is built upon community involvement as an open-source library, signifying its reliance on the help of volunteers to evolve. If one is considering contributing, know that they are more than welcome. For instance, a valuable contribution would be adding a new dataset to the library or implementing a new feature.

\section{Conclusion}

This contribution represents the largest reproducibility study in \gls{eeg}-based \gls{bci}, leveraging the \gls{moabb} library. By utilizing openly available data collected from different hardware sources in varied formats and structures, a systematic benchmarking process was undertaken. Machine learning pipelines from published works were re-implemented in a unified and open framework, aligning with the established standards of the machine learning community. This effort extends to \gls{dl} pipelines, considering the rapidly evolving processing standards for time series data.

The study's strength lies in the extensive number of subjects analyzed across diverse datasets, enabling robust assessments through meta-analysis statistical techniques. The pipelines undergo evaluation using 5-fold cross-validation, employing the \gls{auc} metric for binary classification tasks and accuracy for datasets with multiple classes. Furthermore, the environmental impact of the pipelines is assessed and factored into the reported results.

The primary outcome is a comprehensive and meticulous benchmarking of prominent pipelines from within the \gls{bci} literature. Resources are provided to reproduce these results and facilitate comparisons with future works, including result tables in the annexes and on a dedicated online platform to streamline comparisons and avoid unnecessary duplications. The Riemannian pipelines demonstrate the highest accuracy, whereas \gls{dl} pipelines, while achieving admirable accuracy with extensive trial data, show limitations across most datasets. Although data augmentation techniques and advanced parameterization can enhance their performance, improvements are still required for these off-the-shelf pipelines.

As the benchmark incorporates various pipelines and datasets, recommendations can be formulated regarding the optimal number of trials or channels for designing \gls{bci} experiments to achieve peak performance. A detailed analysis of all considered datasets is presented, aiding practitioners in tailoring their experimental designs or selecting specific datasets for evaluating novel pipelines.

Future development avenues are outlined along two key directions. Firstly, benchmarks could progress towards evaluations that mirror real-world experimental conditions, aiming to narrow the disparity between offline assessment and practical online \gls{bci} applications. Secondly, the integration of novel \gls{bci} paradigms like CVEP and transfer learning approaches across datasets is suggested for further exploration and integration.

\section{Acknowledgements}

SC, BA and SS were supported by DATAIA Convergence Institute as part of the ``Programme d’Investissement d’Avenir'', (ANR-17-CONV-0003) operated by LISN-CNRS.
This work was granted access to the HPC resources of IDRIS under the allocation 2023-AD011014322 made by GENCI.
\bibliography{biblio}

\begin{thebibliography}{128}
\providecommand{\natexlab}[1]{#1}
\providecommand{\url}[1]{\texttt{#1}}
\expandafter\ifx\csname urlstyle\endcsname\relax
  \providecommand{\doi}[1]{doi: #1}\else
  \providecommand{\doi}{doi: \begingroup \urlstyle{rm}\Url}\fi

\bibitem[Abadi et~al.(2015)Abadi, Agarwal, Barham, Brevdo, Chen, Citro,
  Corrado, Davis, Dean, Devin, Ghemawat, Goodfellow, Harp, Irving, Isard, Jia,
  Jozefowicz, Kaiser, Kudlur, Levenberg, Man\'{e}, Monga, Moore, Murray, Olah,
  Schuster, Shlens, Steiner, Sutskever, Talwar, Tucker, Vanhoucke, Vasudevan,
  Vi\'{e}gas, Vinyals, Warden, Wattenberg, Wicke, Yu, and
  Zheng]{tensorflow2015}
M.~Abadi, A.~Agarwal, P.~Barham, E.~Brevdo, Z.~Chen, C.~Citro, G.~S. Corrado,
  A.~Davis, J.~Dean, M.~Devin, S.~Ghemawat, I.~Goodfellow, A.~Harp, G.~Irving,
  M.~Isard, Y.~Jia, R.~Jozefowicz, L.~Kaiser, M.~Kudlur, J.~Levenberg,
  D.~Man\'{e}, R.~Monga, S.~Moore, D.~Murray, C.~Olah, M.~Schuster, J.~Shlens,
  B.~Steiner, I.~Sutskever, K.~Talwar, P.~Tucker, V.~Vanhoucke, V.~Vasudevan,
  F.~Vi\'{e}gas, O.~Vinyals, P.~Warden, M.~Wattenberg, M.~Wicke, Y.~Yu, and
  X.~Zheng.
\newblock {TensorFlow}: Large-scale machine learning on heterogeneous systems,
  2015.
\newblock URL \url{https://www.tensorflow.org/}.

\bibitem[Ang et~al.(2008)Ang, Chin, Zhang, and Guan]{ang2008filter}
K.~K. Ang, Z.~Y. Chin, H.~Zhang, and C.~Guan.
\newblock Filter bank common spatial pattern ({FBCSP}) in brain-computer
  interface.
\newblock In \emph{IEEE IJCNN}, pages 2390--2397, 2008.

\bibitem[Aric{\`o} et~al.(2014)Aric{\`o}, Aloise, Schettini, Salinari, Mattia,
  and Cincotti]{arico2014influence}
P.~Aric{\`o}, F.~Aloise, F.~Schettini, S.~Salinari, D.~Mattia, and F.~Cincotti.
\newblock Influence of {P300} latency jitter on event related potential-based
  brain--computer interface performance.
\newblock \emph{Journal of neural engineering}, 11\penalty0 (3):\penalty0
  035008, 2014.

\bibitem[Aristimunha et~al.(2023{\natexlab{a}})Aristimunha, Carrara, Guetschel,
  Sedlar, Rodrigues, Sosulski, Narayanan, Bjareholt, Quentin, Schirrmeister,
  Kalunga, Darmet, Gregoire, Abdul~Hussain, Gatti, Goncharenko, Thielen,
  Moreau, Roy, Jayaram, Barachant, and Chevallier]{moabb-software}
B.~Aristimunha, I.~Carrara, P.~Guetschel, S.~Sedlar, P.~Rodrigues, J.~Sosulski,
  D.~Narayanan, E.~Bjareholt, B.~Quentin, R.~T. Schirrmeister, E.~Kalunga,
  L.~Darmet, C.~Gregoire, A.~Abdul~Hussain, R.~Gatti, V.~Goncharenko,
  J.~Thielen, T.~Moreau, Y.~Roy, V.~Jayaram, A.~Barachant, and S.~Chevallier.
\newblock {Mother of all BCI Benchmarks}, 2023{\natexlab{a}}.
\newblock URL \url{https://github.com/NeuroTechX/moabb}.

\bibitem[Aristimunha et~al.(2023{\natexlab{b}})Aristimunha, de~Camargo, Pinaya,
  Chevallier, Gramfort, and Rommel]{aristimunha2023evaluating}
B.~Aristimunha, R.~Y. de~Camargo, W.~H.~L. Pinaya, S.~Chevallier, A.~Gramfort,
  and C.~Rommel.
\newblock {Evaluating the structure of cognitive tasks with transfer learning}.
\newblock \emph{arXiv preprint arXiv:2308.02408}, 2023{\natexlab{b}}.

\bibitem[Baker(2016)]{Baker2016}
M.~Baker.
\newblock 1,500 scientists lift the lid on reproducibility.
\newblock \emph{Nature}, 533\penalty0 (7604):\penalty0 452--454, May 2016.

\bibitem[Banville et~al.(2021)Banville, Chehab, Hyv{\"a}rinen, Engemann, and
  Gramfort]{banville2021uncovering}
H.~Banville, O.~Chehab, A.~Hyv{\"a}rinen, D.-A. Engemann, and A.~Gramfort.
\newblock Uncovering the structure of clinical {EEG} signals with
  self-supervised learning.
\newblock \emph{Journal of Neural Engineering}, 18\penalty0 (4):\penalty0
  046020, 2021.

\bibitem[Barachant(2012)]{barachant2012commande}
A.~Barachant.
\newblock \emph{Commande robuste d'un effecteur par une interface cerveau
  machine {EEG} asynchrone}.
\newblock PhD thesis, Grenoble, 2012.

\bibitem[Barachant(2014)]{barachant2014meg}
A.~Barachant.
\newblock {MEG} decoding using {Riemannian} geometry and unsupervised
  classification.
\newblock \emph{Grenoble University: Grenoble, France}, 2014.

\bibitem[Barachant and Congedo(2014)]{barachant2014plug}
A.~Barachant and M.~Congedo.
\newblock A plug\&play {P300 BCI} using information geometry.
\newblock \emph{arXiv preprint arXiv:1409.0107}, 2014.

\bibitem[Barachant et~al.(2010{\natexlab{a}})Barachant, Bonnet, Congedo, and
  Jutten]{5662067}
A.~Barachant, S.~Bonnet, M.~Congedo, and C.~Jutten.
\newblock Common spatial pattern revisited by riemannian geometry.
\newblock In \emph{IEEE International Workshop on Multimedia Signal
  Processing}, pages 472--476, 2010{\natexlab{a}}.
\newblock \doi{10.1109/MMSP.2010.5662067}.

\bibitem[Barachant et~al.(2010{\natexlab{b}})Barachant, Bonnet, Congedo, and
  Jutten]{barachant2010riemannian}
A.~Barachant, S.~Bonnet, M.~Congedo, and C.~Jutten.
\newblock Riemannian geometry applied to {BCI} classification.
\newblock \emph{Lva/Ica}, 10:\penalty0 629--636, 2010{\natexlab{b}}.

\bibitem[Barachant et~al.(2011)Barachant, Bonnet, Congedo, and
  Jutten]{barachant2011multiclass}
A.~Barachant, S.~Bonnet, M.~Congedo, and C.~Jutten.
\newblock Multiclass brain--computer interface classification by {Riemannian}
  geometry.
\newblock \emph{IEEE Transactions on Biomedical Engineering}, 59\penalty0
  (4):\penalty0 920--928, 2011.

\bibitem[Barachant et~al.(2023)Barachant, Barthélemy, King, Gramfort,
  Chevallier, Rodrigues, Olivetti, Goncharenko, vom Berg, Reguig, Lebeurrier,
  Bjäreholt, Yamamoto, Clisson, and Corsi]{pyriemann}
A.~Barachant, Q.~Barthélemy, J.-R. King, A.~Gramfort, S.~Chevallier, P.~L.~C.
  Rodrigues, E.~Olivetti, V.~Goncharenko, G.~W. vom Berg, G.~Reguig,
  A.~Lebeurrier, E.~Bjäreholt, M.~S. Yamamoto, P.~Clisson, and M.-C. Corsi.
\newblock pyriemann, June 2023.
\newblock URL \url{https://doi.org/10.5281/zenodo.593816}.

\bibitem[Barth{\'e}lemy et~al.(2023)Barth{\'e}lemy, Chevallier, Bertrand-Lalo,
  and Clisson]{barthelemy2023end}
Q.~Barth{\'e}lemy, S.~Chevallier, R.~Bertrand-Lalo, and P.~Clisson.
\newblock End-to-end {P300 BCI} using {B}ayesian accumulation of riemannian
  probabilities.
\newblock \emph{Brain-Computer Interfaces}, 10\penalty0 (1):\penalty0 50--61,
  2023.

\bibitem[Bianchi et~al.(2021)Bianchi, Liti, Liuzzi, Piccialli, and
  Salvatore]{bianchi2021improving}
L.~Bianchi, C.~Liti, G.~Liuzzi, V.~Piccialli, and C.~Salvatore.
\newblock Improving {P300} speller performance by means of optimization and
  machine learning.
\newblock \emph{Annals of Operations Research}, pages 1--39, 2021.

\bibitem[Bin et~al.(2009{\natexlab{a}})Bin, Gao, Wang, Hong, and
  Gao]{bin2009vep}
G.~Bin, X.~Gao, Y.~Wang, B.~Hong, and S.~Gao.
\newblock {VEP}-based brain-computer interfaces: time, frequency, and code
  modulations [research frontier].
\newblock \emph{IEEE Computational Intelligence Magazine}, 4\penalty0
  (4):\penalty0 22--26, 2009{\natexlab{a}}.

\bibitem[Bin et~al.(2009{\natexlab{b}})Bin, Gao, Yan, Hong, and
  Gao]{bin2009online}
G.~Bin, X.~Gao, Z.~Yan, B.~Hong, and S.~Gao.
\newblock An online multi-channel {SSVEP}-based brain--computer interface using
  a canonical correlation analysis method.
\newblock \emph{Journal of neural engineering}, 6\penalty0 (4):\penalty0
  046002, 2009{\natexlab{b}}.

\bibitem[Blankertz et~al.(2007)Blankertz, Tomioka, Lemm, Kawanabe, and
  Muller]{blankertz2007optimizing}
B.~Blankertz, R.~Tomioka, S.~Lemm, M.~Kawanabe, and K.-R. Muller.
\newblock Optimizing spatial filters for robust {EEG} single-trial analysis.
\newblock \emph{IEEE Signal processing magazine}, 25\penalty0 (1):\penalty0
  41--56, 2007.

\bibitem[Boumal(2023)]{boumal2023intromanifolds}
N.~Boumal.
\newblock \emph{An introduction to optimization on smooth manifolds}.
\newblock Cambridge University Press, 2023.

\bibitem[Cardona-{\'A}lvarez et~al.(2023)Cardona-{\'A}lvarez, {\'A}lvarez-Meza,
  C{\'a}rdenas-Pe{\~n}a, Casta{\~n}o-Duque, and
  Castellanos-Dominguez]{cardona2023novel}
Y.~N. Cardona-{\'A}lvarez, A.~M. {\'A}lvarez-Meza, D.~A. C{\'a}rdenas-Pe{\~n}a,
  G.~A. Casta{\~n}o-Duque, and G.~Castellanos-Dominguez.
\newblock A novel {OpenBCI} framework for {EEG}-based neurophysiological
  experiments.
\newblock \emph{Sensors}, 23\penalty0 (7):\penalty0 3763, 2023.

\bibitem[Carrara and Papadopoulo(2023)]{carrara2023classification}
I.~Carrara and T.~Papadopoulo.
\newblock Classification of {BCI-EEG} based on augmented covariance matrix.
\newblock \emph{arXiv preprint arXiv:2302.04508}, 2023.

\bibitem[Carrara and Papadopoulo(2024)]{carrara2024pseudo}
I.~Carrara and T.~Papadopoulo.
\newblock Pseudo-online framework for {BCI} evaluation: a {MOABB} perspective
  using various {MI} and {SSVEP} datasets.
\newblock \emph{Journal of Neural Engineering}, 21\penalty0 (1):\penalty0
  016003, 2024.

\bibitem[Cattan et~al.(2019)Cattan, Andreev, Rodrigues, and
  Congedo]{cattan2019dataset}
G.~Cattan, A.~Andreev, P.~Rodrigues, and M.~Congedo.
\newblock Dataset of an {EEG-based BCI} experiment in virtual reality and on a
  personal computer.
\newblock \emph{arXiv preprint arXiv:1903.11297}, 2019.

\bibitem[Cawley and Talbot(2010)]{cawley2010over}
G.~C. Cawley and N.~L. Talbot.
\newblock On over-fitting in model selection and subsequent selection bias in
  performance evaluation.
\newblock \emph{The Journal of Machine Learning Research}, 11:\penalty0
  2079--2107, 2010.

\bibitem[Chen et~al.(2020)Chen, Kornblith, Norouzi, and Hinton]{chen2020simple}
T.~Chen, S.~Kornblith, M.~Norouzi, and G.~Hinton.
\newblock A simple framework for contrastive learning of visual
  representations.
\newblock In \emph{International conference on machine learning}, pages
  1597--1607. PMLR, 2020.

\bibitem[Chen et~al.(2022)Chen, Teng, Chen, Pan, and Geyer]{chen2022toward}
X.~Chen, X.~Teng, H.~Chen, Y.~Pan, and P.~Geyer.
\newblock Toward reliable signals decoding for electroencephalogram: A
  benchmark study to {EEGNeX}.
\newblock \emph{arXiv preprint arXiv:2207.12369}, 2022.

\bibitem[Chevallier et~al.(2018{\natexlab{a}})Chevallier, Bao, Hammami,
  Marlats, Mayaud, Annane, Lofaso, and Azabou]{chevallier2018brain}
S.~Chevallier, G.~Bao, M.~Hammami, F.~Marlats, L.~Mayaud, D.~Annane, F.~Lofaso,
  and E.~Azabou.
\newblock Brain-machine interface for mechanical ventilation using
  respiratory-related evoked potential.
\newblock In \emph{Artificial Neural Networks and Machine Learning--ICANN 2018:
  27th International Conference on Artificial Neural Networks, Rhodes, Greece,
  October 4-7, 2018, Proceedings, Part III 27}, pages 662--671. Springer,
  2018{\natexlab{a}}.

\bibitem[Chevallier et~al.(2018{\natexlab{b}})Chevallier, Kalunga,
  Barth{\'e}lemy, and Yger]{chevallier2018riemannian}
S.~Chevallier, E.~Kalunga, Q.~Barth{\'e}lemy, and F.~Yger.
\newblock Riemannian classification for {SSVEP based BCI}: offline versus
  online implementations.
\newblock In \emph{Brain--Computer Interfaces Handbook: Technological and
  Theoretical Advances}. Taylor \& Francis, 2018{\natexlab{b}}.

\bibitem[Cho et~al.(2017)Cho, Ahn, Ahn, Kwon, and Jun]{cho2017eeg}
H.~Cho, M.~Ahn, S.~Ahn, M.~Kwon, and S.~C. Jun.
\newblock {EEG} datasets for motor imagery brain--computer interface.
\newblock \emph{GigaScience}, 6\penalty0 (7):\penalty0 gix034, 2017.

\bibitem[Clisson et~al.(2019)Clisson, Bertrand-Lalo, Congedo, Victor-Thomas,
  and Chatel-Goldman]{clisson2019timeflux}
P.~Clisson, R.~Bertrand-Lalo, M.~Congedo, G.~Victor-Thomas, and
  J.~Chatel-Goldman.
\newblock Timeflux: an open-source framework for the acquisition and near
  real-time processing of signal streams.
\newblock In \emph{BCI 2019-8th International Brain-Computer Interface
  Conference}, 2019.

\bibitem[Congedo et~al.(2011)Congedo, Goyat, Tarrin, Ionescu, Varnet, Rivet,
  Phlypo, Jrad, Acquadro, and Jutten]{congedo2011brain}
M.~Congedo, M.~Goyat, N.~Tarrin, G.~Ionescu, L.~Varnet, B.~Rivet, R.~Phlypo,
  N.~Jrad, M.~Acquadro, and C.~Jutten.
\newblock Brain invaders: a prototype of an open-source {P300}-based video game
  working with the {OpenViBE} platform.
\newblock In \emph{BCI 2011-5th International Brain-Computer Interface
  Conference}, pages 280--283, 2011.

\bibitem[Congedo et~al.(2017)Congedo, Barachant, and
  Bhatia]{congedo2017riemannian}
M.~Congedo, A.~Barachant, and R.~Bhatia.
\newblock Riemannian geometry for {EEG}-based brain-computer interfaces; a
  primer and a review.
\newblock \emph{Brain-Computer Interfaces}, 4\penalty0 (3):\penalty0 155--174,
  2017.

\bibitem[Corsi et~al.(2022)Corsi, Chevallier, Fallani, and
  Yger]{corsi2022functional}
M.-C. Corsi, S.~Chevallier, F.~D.~V. Fallani, and F.~Yger.
\newblock Functional connectivity ensemble method to enhance {BCI} performance
  ({FUCONE}).
\newblock \emph{IEEE Transactions on Biomedical Engineering}, 69\penalty0
  (9):\penalty0 2826--2838, 2022.

\bibitem[Courty et~al.(2024)Courty, Schmidt, Goyal-Kamal, MarionCoutarel, Feld,
  Lecourt, LiamConnell, SabAmine, kngoyal, inimaz, Léval, Blanche, Cruveiller,
  ouminasara, Zhao, Joshi, Bogroff, Saboni, de~Lavoreille, Laskaris, Abati,
  Blank, Wang, Catovic, alencon, Stechly, JPW, MinervaBooks, Carkaci, and
  Crall]{codecarbon}
B.~Courty, V.~Schmidt, Goyal-Kamal, MarionCoutarel, B.~Feld, J.~Lecourt,
  LiamConnell, SabAmine, kngoyal, inimaz, M.~Léval, L.~Blanche, A.~Cruveiller,
  ouminasara, F.~Zhao, A.~Joshi, A.~Bogroff, A.~Saboni, H.~de~Lavoreille,
  N.~Laskaris, E.~Abati, D.~Blank, Z.~Wang, A.~Catovic, alencon, M.~Stechly,
  JPW, MinervaBooks, N.~Carkaci, and J.~Crall.
\newblock Codecarbon, 2024.

\bibitem[Craik et~al.(2019)Craik, He, and Contreras-Vidal]{craik2019deep}
A.~Craik, Y.~He, and J.~L. Contreras-Vidal.
\newblock Deep learning for electroencephalogram {EEG} classification tasks: a
  review.
\newblock \emph{Journal of neural engineering}, 16\penalty0 (3):\penalty0
  031001, 2019.

\bibitem[Delorme(2023)]{delorme2023eeg}
A.~Delorme.
\newblock {EEG} is better left alone.
\newblock \emph{Scientific reports}, 13\penalty0 (1):\penalty0 2372, 2023.

\bibitem[Faller et~al.(2012)Faller, Vidaurre, Solis-Escalante, Neuper, and
  Scherer]{faller2012autocalibration}
J.~Faller, C.~Vidaurre, T.~Solis-Escalante, C.~Neuper, and R.~Scherer.
\newblock Autocalibration and recurrent adaptation: Towards a plug and play
  online {ERD-BCI}.
\newblock \emph{IEEE Transactions on Neural Systems and Rehabilitation
  Engineering}, 20\penalty0 (3):\penalty0 313--319, 2012.

\bibitem[Farwell and Donchin(1988)]{farwell1988talking}
L.~A. Farwell and E.~Donchin.
\newblock Talking off the top of your head: toward a mental prosthesis
  utilizing event-related brain potentials.
\newblock \emph{Electroencephalography and clinical Neurophysiology},
  70\penalty0 (6):\penalty0 510--523, 1988.

\bibitem[Frey(2016)]{frey2016comparison}
J.~Frey.
\newblock Comparison of an open-hardware electroencephalography amplifier with
  medical grade device in brain-computer interface applications.
\newblock In \emph{PhyCS-International Conference on Physiological Computing
  Systems}. SCITEPRESS, 2016.

\bibitem[Gao et~al.(2014)Gao, Wang, Gao, and Hong]{gao2014visual}
S.~Gao, Y.~Wang, X.~Gao, and B.~Hong.
\newblock Visual and auditory brain--computer interfaces.
\newblock \emph{IEEE Transactions on Biomedical Engineering}, 61\penalty0
  (5):\penalty0 1436--1447, 2014.

\bibitem[Garcia~Badaracco(2020)]{scikeras}
A.~Garcia~Badaracco.
\newblock \emph{{SciKeras}}, 2020.
\newblock URL \url{https://github.com/adriangb/scikeras}.

\bibitem[Gramfort et~al.(2013)Gramfort, Luessi, Larson, Engemann, Strohmeier,
  Brodbeck, Goj, Jas, Brooks, Parkkonen, and
  H{\"a}m{\"a}l{\"a}inen]{GramfortEtAl2013a}
A.~Gramfort, M.~Luessi, E.~Larson, D.~A. Engemann, D.~Strohmeier, C.~Brodbeck,
  R.~Goj, M.~Jas, T.~Brooks, L.~Parkkonen, and M.~S. H{\"a}m{\"a}l{\"a}inen.
\newblock {MEG} and {EEG} data analysis with {{MNE}}-{{Python}}.
\newblock \emph{Frontiers in Neuroscience}, 7\penalty0 (267):\penalty0 1--13,
  2013.
\newblock \doi{10.3389/fnins.2013.00267}.

\bibitem[Grosse-Wentrup et~al.(2009)Grosse-Wentrup, Liefhold, Gramann, and
  Buss]{grosse2009beamforming}
M.~Grosse-Wentrup, C.~Liefhold, K.~Gramann, and M.~Buss.
\newblock Beamforming in noninvasive brain--computer interfaces.
\newblock \emph{IEEE Transactions on Biomedical Engineering}, 56\penalty0
  (4):\penalty0 1209--1219, 2009.

\bibitem[Guetschel and Tangermann(2023)]{guetschel23transfer}
P.~Guetschel and M.~Tangermann.
\newblock Transfer learning between motor imagery datasets using deep learning
  - validation of framework and comparison of datasets, Nov. 2023.

\bibitem[Guger et~al.(2009)Guger, Daban, Sellers, Holzner, Krausz, Carabalona,
  Gramatica, and Edlinger]{guger2009many}
C.~Guger, S.~Daban, E.~Sellers, C.~Holzner, G.~Krausz, R.~Carabalona,
  F.~Gramatica, and G.~Edlinger.
\newblock How many people are able to control a {P300}-based brain--computer
  interface ({BCI})?
\newblock \emph{Neuroscience letters}, 462\penalty0 (1):\penalty0 94--98, 2009.

\bibitem[Harris et~al.(2020)Harris, Millman, van~der Walt, Gommers, Virtanen,
  Cournapeau, Wieser, Taylor, Berg, Smith, Kern, Picus, Hoyer, van Kerkwijk,
  Brett, Haldane, del R{\'{i}}o, Wiebe, Peterson, G{\'{e}}rard-Marchant,
  Sheppard, Reddy, Weckesser, Abbasi, Gohlke, and Oliphant]{harris2020array}
C.~R. Harris, K.~J. Millman, S.~J. van~der Walt, R.~Gommers, P.~Virtanen,
  D.~Cournapeau, E.~Wieser, J.~Taylor, S.~Berg, N.~J. Smith, R.~Kern, M.~Picus,
  S.~Hoyer, M.~H. van Kerkwijk, M.~Brett, A.~Haldane, J.~F. del R{\'{i}}o,
  M.~Wiebe, P.~Peterson, P.~G{\'{e}}rard-Marchant, K.~Sheppard, T.~Reddy,
  W.~Weckesser, H.~Abbasi, C.~Gohlke, and T.~E. Oliphant.
\newblock Array programming with {NumPy}.
\newblock \emph{Nature}, 585\penalty0 (7825):\penalty0 357--362, Sept. 2020.

\bibitem[Hedges and Olkin(2014)]{hedges2014statistical}
L.~V. Hedges and I.~Olkin.
\newblock \emph{Statistical methods for meta-analysis}.
\newblock Academic press, 2014.

\bibitem[Hoffmann et~al.(2008)Hoffmann, Vesin, Ebrahimi, and
  Diserens]{hoffmann2008efficient}
U.~Hoffmann, J.-M. Vesin, T.~Ebrahimi, and K.~Diserens.
\newblock An efficient {P300}-based brain--computer interface for disabled
  subjects.
\newblock \emph{Journal of Neuroscience methods}, 167\penalty0 (1):\penalty0
  115--125, 2008.

\bibitem[H{\"u}bner et~al.(2017)H{\"u}bner, Verhoeven, Schmid, M{\"u}ller,
  Tangermann, and Kindermans]{hubner2017learning}
D.~H{\"u}bner, T.~Verhoeven, K.~Schmid, K.-R. M{\"u}ller, M.~Tangermann, and
  P.-J. Kindermans.
\newblock Learning from label proportions in brain-computer interfaces: Online
  unsupervised learning with guarantees.
\newblock \emph{PloS one}, 12\penalty0 (4):\penalty0 e0175856, 2017.

\bibitem[H{\"u}ebner et~al.(2018)H{\"u}ebner, Verhoeven, M{\"u}eller,
  Kindermans, and Tangermann]{huebner2018unsupervised}
D.~H{\"u}ebner, T.~Verhoeven, K.-R. M{\"u}eller, P.-J. Kindermans, and
  M.~Tangermann.
\newblock Unsupervised learning for brain-computer interfaces based on
  event-related potentials: Review and online comparison.
\newblock \emph{IEEE Computational Intelligence Magazine}, 13\penalty0
  (2):\penalty0 66--77, 2018.

\bibitem[Ingolfsson et~al.(2020)Ingolfsson, Hersche, Wang, Kobayashi,
  Cavigelli, and Benini]{ingolfsson2020eegtcnet}
T.~M. Ingolfsson, M.~Hersche, X.~Wang, N.~Kobayashi, L.~Cavigelli, and
  L.~Benini.
\newblock {EEG-TCNet}: An accurate temporal convolutional network for embedded
  motor-imagery brain--machine interfaces.
\newblock In \emph{IEEE International Conference on Systems, Man, and
  Cybernetics (SMC)}, pages 2958--2965. IEEE, 2020.

\bibitem[Jay et~al.(2023)Jay, Ostapenco, Lef{\`e}vre, Trystram, Orgerie, and
  Fichel]{jay2023experimental}
M.~Jay, V.~Ostapenco, L.~Lef{\`e}vre, D.~Trystram, A.-C. Orgerie, and
  B.~Fichel.
\newblock An experimental comparison of software-based power meters: focus on
  {CPU} and {GPU}.
\newblock In \emph{IEEE/ACM international symposium on cluster, cloud and
  internet computing}, pages 1--13, 2023.

\bibitem[Jayaram and Barachant(2018)]{jayaram2018moabb}
V.~Jayaram and A.~Barachant.
\newblock {MOABB}: trustworthy algorithm benchmarking for bcis.
\newblock \emph{Journal of neural engineering}, 15\penalty0 (6):\penalty0
  066011, 2018.

\bibitem[Jayaram et~al.(2016)Jayaram, Alamgir, Altun, Scholkopf, and
  Grosse-Wentrup]{jayaram2016transfer}
V.~Jayaram, M.~Alamgir, Y.~Altun, B.~Scholkopf, and M.~Grosse-Wentrup.
\newblock Transfer learning in brain-computer interfaces.
\newblock \emph{IEEE Computational Intelligence Magazine}, 11\penalty0
  (1):\penalty0 20--31, 2016.

\bibitem[Kalunga et~al.(2016)Kalunga, Chevallier, Barth{\'e}lemy, Djouani,
  Monacelli, and Hamam]{kalunga2016online}
E.~K. Kalunga, S.~Chevallier, Q.~Barth{\'e}lemy, K.~Djouani, E.~Monacelli, and
  Y.~Hamam.
\newblock Online {SSVEP-based BCI} using {Riemannian} geometry.
\newblock \emph{Neurocomputing}, 191:\penalty0 55--68, 2016.

\bibitem[Kalunga et~al.(2018)Kalunga, Chevallier, and
  Barth{\'e}lemy]{kalunga2018transfer}
E.~K. Kalunga, S.~Chevallier, and Q.~Barth{\'e}lemy.
\newblock Transfer learning for {SSVEP-based BCI} using riemannian similarities
  between users.
\newblock In \emph{EUSIPCO}, pages 1685--1689, 2018.

\bibitem[Khazem et~al.(2021)Khazem, Chevallier, Barth{\'e}lemy, Haroun, and
  No{\^u}s]{khazem2021minimizing}
S.~Khazem, S.~Chevallier, Q.~Barth{\'e}lemy, K.~Haroun, and C.~No{\^u}s.
\newblock Minimizing subject-dependent calibration for {BCI} with {Riemannian}
  transfer learning.
\newblock In \emph{International IEEE/EMBS Conference on Neural Engineering
  (NER)}, pages 523--526. IEEE, 2021.

\bibitem[Kindermans et~al.(2012)Kindermans, Verschore, Verstraeten, and
  Schrauwen]{kindermans2012p300}
P.-J. Kindermans, H.~Verschore, D.~Verstraeten, and B.~Schrauwen.
\newblock A {P300} {BCI} for the masses: Prior information enables instant
  unsupervised spelling.
\newblock \emph{Advances in neural information processing systems}, 25, 2012.

\bibitem[Kingma and Ba(2014)]{adam}
D.~P. Kingma and J.~Ba.
\newblock Adam: A method for stochastic optimization.
\newblock \emph{arXiv preprint arXiv:1412.6980}, 2014.

\bibitem[Koles et~al.(1990)Koles, Lazar, and Zhou]{koles1990spatial}
Z.~J. Koles, M.~S. Lazar, and S.~Z. Zhou.
\newblock Spatial patterns underlying population differences in the background
  {EEG}.
\newblock \emph{Brain topography}, 2:\penalty0 275--284, 1990.

\bibitem[Korczowski et~al.(2019{\natexlab{a}})Korczowski, Cederhout, Andreev,
  Cattan, Rodrigues, Gautheret, and Congedo]{korczowski2019brain_2015_a}
L.~Korczowski, M.~Cederhout, A.~Andreev, G.~Cattan, P.~L.~C. Rodrigues,
  V.~Gautheret, and M.~Congedo.
\newblock \emph{Brain Invaders calibration-less {P300}-based {BCI} with
  modulation of flash duration Dataset (bi2015a)}, 2019{\natexlab{a}}.

\bibitem[Korczowski et~al.(2019{\natexlab{b}})Korczowski, Cederhout, Andreev,
  Cattan, Rodrigues, Gautheret, and Congedo]{korczowski2019brain_2015_b}
L.~Korczowski, M.~Cederhout, A.~Andreev, G.~Cattan, P.~L.~C. Rodrigues,
  V.~Gautheret, and M.~Congedo.
\newblock \emph{Brain Invaders Cooperative versus Competitive: Multi-User
  {P300}-based Brain-Computer Interface Dataset (bi2015b)}, 2019{\natexlab{b}}.

\bibitem[Korczowski et~al.(2019{\natexlab{c}})Korczowski, Ostaschenko, Andreev,
  Cattan, Rodrigues, Gautheret, and Congedo]{korczowski2019brain_a}
L.~Korczowski, E.~Ostaschenko, A.~Andreev, G.~Cattan, P.~L.~C. Rodrigues,
  V.~Gautheret, and M.~Congedo.
\newblock \emph{Brain Invaders calibration-less {P300-based BCI} using dry
  {EEG} electrodes Dataset (bi2014a)}, 2019{\natexlab{c}}.

\bibitem[Korczowski et~al.(2019{\natexlab{d}})Korczowski, Ostaschenko, Andreev,
  Cattan, Rodrigues, Gautheret, and Congedo]{korczowski2019brain_b}
L.~Korczowski, E.~Ostaschenko, A.~Andreev, G.~Cattan, P.~L.~C. Rodrigues,
  V.~Gautheret, and M.~Congedo.
\newblock \emph{Brain Invaders Solo versus Collaboration: Multi-User
  {P300}-based Brain-Computer Interface Dataset (bi2014b)}, 2019{\natexlab{d}}.

\bibitem[Krizhevsky et~al.(2017)Krizhevsky, Sutskever, and
  Hinton]{krizhevsky2017imagenet}
A.~Krizhevsky, I.~Sutskever, and G.~E. Hinton.
\newblock Imagenet classification with deep convolutional neural networks.
\newblock \emph{Communications of the ACM}, 60\penalty0 (6):\penalty0 84--90,
  2017.

\bibitem[Lawhern et~al.(2018)Lawhern, Solon, Waytowich, Gordon, Hung, and
  Lance]{lawhern2018eegnet}
V.~J. Lawhern, A.~J. Solon, N.~R. Waytowich, S.~M. Gordon, C.~P. Hung, and
  B.~J. Lance.
\newblock {EEGNet}: a compact convolutional neural network for {EEG}-based
  brain--computer interfaces.
\newblock \emph{Journal of neural engineering}, 15\penalty0 (5):\penalty0
  056013, 2018.

\bibitem[LeCun et~al.(2002)LeCun, Bottou, Orr, and
  M{\"u}ller]{lecun2002efficient}
Y.~LeCun, L.~Bottou, G.~B. Orr, and K.-R. M{\"u}ller.
\newblock Efficient backprop.
\newblock In \emph{Neural networks: Tricks of the trade}, pages 9--50.
  Springer, 2002.

\bibitem[Lee et~al.(2019)Lee, Kwon, Kim, Kim, Lee, Williamson, Fazli, and
  Lee]{lee2019eeg}
M.-H. Lee, O.-Y. Kwon, Y.-J. Kim, H.-K. Kim, Y.-E. Lee, J.~Williamson,
  S.~Fazli, and S.-W. Lee.
\newblock {EEG} dataset and openbmi toolbox for three {BCI} paradigms: An
  investigation into {BCI} illiteracy.
\newblock \emph{GigaScience}, 8\penalty0 (5):\penalty0 giz002, 2019.

\bibitem[Leeb et~al.(2007)Leeb, Lee, Keinrath, Scherer, Bischof, and
  Pfurtscheller]{leeb2007brain}
R.~Leeb, F.~Lee, C.~Keinrath, R.~Scherer, H.~Bischof, and G.~Pfurtscheller.
\newblock Brain--computer communication: motivation, aim, and impact of
  exploring a virtual apartment.
\newblock \emph{IEEE Transactions on Neural Systems and Rehabilitation
  Engineering}, 15\penalty0 (4):\penalty0 473--482, 2007.

\bibitem[Ligozat et~al.(2022)Ligozat, Lefevre, Bugeau, and
  Combaz]{ligozat2022unraveling}
A.-L. Ligozat, J.~Lefevre, A.~Bugeau, and J.~Combaz.
\newblock Unraveling the hidden environmental impacts of {AI} solutions for
  environment life cycle assessment of {AI} solutions.
\newblock \emph{Sustainability}, 14\penalty0 (9):\penalty0 5172, 2022.

\bibitem[Lin et~al.(2006)Lin, Zhang, Wu, and Gao]{lin2006frequency}
Z.~Lin, C.~Zhang, W.~Wu, and X.~Gao.
\newblock Frequency recognition based on canonical correlation analysis for
  {SSVEP}-based {BCI}s.
\newblock \emph{IEEE transactions on biomedical engineering}, 53\penalty0
  (12):\penalty0 2610--2614, 2006.

\bibitem[Lotte and Guan(2010)]{lotte2010regularizing}
F.~Lotte and C.~Guan.
\newblock Regularizing common spatial patterns to improve {BCI} designs:
  unified theory and new algorithms.
\newblock \emph{IEEE Transactions on biomedical Engineering}, 58\penalty0
  (2):\penalty0 355--362, 2010.

\bibitem[Lotte et~al.(2007)Lotte, Congedo, L{\'e}cuyer, Lamarche, and
  Arnaldi]{lotte2007review}
F.~Lotte, M.~Congedo, A.~L{\'e}cuyer, F.~Lamarche, and B.~Arnaldi.
\newblock A review of classification algorithms for {EEG}-based brain--computer
  interfaces.
\newblock \emph{Journal of neural engineering}, 4\penalty0 (2):\penalty0 R1,
  2007.

\bibitem[Lotte et~al.(2018)Lotte, Bougrain, Cichocki, Clerc, Congedo,
  Rakotomamonjy, and Yger]{lotte2018review}
F.~Lotte, L.~Bougrain, A.~Cichocki, M.~Clerc, M.~Congedo, A.~Rakotomamonjy, and
  F.~Yger.
\newblock A review of classification algorithms for {EEG}-based brain--computer
  interfaces: a 10 year update.
\newblock \emph{Journal of neural engineering}, 15\penalty0 (3):\penalty0
  031005, 2018.

\bibitem[Luccioni et~al.(2023)Luccioni, Viguier, and
  Ligozat]{luccioni2023estimating}
A.~S. Luccioni, S.~Viguier, and A.-L. Ligozat.
\newblock Estimating the carbon footprint of bloom, a 176b parameter language
  model.
\newblock \emph{Journal of Machine Learning Research}, 24\penalty0
  (253):\penalty0 1--15, 2023.

\bibitem[Luck(2014)]{luck_2014}
S.~J. Luck.
\newblock \emph{An introduction to the event-related potential technique}.
\newblock The MIT Press, 2014.

\bibitem[Mart{\'\i}nez-Cagigal et~al.(2021)Mart{\'\i}nez-Cagigal, Thielen,
  Santamaria-Vazquez, P{\'e}rez-Velasco, Desain, and
  Hornero]{martinez2021brain}
V.~Mart{\'\i}nez-Cagigal, J.~Thielen, E.~Santamaria-Vazquez,
  S.~P{\'e}rez-Velasco, P.~Desain, and R.~Hornero.
\newblock Brain-computer interfaces based on code-modulated visual evoked
  potentials (c-{VEP}): A literature review.
\newblock \emph{Journal of Neural Engineering}, 18\penalty0 (6):\penalty0
  061002, 2021.

\bibitem[Moakher(2005)]{moakher2005differential}
M.~Moakher.
\newblock A differential geometric approach to the geometric mean of symmetric
  positive-definite matrices.
\newblock \emph{SIAM journal on matrix analysis and applications}, 26\penalty0
  (3):\penalty0 735--747, 2005.

\bibitem[Moreau et~al.(2022)Moreau, Massias, Gramfort, Ablin, Bannier,
  Charlier, Dagr{\'e}ou, {la Tour}, Durif, Dantas, Klopfenstein, Larsson, Lai,
  Lefort, Mal{\'e}zieux, Moufad, Nguyen, Rakotomamonjy, Ramzi, Salmon, and
  Vaiter]{Moreau2022}
T.~Moreau, M.~Massias, A.~Gramfort, P.~Ablin, P.-A. Bannier, B.~Charlier,
  M.~Dagr{\'e}ou, T.~D. {la Tour}, G.~Durif, C.~F. Dantas, Q.~Klopfenstein,
  J.~Larsson, E.~Lai, T.~Lefort, B.~Mal{\'e}zieux, B.~Moufad, B.~T. Nguyen,
  A.~Rakotomamonjy, Z.~Ramzi, J.~Salmon, and S.~Vaiter.
\newblock Benchopt: {{Reproducible}}, efficient and collaborative optimization
  benchmarks.
\newblock In \emph{Advances in {{Neural Information Processing Systems}}
  ({{NeurIPS}})}, volume~36, New-Orleans, LA, USA, Nov. 2022. Curran
  Associates, Inc.

\bibitem[M{\"u}ller-Gerking et~al.(1999)M{\"u}ller-Gerking, Pfurtscheller, and
  Flyvbjerg]{muller1999designing}
J.~M{\"u}ller-Gerking, G.~Pfurtscheller, and H.~Flyvbjerg.
\newblock Designing optimal spatial filters for single-trial {EEG}
  classification in a movement task.
\newblock \emph{Clinical neurophysiology}, 110\penalty0 (5):\penalty0 787--798,
  1999.

\bibitem[Nagel and Sp{\"u}ler(2019)]{nagel2019world}
S.~Nagel and M.~Sp{\"u}ler.
\newblock World’s fastest brain-computer interface: combining {EEG2Code} with
  deep learning.
\newblock \emph{PloS one}, 14\penalty0 (9):\penalty0 e0221909, 2019.

\bibitem[Nakanishi et~al.(2014)Nakanishi, Wang, Wang, Mitsukura, and
  Jung]{nakanishi2014enhancing}
M.~Nakanishi, Y.~Wang, Y.-T. Wang, Y.~Mitsukura, and T.-P. Jung.
\newblock Enhancing unsupervised canonical correlation analysis-based frequency
  detection of {SSVEP}s by incorporating background {EEG}.
\newblock In \emph{2014 36th Annual International Conference of the IEEE
  Engineering in Medicine and Biology Society}, pages 3053--3056. IEEE, 2014.

\bibitem[Nakanishi et~al.(2015)Nakanishi, Wang, Wang, and
  Jung]{nakanishi2015comparison}
M.~Nakanishi, Y.~Wang, Y.-T. Wang, and T.-P. Jung.
\newblock A comparison study of canonical correlation analysis based methods
  for detecting steady-state visual evoked potentials.
\newblock \emph{PloS one}, 10\penalty0 (10):\penalty0 e0140703, 2015.

\bibitem[Nakanishi et~al.(2017)Nakanishi, Wang, Chen, Wang, Gao, and
  Jung]{nakanishi2017enhancing}
M.~Nakanishi, Y.~Wang, X.~Chen, Y.-T. Wang, X.~Gao, and T.-P. Jung.
\newblock Enhancing detection of {SSVEP}s for a high-speed brain speller using
  task-related component analysis.
\newblock \emph{IEEE Transactions on Biomedical Engineering}, 65\penalty0
  (1):\penalty0 104--112, 2017.

\bibitem[Nam et~al.(2018)Nam, Nijholt, and Lotte]{nam2018brain}
C.~S. Nam, A.~Nijholt, and F.~Lotte.
\newblock \emph{Brain--computer interfaces handbook: technological and
  theoretical advances}.
\newblock CRC Press, 2018.

\bibitem[Oikonomou et~al.(2016)Oikonomou, Liaros, Georgiadis, Chatzilari, Adam,
  Nikolopoulos, and Kompatsiaris]{oikonomou2016comparative}
V.~P. Oikonomou, G.~Liaros, K.~Georgiadis, E.~Chatzilari, K.~Adam,
  S.~Nikolopoulos, and I.~Kompatsiaris.
\newblock Comparative evaluation of state-of-the-art algorithms for
  {SSVEP-based BCIs}.
\newblock \emph{arXiv preprint arXiv:1602.00904}, 2016.

\bibitem[Paszke et~al.(2019)Paszke, Gross, Massa, Lerer, Bradbury, Chanan,
  Killeen, Lin, Gimelshein, Antiga, Desmaison, Kopf, Yang, DeVito, Raison,
  Tejani, Chilamkurthy, Steiner, Fang, Bai, and Chintala]{pytorch}
A.~Paszke, S.~Gross, F.~Massa, A.~Lerer, J.~Bradbury, G.~Chanan, T.~Killeen,
  Z.~Lin, N.~Gimelshein, L.~Antiga, A.~Desmaison, A.~Kopf, E.~Yang, Z.~DeVito,
  M.~Raison, A.~Tejani, S.~Chilamkurthy, B.~Steiner, L.~Fang, J.~Bai, and
  S.~Chintala.
\newblock {PyTorch}: An imperative style, high-performance deep learning
  library.
\newblock In \emph{Advances in Neural Information Processing Systems 32}, pages
  8024--8035, 2019.

\bibitem[Pedregosa et~al.(2011)Pedregosa, Varoquaux, Gramfort, Michel, Thirion,
  Grisel, Blondel, Prettenhofer, Weiss, Dubourg, Vanderplas, Passos,
  Cournapeau, Brucher, Perrot, and Duchesnay]{scikit-learn}
F.~Pedregosa, G.~Varoquaux, A.~Gramfort, V.~Michel, B.~Thirion, O.~Grisel,
  M.~Blondel, P.~Prettenhofer, R.~Weiss, V.~Dubourg, J.~Vanderplas, A.~Passos,
  D.~Cournapeau, M.~Brucher, M.~Perrot, and E.~Duchesnay.
\newblock Scikit-learn: Machine learning in {P}ython.
\newblock \emph{Journal of Machine Learning Research}, 12:\penalty0 2825--2830,
  2011.

\bibitem[Peirce et~al.(2022)Peirce, Hirst, and MacAskill]{peirce2022building}
J.~Peirce, R.~Hirst, and M.~MacAskill.
\newblock \emph{Building experiments in {PsychoPy}}.
\newblock Sage, 2022.

\bibitem[Pernet et~al.(2019)Pernet, Appelhoff, Gorgolewski, Flandin, Phillips,
  Delorme, and Oostenveld]{pernet2019eeg}
C.~R. Pernet, S.~Appelhoff, K.~J. Gorgolewski, G.~Flandin, C.~Phillips,
  A.~Delorme, and R.~Oostenveld.
\newblock {EEG-BIDS}, an extension to the brain imaging data structure for
  electroencephalography.
\newblock \emph{Scientific data}, 6\penalty0 (1):\penalty0 103, 2019.

\bibitem[Ramoser et~al.(2000)Ramoser, Muller-Gerking, and
  Pfurtscheller]{ramoser2000optimal}
H.~Ramoser, J.~Muller-Gerking, and G.~Pfurtscheller.
\newblock Optimal spatial filtering of single trial {EEG} during imagined hand
  movement.
\newblock \emph{IEEE transactions on rehabilitation engineering}, 8\penalty0
  (4):\penalty0 441--446, 2000.

\bibitem[Renard et~al.(2010)Renard, Lotte, Gibert, Congedo, Maby, Delannoy,
  Bertrand, and L{\'e}cuyer]{renard2010openvibe}
Y.~Renard, F.~Lotte, G.~Gibert, M.~Congedo, E.~Maby, V.~Delannoy, O.~Bertrand,
  and A.~L{\'e}cuyer.
\newblock {Openvibe}: An open-source software platform to design, test, and use
  brain--computer interfaces in real and virtual environments.
\newblock \emph{Presence}, 19\penalty0 (1):\penalty0 35--53, 2010.

\bibitem[Riccio et~al.(2013)Riccio, Simione, Schettini, Pizzimenti, Inghilleri,
  Belardinelli, Mattia, and Cincotti]{riccio2013attention}
A.~Riccio, L.~Simione, F.~Schettini, A.~Pizzimenti, M.~Inghilleri, M.~O.
  Belardinelli, D.~Mattia, and F.~Cincotti.
\newblock Attention and {P300-based BCI} performance in people with amyotrophic
  lateral sclerosis.
\newblock \emph{Frontiers in human neuroscience}, 7:\penalty0 732, 2013.

\bibitem[Rivet et~al.(2009)Rivet, Souloumiac, Attina, and
  Gibert]{rivet2009xdawn}
B.~Rivet, A.~Souloumiac, V.~Attina, and G.~Gibert.
\newblock {xDAWN} algorithm to enhance evoked potentials: application to
  brain--computer interface.
\newblock \emph{IEEE Transactions on Biomedical Engineering}, 56\penalty0
  (8):\penalty0 2035--2043, 2009.

\bibitem[RMJ(1949)]{rmj1949american}
S.~RMJ.
\newblock The american soldier, vol. 1: Adjustment during army life, 1949.

\bibitem[Rommel et~al.(2022)Rommel, Paillard, Moreau, and
  Gramfort]{rommel2022data}
C.~Rommel, J.~Paillard, T.~Moreau, and A.~Gramfort.
\newblock Data augmentation for learning predictive models on {EEG}: a
  systematic comparison.
\newblock \emph{Journal of Neural Engineering}, 19\penalty0 (6):\penalty0
  066020, 2022.

\bibitem[Roy(2022)]{roy2022neuroergonomics}
R.~N. Roy.
\newblock \emph{Neuroergonomics and physiological computing contributions to
  human-machine interaction}.
\newblock PhD thesis, Universit{\'e} Paul Sabatier, 2022.

\bibitem[Roy et~al.(2019)Roy, Banville, Albuquerque, Gramfort, Falk, and
  Faubert]{roy2019deep}
Y.~Roy, H.~Banville, I.~Albuquerque, A.~Gramfort, T.~H. Falk, and J.~Faubert.
\newblock Deep learning-based electroencephalography analysis: a systematic
  review.
\newblock \emph{Journal of neural engineering}, 16\penalty0 (5):\penalty0
  051001, 2019.

\bibitem[Salami et~al.(2022)Salami, Andreu-Perez, and
  Gillmeister]{salami2022eeg}
A.~Salami, J.~Andreu-Perez, and H.~Gillmeister.
\newblock {EEG-ITNet}: An explainable inception temporal convolutional network
  for motor imagery classification.
\newblock \emph{IEEE Access}, 10:\penalty0 36672--36685, 2022.

\bibitem[Schalk et~al.(2004)Schalk, McFarland, Hinterberger, Birbaumer, and
  Wolpaw]{schalk2004bci2000}
G.~Schalk, D.~J. McFarland, T.~Hinterberger, N.~Birbaumer, and J.~R. Wolpaw.
\newblock {BCI2000}: a general-purpose brain-computer interface ({BCI}) system.
\newblock \emph{IEEE Transactions on biomedical engineering}, 51\penalty0
  (6):\penalty0 1034--1043, 2004.

\bibitem[Scherer et~al.(2015)Scherer, Faller, Friedrich, Opisso, Costa,
  K{\"u}bler, and M{\"u}ller-Putz]{scherer2015individually}
R.~Scherer, J.~Faller, E.~V. Friedrich, E.~Opisso, U.~Costa, A.~K{\"u}bler, and
  G.~R. M{\"u}ller-Putz.
\newblock Individually adapted imagery improves brain-computer interface
  performance in end-users with disability.
\newblock \emph{PloS one}, 10\penalty0 (5):\penalty0 e0123727, 2015.

\bibitem[Schirrmeister et~al.(2017)Schirrmeister, Springenberg, Fiederer,
  Glasstetter, Eggensperger, Tangermann, Hutter, Burgard, and
  Ball]{schirrmeister2017deep}
R.~T. Schirrmeister, J.~T. Springenberg, L.~D.~J. Fiederer, M.~Glasstetter,
  K.~Eggensperger, M.~Tangermann, F.~Hutter, W.~Burgard, and T.~Ball.
\newblock Deep learning with convolutional neural networks for {EEG} decoding
  and visualization.
\newblock \emph{Human brain mapping}, 38\penalty0 (11):\penalty0 5391--5420,
  2017.

\bibitem[Schneider et~al.(2019)Schneider, Baevski, Collobert, and
  Auli]{schneider2019wav2vec}
S.~Schneider, A.~Baevski, R.~Collobert, and M.~Auli.
\newblock wav2vec: Unsupervised pre-training for speech recognition.
\newblock \emph{arXiv preprint arXiv:1904.05862}, 2019.

\bibitem[Shin et~al.(2016)Shin, von L{\"u}hmann, Blankertz, Kim, Jeong, Hwang,
  and M{\"u}ller]{shin2016open}
J.~Shin, A.~von L{\"u}hmann, B.~Blankertz, D.-W. Kim, J.~Jeong, H.-J. Hwang,
  and K.-R. M{\"u}ller.
\newblock Open access dataset for {EEG+ NIRS} single-trial classification.
\newblock \emph{IEEE Transactions on Neural Systems and Rehabilitation
  Engineering}, 25\penalty0 (10):\penalty0 1735--1745, 2016.

\bibitem[Sosulski and Tangermann(2019)]{sosulski2019spatial}
J.~Sosulski and M.~Tangermann.
\newblock Spatial filters for auditory evoked potentials transfer between
  different experimental conditions.
\newblock In \emph{GBCIC}, 2019.

\bibitem[Sosulski and Tangermann(2023)]{sosulski2023umm}
J.~Sosulski and M.~Tangermann.
\newblock {UMM}: Unsupervised mean-difference maximization.
\newblock \emph{arXiv preprint arXiv:2306.11830}, 2023.

\bibitem[Sosulski et~al.(2021)Sosulski, H{\"u}bner, Klein, and
  Tangermann]{sosulski2021online}
J.~Sosulski, D.~H{\"u}bner, A.~Klein, and M.~Tangermann.
\newblock Online optimization of stimulation speed in an auditory
  brain-computer interface under time constraints.
\newblock \emph{arXiv preprint arXiv:2109.06011}, 2021.

\bibitem[Steyrl et~al.(2016)Steyrl, Scherer, Faller, and
  M{\"u}ller-Putz]{steyrl2016random}
D.~Steyrl, R.~Scherer, J.~Faller, and G.~R. M{\"u}ller-Putz.
\newblock Random forests in non-invasive sensorimotor rhythm brain-computer
  interfaces: a practical and convenient non-linear classifier.
\newblock \emph{Biomedical Engineering/Biomedizinische Technik}, 61\penalty0
  (1):\penalty0 77--86, 2016.

\bibitem[Sun et~al.(2022)Sun, Zheng, Pei, Gao, and Wang]{sun2022120}
Q.~Sun, L.~Zheng, W.~Pei, X.~Gao, and Y.~Wang.
\newblock A 120-target brain-computer interface based on code-modulated visual
  evoked potentials.
\newblock \emph{Journal of Neuroscience Methods}, 375:\penalty0 109597, 2022.

\bibitem[Tangermann et~al.(2012)Tangermann, M{\"u}ller, Aertsen, Birbaumer,
  Braun, Brunner, Leeb, Mehring, Miller, Mueller-Putz,
  et~al.]{tangermann2012review}
M.~Tangermann, K.-R. M{\"u}ller, A.~Aertsen, N.~Birbaumer, C.~Braun,
  C.~Brunner, R.~Leeb, C.~Mehring, K.~J. Miller, G.~Mueller-Putz, et~al.
\newblock Review of the {BCI} competition {IV}.
\newblock \emph{Frontiers in neuroscience}, page~55, 2012.

\bibitem[Thielen et~al.(2015)Thielen, van~den Broek, Farquhar, and
  Desain]{thielen2015broad}
J.~Thielen, P.~van~den Broek, J.~Farquhar, and P.~Desain.
\newblock Broad-band visually evoked potentials: re (con) volution in
  brain-computer interfacing.
\newblock \emph{PloS one}, 10\penalty0 (7):\penalty0 e0133797, 2015.

\bibitem[Thielen et~al.(2021)Thielen, Marsman, Farquhar, and
  Desain]{thielen2021full}
J.~Thielen, P.~Marsman, J.~Farquhar, and P.~Desain.
\newblock From full calibration to zero training for a code-modulated visual
  evoked potentials for brain--computer interface.
\newblock \emph{Journal of Neural Engineering}, 18\penalty0 (5):\penalty0
  056007, 2021.

\bibitem[Tietz et~al.(2017)Tietz, Fan, Nouri, Bossan, and {skorch
  Developers}]{skorch}
M.~Tietz, T.~J. Fan, D.~Nouri, B.~Bossan, and {skorch Developers}.
\newblock \emph{skorch: A scikit-learn compatible neural network library that
  wraps {PyTorch}}, July 2017.
\newblock URL \url{https://skorch.readthedocs.io/en/stable/}.

\bibitem[Vaineau et~al.(2019)Vaineau, Barachant, Andreev, Rodrigues, Cattan,
  and Congedo]{vaineau2019brain}
E.~Vaineau, A.~Barachant, A.~Andreev, P.~C. Rodrigues, G.~Cattan, and
  M.~Congedo.
\newblock Brain invaders adaptive versus non-adaptive {P300} brain-computer
  interface dataset.
\newblock \emph{arXiv preprint arXiv:1904.09111}, 2019.

\bibitem[Van~Veen et~al.(2019)Van~Veen, Barachant, Andreev, Cattan, Rodrigues,
  and Congedo]{van2019building}
G.~Van~Veen, A.~Barachant, A.~Andreev, G.~Cattan, P.~C. Rodrigues, and
  M.~Congedo.
\newblock Building brain invaders: {EEG} data of an experimental validation.
\newblock \emph{arXiv preprint arXiv:1905.05182}, 2019.

\bibitem[Vaswani et~al.(2017)Vaswani, Shazeer, Parmar, Uszkoreit, Jones, Gomez,
  Kaiser, and Polosukhin]{vaswani2017attention}
A.~Vaswani, N.~Shazeer, N.~Parmar, J.~Uszkoreit, L.~Jones, A.~N. Gomez,
  {\L}.~Kaiser, and I.~Polosukhin.
\newblock Attention is all you need.
\newblock \emph{Advances in neural information processing systems}, 30, 2017.

\bibitem[Virtanen et~al.(2020)Virtanen, Gommers, Oliphant, Haberland, Reddy,
  Cournapeau, Burovski, Peterson, Weckesser, Bright, {van der Walt}, Brett,
  Wilson, Millman, Mayorov, Nelson, Jones, Kern, Larson, Carey, Polat, Feng,
  Moore, {VanderPlas}, Laxalde, Perktold, Cimrman, Henriksen, Quintero, Harris,
  Archibald, Ribeiro, Pedregosa, {van Mulbregt}, and {SciPy 1.0
  Contributors}]{2020SciPy-NMeth}
P.~Virtanen, R.~Gommers, T.~E. Oliphant, M.~Haberland, T.~Reddy, D.~Cournapeau,
  E.~Burovski, P.~Peterson, W.~Weckesser, J.~Bright, S.~J. {van der Walt},
  M.~Brett, J.~Wilson, K.~J. Millman, N.~Mayorov, A.~R.~J. Nelson, E.~Jones,
  R.~Kern, E.~Larson, C.~J. Carey, {\.I}.~Polat, Y.~Feng, E.~W. Moore,
  J.~{VanderPlas}, D.~Laxalde, J.~Perktold, R.~Cimrman, I.~Henriksen, E.~A.
  Quintero, C.~R. Harris, A.~M. Archibald, A.~H. Ribeiro, F.~Pedregosa, P.~{van
  Mulbregt}, and {SciPy 1.0 Contributors}.
\newblock {SciPy} 1.0: Fundamental algorithms for scientific computing in
  python.
\newblock \emph{Nature Methods}, 17:\penalty0 261--272, 2020.

\bibitem[Wan et~al.(2021)Wan, Yang, Huang, Zeng, and Liu]{transferEEG}
Z.~Wan, R.~Yang, M.~Huang, N.~Zeng, and X.~Liu.
\newblock A review on transfer learning in {EEG} signal analysis.
\newblock \emph{Neurocomputing}, 421:\penalty0 1--14, 01 2021.

\bibitem[Wang et~al.(2016)Wang, Chen, Gao, and Gao]{wang2016benchmark}
Y.~Wang, X.~Chen, X.~Gao, and S.~Gao.
\newblock A benchmark dataset for {SSVEP}-based brain--computer interfaces.
\newblock \emph{IEEE Transactions on Neural Systems and Rehabilitation
  Engineering}, 25\penalty0 (10):\penalty0 1746--1752, 2016.

\bibitem[Wei et~al.(2022)Wei, Faisal, Grosse-Wentrup, Gramfort, Chevallier,
  Jayaram, Jeunet, Bakas, Ludwig, Barmpas, et~al.]{wei2022beetl}
X.~Wei, A.~A. Faisal, M.~Grosse-Wentrup, A.~Gramfort, S.~Chevallier,
  V.~Jayaram, C.~Jeunet, S.~Bakas, S.~Ludwig, K.~Barmpas, et~al.
\newblock 2021 {BEETL} competition: Advancing transfer learning for subject
  independence \& heterogenous {EEG} data sets.
\newblock In \emph{NeurIPS 2021 Competitions and Demonstrations Track}, pages
  205--219. PMLR, 2022.

\bibitem[Wilcoxon(1992)]{wilcoxon1992individual}
F.~Wilcoxon.
\newblock Individual comparisons by ranking methods.
\newblock In \emph{Breakthroughs in Statistics: Methodology and Distribution},
  pages 196--202. Springer, 1992.

\bibitem[Wu and Yao(2007)]{wu2007influence}
Z.~Wu and D.~Yao.
\newblock The influence of cognitive tasks on different frequencies
  steady-state visual evoked potentials.
\newblock \emph{Brain topography}, 20:\penalty0 97--104, 2007.

\bibitem[Yger et~al.(2016)Yger, Berar, and Lotte]{yger2016riemannian}
F.~Yger, M.~Berar, and F.~Lotte.
\newblock Riemannian approaches in brain-computer interfaces: a review.
\newblock \emph{IEEE Transactions on Neural Systems and Rehabilitation
  Engineering}, 25\penalty0 (10):\penalty0 1753--1762, 2016.

\bibitem[Yi et~al.(2014)Yi, Qiu, Wang, Qi, Zhang, Zhou, He, and
  Ming]{yi2014evaluation}
W.~Yi, S.~Qiu, K.~Wang, H.~Qi, L.~Zhang, P.~Zhou, F.~He, and D.~Ming.
\newblock Evaluation of {EEG} oscillatory patterns and cognitive process during
  simple and compound limb motor imagery.
\newblock \emph{PloS one}, 9\penalty0 (12):\penalty0 e114853, 2014.

\bibitem[Zhang et~al.(2014)Zhang, Zhou, Jin, Wang, and
  Cichocki]{zhang2014frequency}
Y.~Zhang, G.~Zhou, J.~Jin, X.~Wang, and A.~Cichocki.
\newblock Frequency recognition in {SSVEP-based BCI} using multiset canonical
  correlation analysis.
\newblock \emph{International journal of neural systems}, 24\penalty0
  (04):\penalty0 1450013, 2014.

\bibitem[Zhang et~al.(2024)Zhang, hua Zhong, and Liu]{torcheeg}
Z.~Zhang, S.~hua Zhong, and Y.~Liu.
\newblock {TorchEEG}, 2024.
\newblock URL \url{https://torcheeg.readthedocs.io}.

\bibitem[Zhou et~al.(2016)Zhou, Wu, Lv, Zhang, and Guo]{zhou2016fully}
B.~Zhou, X.~Wu, Z.~Lv, L.~Zhang, and X.~Guo.
\newblock A fully automated trial selection method for optimization of motor
  imagery based brain-computer interface.
\newblock \emph{PloS one}, 11\penalty0 (9):\penalty0 e0162657, 2016.

\end{thebibliography}

\appendix

\section{Detailed pipelines evaluation}

This section provides details regarding the automatic parametrization and the evaluation conducted in the benchmark. 

\subsection{Specialized grid search}
\label{app:gridsearch}

The parametrization of evaluated pipelines should be generic to ensure a fair evaluation and automatic to avoid information leakage. 
A dictionary structure containing the parameter to search for each element of the machine learning pipeline is implemented to ensure this process when an evaluation is launched.
As shown below, the \texttt{param\_grid} structure, with names matching the \texttt{pipeline} structure, could be passed to the  function \texttt{evaluation.process()}.

% \begin{listing}[H]    
%     \vspace*{-4mm}
\begin{lstlisting}[language=Python]
    pipelines = {}
    pipelines["GridSearchEN"] = Pipeline(
        steps=[
            ("Covariances", Covariances("cov")),
            ("Tangent_Space", 
                TangentSpace(metric="riemann")),
            (
                "LogistReg",
                LogisticRegression(
                    penalty="elasticnet",
                    l1_ratio=0.70,
                    intercept_scaling=1000.0,
                    solver="saga",
                    max_iter=1000,
                ),
            ),
        ]
    )
    
    param_grid = {}
    param_grid["GridSearchEN"] = {
        "LogistReg__l1_ratio": 
            [0.15, 0.30, 0.45, 0.60, 0.75],
    }
    
    evaluation = WithinSessionEvaluation(
        paradigm=paradigm,
        datasets=dataset,
        overwrite=True,
        random_state=42,
        hdf5_path=path,
        n_jobs=-1,
    )
    result = evaluation.process(pipelines, 
        param_grid)
    \end{lstlisting} 
%     \label{evaluation_process}    
% \end{listing}

\begin{table*}[!ht]
\begin{center}
    \caption{Parameter used in Grid Search. For the ACM+TS+SVM pipeline, we reduce the hyperparameter search due to computational constraint (both order and lag to $[1-5]$ for datasets with more than 60 electrodes - Cho2017, Lee2019-MI, PhysionetMI and Weibo2014. While we select parameters to $[1-3]$ for datasets with more than 100 electrodes - Schirrmeister2017 and GrosseWentrup2009)}
\label{table:gridsearch_parameter}
\resizebox{\linewidth}{!}{\begin{tabular}{c|c|c}
  \hline
  \textbf{Pipeline}  & \textbf{Parameter}  &  \textbf{Value} \\ \hline \hline
    \gls{csp} + SVMGrid & csp\_{nfilter} & [2 - 8] \\ 
     & svc\_{C} & [0.5, 1, 1.5] \\ 
     & svc\_{kernel} & ["rbf", "linear"] \\ \hline
     EnGrid &  logisticregression\_{l1 ratio} & [0.20, 0.30, 0.45, 0.65, 0.75] \\ \hline
    LogVarGrid & svc\_{C} & [0.01, 0.05, 0.1, 0.5, 1, 5, 10, 50, 100] \\ \hline
    \gls{tssvm} Grid & svc\_{C} & [0.5, 1, 1.5] \\ 
     & svc\_{kernel} & ["rbf", "linear"] \\ \hline
     ACM + TANG + \gls{svm} Grid & augmenteddataset\_{order} & [1 - 10] \\ 
     & augmenteddataset\_{lag} & [1 - 10] \\ 
     & svc\_{C} & [0.5, 1, 1.5] \\ 
     & svc\_{kernel} & ["rbf", "linear"] \\ \hline
\end{tabular}
}
\end{center}
\end{table*}

\subsection{Other evaluation types}

\begin{figure}
    \centering
    \includegraphics[scale=.73]{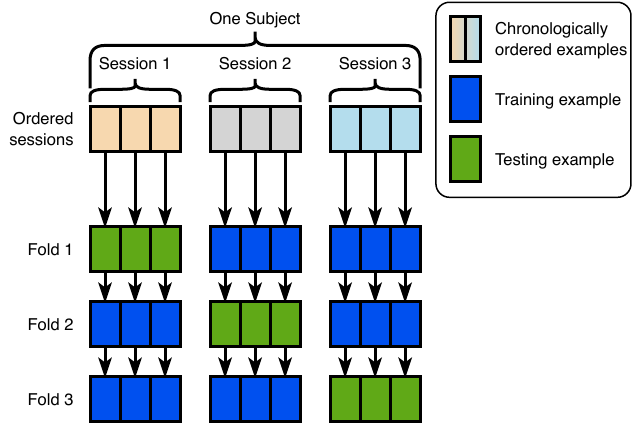}
    \caption{Cross-session evaluation}
    \label{fig:cross-session}
\end{figure}

\begin{figure}
    \centering
    \includegraphics[scale=.73]{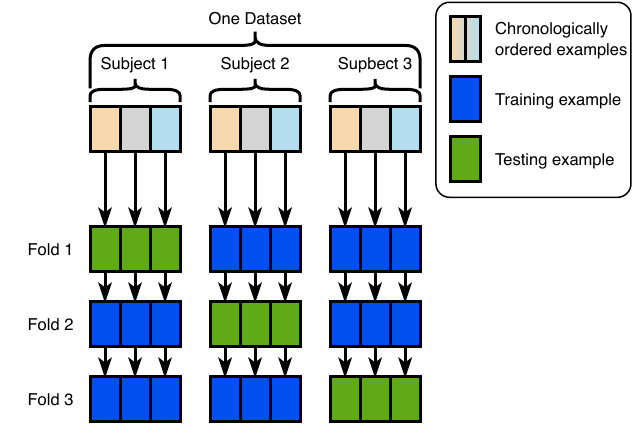}
    \caption{Cross-subject evaluation}
    \label{fig:cross-subject}
\end{figure}

The proposed benchmark focus on within-session evaluation, but it is possible to conduct other evaluation types like cross-session or cross-subject. The structure of the cross-session evaluation is shown in~\autoref{fig:cross-session} and \autoref{fig:cross-subject} shows the structure of the cross-subject one.

\section{Machine learning algorithms}
\label{app:pipelines}

This section describes the machine learning pipelines and the choices of implementation. 

\subsection{CSP-baseline pipelines}
\label{app:csp}

Mathematically, the \gls{csp} algorithm seeks to find spatial filters by solving a generalized eigenvalue problem. Let $X_1$ and $X_2$ be the data matrices of the band-pass filtered \gls{eeg} signals for two conditions, each with dimensions of (time $\times$ channel). The \gls{csp} algorithm constructs discriminative and common activity matrices, $S_d$ and $S_c$, respectively. These matrices are defined as follows: 
\begin{align} S_d = \Sigma_1 - \Sigma_2  \quad \text{and} \quad S_c = \Sigma_1 + \Sigma_2, \label{eq:Sc} \end{align} 

where $\Sigma_1 = X_1^TX_1$ and $\Sigma_{2} = X_{2}^{T} X_{2}$ are the estimates of the condition covariance matrices. Then, the objective of \gls{csp} is to find spatial filters $v_j \in \mathbb{R}^C$ that maximize the \gls{eeg} signal bandpower variance between examples from different conditions while simultaneously minimizing its variance between examples from the same condition. This translates into:
\begin{align}
    \text{argmax}_V \frac{V^TS_dV}{V^TS_cV}
\end{align}

which is optimized by solving the following generalised eigenvalue problem~\cite{ramoser2000optimal}:
\begin{align} S_d v = \lambda S_c v, \label{eq:eig_prob} \end{align} 
and selecting the filters $v$ that yield the largest eigenvalues~$\lambda$.

For classification purposes, \gls{csp} utilizes log-variance features extracted from the filtered signals projected onto the \gls{csp} filters. Typically, a small number of patterns (2 to 6) are selected based on the corresponding eigenvalues. The patterns, denoted as $a_j$, provide insights into the specific information captured by the corresponding filters $v_j$. Each filter $v_j$ extracts the activity spanned by pattern, $a_j$ while canceling out other activities spanned by different patterns. This allows for discrimination between different mental states based on the log-variance features. Linear classifiers, such as linear discriminant analysis, are commonly used due to the approximately Gaussian distribution of the log-variance features. In \gls{moabb}, we implemented two different decision head, the \gls{svm} and \gls{lda}, with a \textsc{GridSearch} in some parameters \cite{5662067}.

%TODO: explain filterbank CSP approach

\subsection{CCA-based pipelines}
\label{app:cca}

Based on the work of Hotelling, the \gls{cca} aims at finding a canonical space where the correlation of two sets of variables is maximized. Considering the total covariance matrix $C$ of two sets of variables $x$ and $y$, the within-set covariances matrices are respectively denoted $C_{xx}$ and $C_{yy}$ and the between-sets covariance matrices are $C_{xy}=C^\top_{yx}$. The canonical correlation is defined as:
\begin{equation}
    \rho = \mathrm{max}_{w_x,w_y} \frac{w_x^\top C_{xy} w_y}{\sqrt{w^\top_x C_{xx} w_x w^\top_y C_{yy} w_y }}
\end{equation}
with $w_x$ and $w_y$ the projection vectors that maximizes canonical correlation. The solution is provided by a generalized eigenproblem formulated as $C_{xy}C^{-1}_yy C_{yx}w_x = \lambda^2 C_{xx}w_x$. As $C_{xx}$ and $C_{yy}$ are symmetric positive definite matrices, it is possible to rewrite the problem as symmetric standard eigenproblem $Ax=\lambda x$ that could be solved with any linear algebra library.

In \gls{ssvep}, the \gls{cca} is applied on a set of $x$ \gls{eeg} trials and a set $y$ of reference sinusoid signals as described in~\cite{lin2006frequency}. The frequency of the sine and cosine reference signals should match the stimulus frequency and its harmonics. The obtained $w_x$ could be seen as filter acting on \gls{eeg} signal to enhance the signal components synchronized with  the visual stimulus.

Indeed, the choice reference signals is crucial to obtain robust results and could be complex to parameterize. The multiset canonical correlation analysis (MsetCCA) generalizes the \gls{cca} for multiple references\cite{zhang2014frequency}, with the objective to learn the optimal reference signals without constraint on the sine/cosine shape of the reference set.

The goal of filters learned from \gls{cca} is to enhance the \gls{eeg} activity generated by cortical stimulus response. The cortical signal component unrelated to the task could be filtered out as well, as introduced in task-related component analysis (TRCA)~\cite{nakanishi2017enhancing} for \gls{ssvep}. In this case, an optimization process is designed to find the $w_x$ that maximized the inter-trial covariance of \gls{eeg} signals. 

\subsection{Riemannian pipelines}
\label{ann:riempip}

As described in \autoref{sec:riempip}, the Riemannian pipelines could be directly defined on the manifold. This is the case of the Minimum Distance to Mean (MDM) classifier, that computes the average of each class with Eq.~\eqref{eq:frechet_mean} and use the distance defined in Eq.~\eqref{eq:airm} between an unseen sample and each class center to make a prediction. 

To mitigate the effect of increased dimensions, i.e., for \gls{eeg} recorded with a high number of electrodes, a geodesic filtering could be applied before MDM classification, namely the FgMDM. This filtering step implements a linear discriminant analysis in the tangent space to project all trials on a single hyperplane and the projection back to the manifold.

The second option is to project the samples in the tangent space, vectorize samples that are symmetric matrices and train a classifier on those vectorized data. Algorithms such as logistic regression (LR), logistic regression under $\ell_1$ and $\ell_2$ norm penalties (ElasticNet, EL) or support vector machine (SVM) have been described in the literature.

When considering \gls{mi}, Riemannian classifiers operate directly on covariance matrices estimated from the \gls{eeg} signals. The transient information of \gls{erp} require to first estimate a average \gls{erp} or, better, an average \gls{erp} filtered XDAWN. For \gls{ssvep}, the relevant information is spectral and the covariances need to be estimated on the bandpass filtered signal for each stimulation frequency. All those preprocessing are described in great detail in \cite{yger2016riemannian,chevallier2018riemannian}.

\section{Experimental results}
\label{app:eval}

This section provides a global overview of the pipeline scores for the different paradigms with raincloud plots. The pipelines are grouped by categories: Deep learning, Riemannian, and Raw. The small points correspond to the scores obtained on the individual sessions, with an exception for the last row, i.e. \textit{Average}, where they correspond to the average score over one dataset. The curves above indicate a density estimation of these individual scores. The diamond shapes connected by vertical lines indicate the within-dataset average. Finally, the boxes and horizontal black bars indicate the quartiles.

\begin{figure}
    \centering
    \includegraphics[width=\linewidth]{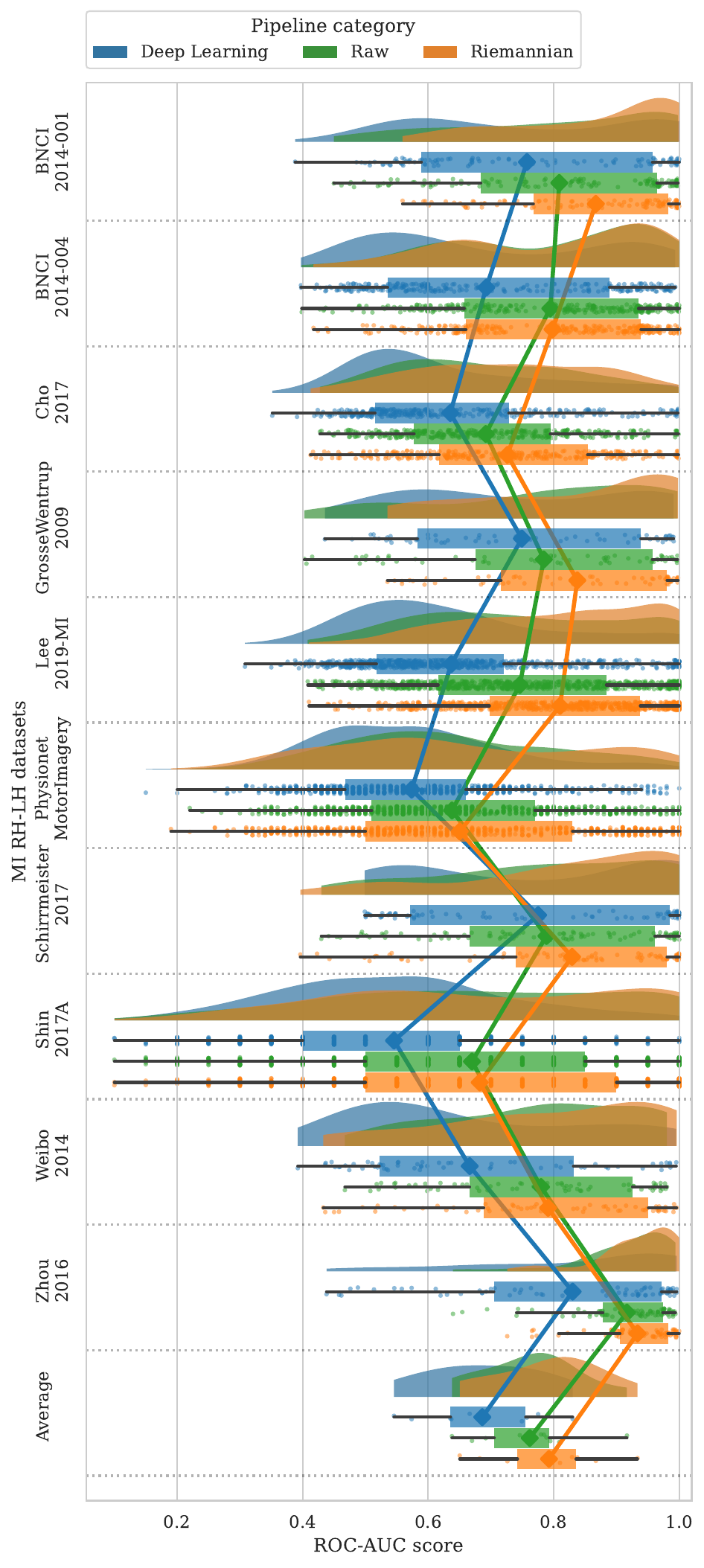}
    \caption{Distributions of ROC-AUC scores on the right hand vs left hand \gls{mi} task of the  pipelines grouped by category.}
    \label{fig-app:LHRH-raincloud}
\end{figure}

\begin{figure}
    \centering
    \includegraphics[width=\linewidth]{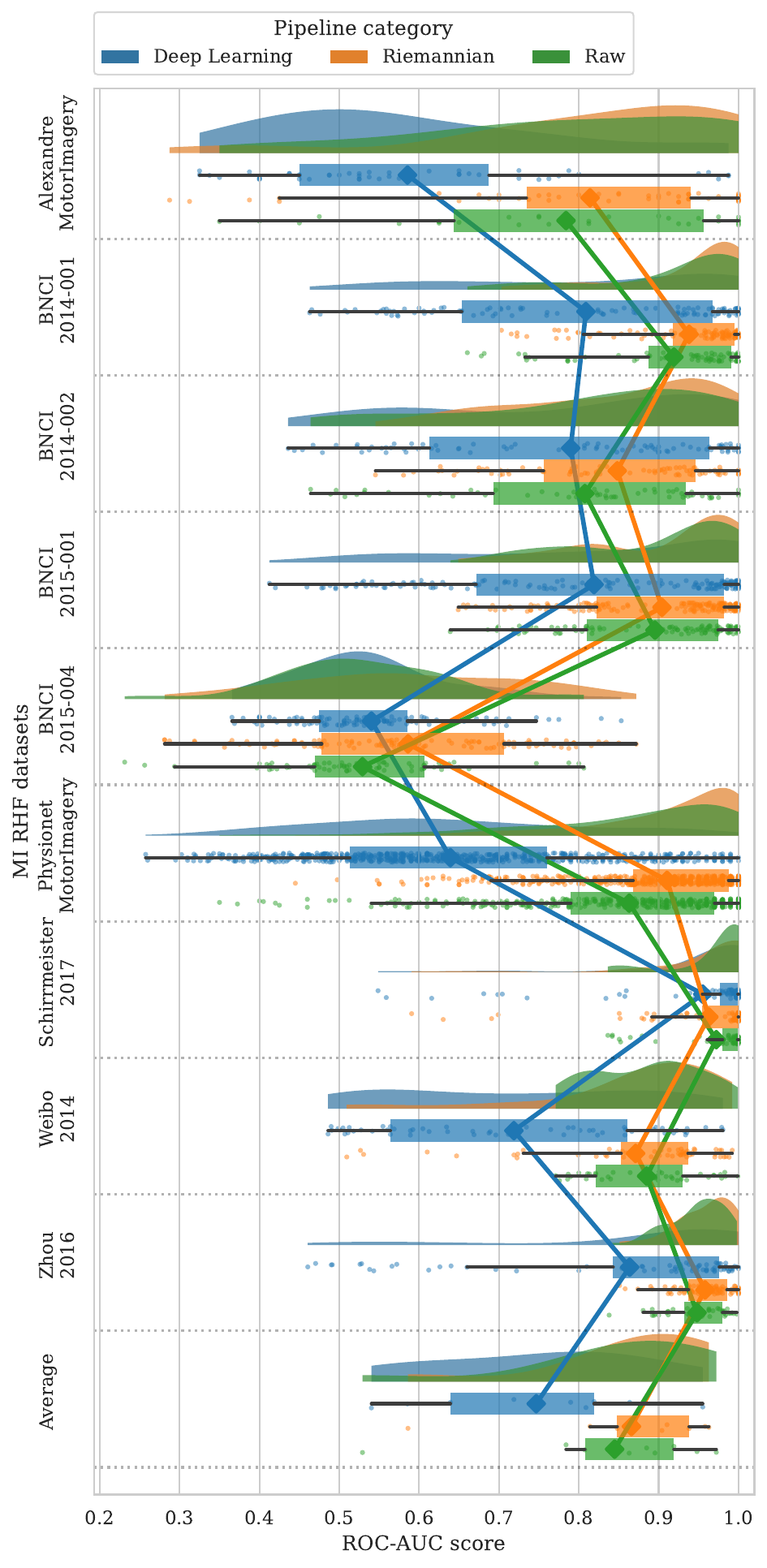}
    \caption{Distributions of ROC-AUC scores on the right hand vs feet \gls{mi} task.}
    \label{fig-app:RHF-raincloud}
\end{figure}

\begin{figure}
    \centering
    \includegraphics[width=\linewidth]{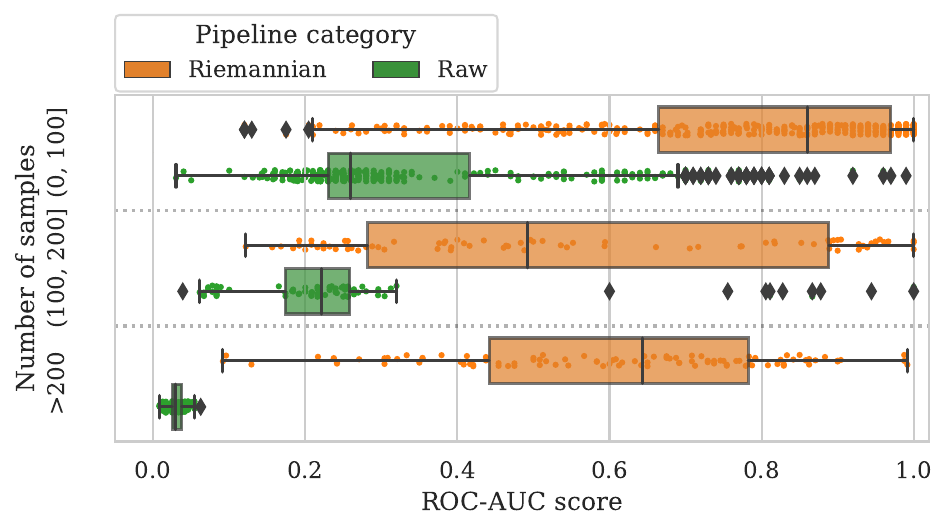}
    \caption{Accuracy scores averaged over all subjects and session of each \gls{ssvep} dataset, per pipeline category.}
    \label{fig-app:SSVEP-raincloud}
\end{figure}

\begin{figure}
    \centering
    \includegraphics[width=\linewidth]{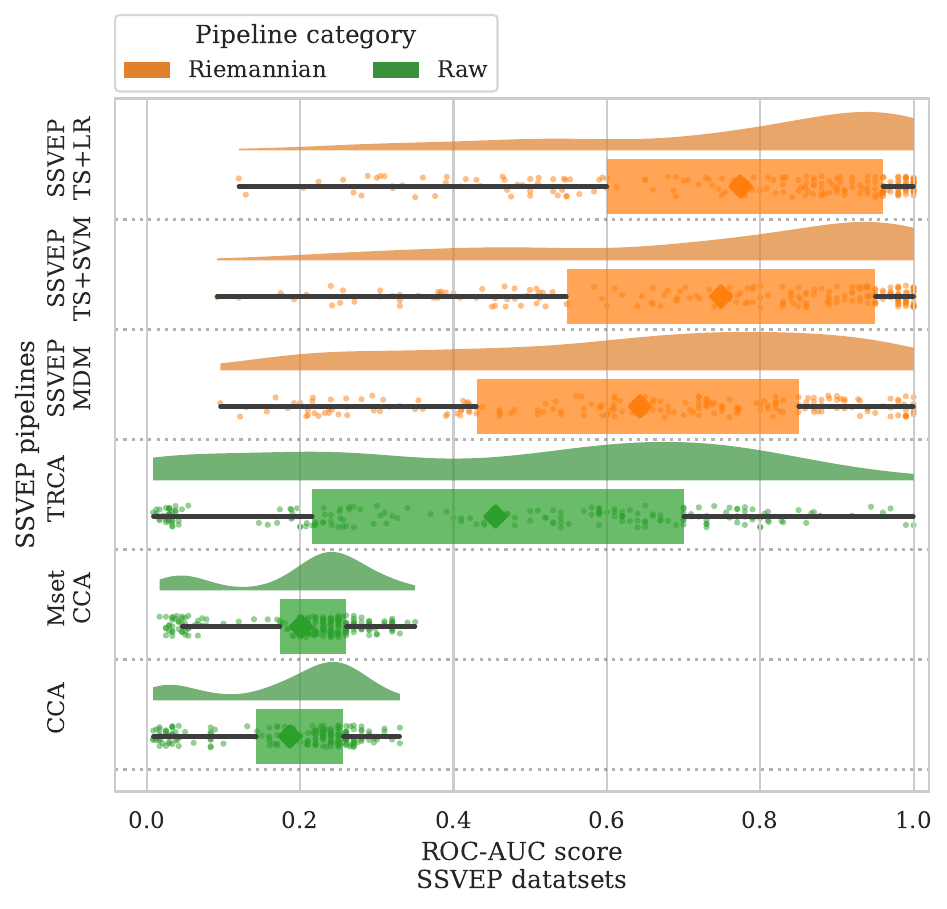}
    \caption{Box-plot representing classification accuracy averaged over all the sessions of all the subjects of all the datasets of the \gls{ssvep} paradigm and over all pipelines of the corresponding category (\textit{Riemannian, Raw}). Box-plots are overlaid with strip-plots, where each point represents the classification accuracy of one within-session evaluation.}
    \label{SSVEP:Riemannia>Raw}
\end{figure}

\begin{figure}
    \centering
    \includegraphics[width=\linewidth]{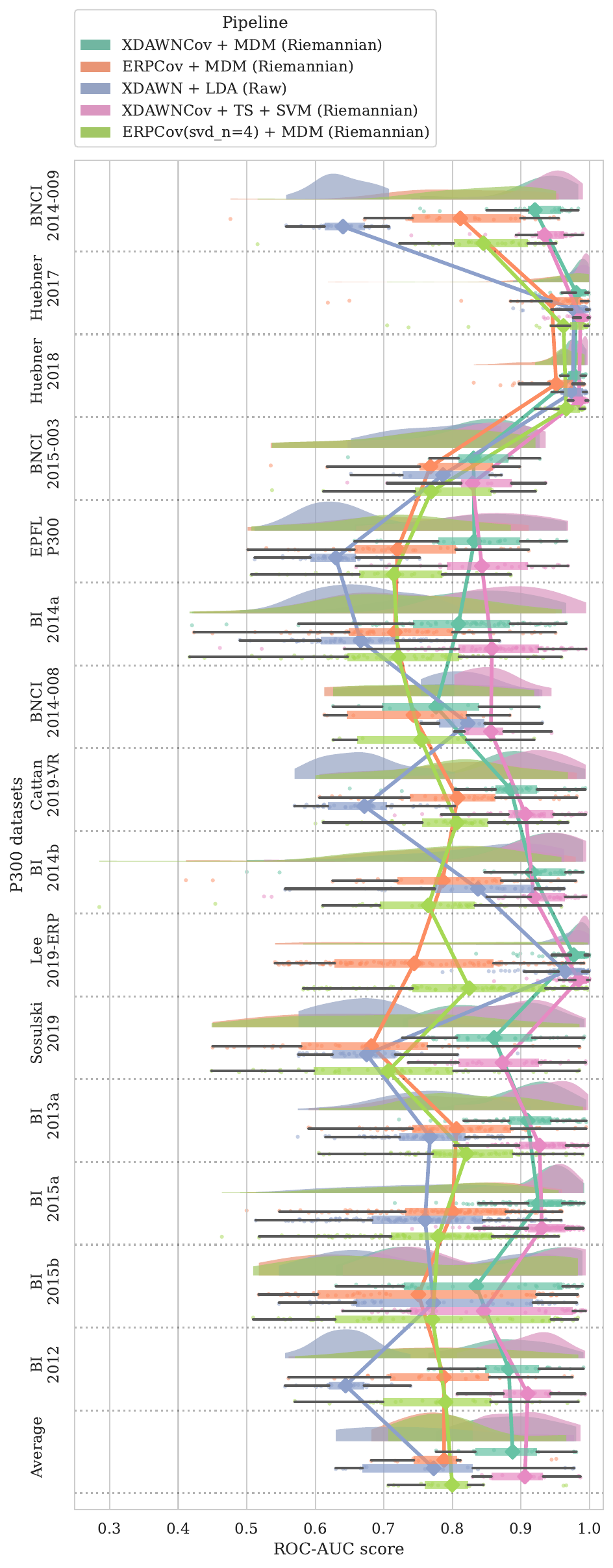}
    \caption{ROC-AUC scores on the \gls{erp} classification task of the different pipelines.}
    \label{fig-app:P300-raincloud}
\end{figure}

\section{Detailed results for each pipeline}
\label{app:res}

This section provides the tables indicating the \gls{roc-auc} for binary classification problems and the accuracy for multi-class tasks. The main objective is to provide a reference benchmark that could be easily reproduced to verify the results or used as-is to compare with new pipelines. It is thus possible to save energy and resources, avoiding reproducing already existing validated results with a simple copy-paste.
To this end, the tables provide here the average result of pipelines across all subjects for a given dataset using within-session evaluation. We did not provide subject-by-subject results for space reasons and we know that this benchmark is meant to evolve with new pipelines, datasets, and evaluation methods. We mirror those tables on a website that will allow further additions and available up-to-date results.
%TODO add web link

\begin{table*}
\centering
\caption{Summary of performances via average on all the motor imagery datasets, for classification using all the labels. Intra-session validation.  Bold numbers represent the best score in each dataset.}
\label{tab:All_agg_dataset}
\begin{adjustbox}{width=1.15\textwidth}
\begin{tabular}{c|cccccc|c}
\toprule
       pipeline &             AlexandreMotorImagery &                      BNCI2014-001 &             PhysionetMotorImagery &                Schirrmeister2017 &                         Weibo2014 &                         Zhou2016 &        Average \\
\midrule
     ACM+TS+SVM &                   69.37$\pm$15.07 & \textbf{77.82}$\pm$\textbf{12.23} &                   55.44$\pm$14.87 &                  82.50$\pm$10.20 & \textbf{63.89}$\pm$\textbf{11.01} & \textbf{85.25}$\pm$\textbf{4.06} &          72.38 \\
        CSP+LDA &                   61.04$\pm$17.22 &                   65.99$\pm$15.47 &                   47.73$\pm$14.35 &                  72.97$\pm$10.42 &                   39.45$\pm$11.87 &                   82.96$\pm$5.20 &          61.69 \\
        CSP+SVM &                   62.92$\pm$16.89 &                   66.88$\pm$15.22 &                   48.52$\pm$14.62 &                  75.89$\pm$10.55 &                   44.08$\pm$11.95 &                   83.08$\pm$5.33 &          63.56 \\
DLCSPauto+shLDA &                   60.63$\pm$17.91 &                   66.31$\pm$15.36 &                   46.85$\pm$14.65 &                  72.82$\pm$10.44 &                   38.84$\pm$11.97 &                   82.06$\pm$5.57 &          61.25 \\
    DeepConvNet &                    37.71$\pm$4.56 &                    35.29$\pm$8.26 &                    27.68$\pm$3.91 &                  56.78$\pm$18.11 &                    24.17$\pm$9.80 &                   55.69$\pm$5.61 &          39.55 \\
       EEGITNet &                    36.04$\pm$3.43 &                    35.55$\pm$6.35 &                    26.15$\pm$4.95 &                  70.44$\pm$14.68 &                    25.78$\pm$8.00 &                  50.68$\pm$16.27 &          40.77 \\
         EEGNeX &                    37.71$\pm$9.64 &                   45.62$\pm$15.29 &                    26.69$\pm$5.64 &                  67.56$\pm$14.15 &                   30.22$\pm$11.02 &                  56.42$\pm$11.29 &          44.03 \\
     EEGNet-8,2 &                    43.96$\pm$8.62 &                   60.46$\pm$20.20 &                    29.04$\pm$7.03 &                  76.99$\pm$13.05 &                   35.35$\pm$14.05 &                   83.34$\pm$3.58 &          54.86 \\
       EEGTCNet &                    34.17$\pm$1.86 &                   41.65$\pm$13.73 &                    25.79$\pm$3.85 &                  71.11$\pm$11.96 &                    17.95$\pm$3.88 &                   37.19$\pm$2.57 &          37.98 \\
      FBCSP+SVM &                   65.00$\pm$17.56 &                   66.53$\pm$12.05 &                   45.49$\pm$12.54 &                   75.94$\pm$8.59 &                   45.21$\pm$10.05 &                   81.99$\pm$4.65 &          63.36 \\
          FgMDM &                   65.63$\pm$15.63 &                   70.14$\pm$15.13 &                   55.04$\pm$14.17 &                  82.97$\pm$10.08 &                    56.94$\pm$9.26 &                   83.07$\pm$4.96 &          68.97 \\
            MDM &                   60.62$\pm$13.69 &                   61.60$\pm$14.20 &                   42.96$\pm$12.98 &                  52.03$\pm$10.11 &                    33.41$\pm$8.67 &                   76.05$\pm$7.10 &          54.45 \\
 ShallowConvNet &                   50.00$\pm$12.94 &                   72.47$\pm$16.50 &                   41.87$\pm$12.50 &                   85.13$\pm$9.57 &                   48.94$\pm$10.36 &                   85.02$\pm$3.78 &          63.91 \\
          TS+EL & \textbf{69.79}$\pm$\textbf{13.75} &                   72.38$\pm$14.85 & \textbf{59.93}$\pm$\textbf{14.07} & \textbf{85.53}$\pm$\textbf{9.40} &                    63.84$\pm$8.77 &                   84.54$\pm$4.93 & \textbf{72.67} \\
          TS+LR &                   69.17$\pm$14.79 &                   71.97$\pm$15.46 &                   58.55$\pm$14.06 &                   84.60$\pm$9.28 &                    62.76$\pm$8.39 &                   84.88$\pm$4.63 &          71.99 \\
         TS+SVM &                   67.92$\pm$12.74 &                   70.76$\pm$15.08 &                   58.46$\pm$15.15 &                   84.41$\pm$9.56 &                    61.47$\pm$9.62 &                   83.66$\pm$4.55 &          71.11 \\
        \midrule
        Average &                             55.73 &                             61.34 &                             43.51 &                            74.85 &                             43.27 &                            74.74 &          58.91 \\
\bottomrule
\end{tabular}\end{adjustbox}
\end{table*}

\begin{table*}
\centering
\caption{Summary of performances via average on all the P300 datasets, for classification using left vs. right motor imagery task. Intra-session validation.  Bold numbers represent the best score in each dataset.}
\label{tab:lhrh_agg_dataset}
\begin{adjustbox}{width=1.15\textwidth}
\begin{tabular}{c|cccccccccc|c}
\toprule
       pipeline &                      BNCI2014-001 &                      BNCI2014-004 &                           Cho2017 &                 GrosseWentrup2009 &                        Lee2019-MI &             PhysionetMotorImagery &                 Schirrmeister2017 &                         Shin2017A &                         Weibo2014 &                         Zhou2016 &        Average \\
\midrule
     ACM+TS+SVM & \textbf{91.71}$\pm$\textbf{10.30} & \textbf{82.67}$\pm$\textbf{15.33} &                   73.56$\pm$14.54 &                   86.60$\pm$15.12 &                   83.05$\pm$13.97 &                   63.55$\pm$21.24 &                   85.82$\pm$13.98 &                   68.97$\pm$23.45 &                   84.78$\pm$13.33 &                   95.03$\pm$4.76 &          81.57 \\
        CSP+LDA &                   82.34$\pm$17.26 &                   80.10$\pm$14.93 &                   71.38$\pm$14.54 &                   76.44$\pm$20.95 &                   76.88$\pm$17.41 &                   65.75$\pm$17.37 &                   77.23$\pm$18.43 & \textbf{72.30}$\pm$\textbf{21.79} &                   80.72$\pm$15.29 &                   93.15$\pm$6.88 &          77.63 \\
        CSP+SVM &                   83.07$\pm$16.53 &                   79.27$\pm$15.68 &                   71.92$\pm$14.25 &                   77.81$\pm$21.27 &                   77.27$\pm$16.73 &                   65.71$\pm$17.90 &                   79.24$\pm$20.07 &                   70.11$\pm$22.19 &                   79.84$\pm$15.86 &                   92.96$\pm$7.86 &          77.72 \\
DLCSPauto+shLDA &                   82.75$\pm$16.69 &                   79.87$\pm$15.11 &                   71.16$\pm$14.53 &                   76.40$\pm$20.83 &                   76.69$\pm$17.23 &                   65.07$\pm$17.68 &                   77.02$\pm$18.48 &                   70.34$\pm$23.30 &                   80.16$\pm$15.23 &                   92.56$\pm$7.21 &           77.2 \\
    DeepConvNet &                   82.07$\pm$15.52 &                   72.36$\pm$18.53 &                   71.67$\pm$12.91 &                   82.38$\pm$15.39 &                   70.65$\pm$15.76 &                   59.57$\pm$16.77 &                   81.23$\pm$17.39 &                   56.03$\pm$19.18 &                   73.64$\pm$15.78 &                   94.42$\pm$6.21 &           74.4 \\
       EEGITNet &                   75.27$\pm$16.37 &                   65.10$\pm$15.32 &                   57.20$\pm$12.21 &                   72.19$\pm$14.71 &                   59.17$\pm$11.72 &                   52.71$\pm$11.11 &                   74.66$\pm$20.52 &                   52.18$\pm$16.78 &                   59.35$\pm$14.06 &                  69.41$\pm$14.66 &          63.72 \\
         EEGNeX &                   66.28$\pm$13.22 &                   66.53$\pm$17.10 &                   53.28$\pm$10.60 &                    57.00$\pm$7.52 &                   55.12$\pm$10.05 &                   51.20$\pm$10.63 &                   68.58$\pm$19.37 &                   49.02$\pm$17.58 &                   57.97$\pm$15.65 &                  61.56$\pm$14.60 &          58.65 \\
     EEGNet-8,2 &                   77.15$\pm$19.33 &                   69.50$\pm$19.50 &                   66.79$\pm$16.34 &                   83.02$\pm$18.08 &                   65.67$\pm$16.43 &                   59.55$\pm$15.95 &                   80.20$\pm$18.13 &                   57.99$\pm$17.28 &                   66.46$\pm$21.78 &                   94.84$\pm$2.83 &          72.12 \\
       EEGTCNet &                   67.46$\pm$20.81 &                   69.70$\pm$19.55 &                   58.34$\pm$12.63 &                   68.45$\pm$16.27 &                   55.68$\pm$12.75 &                   55.90$\pm$12.74 &                   75.62$\pm$22.33 &                   51.26$\pm$16.77 &                   63.16$\pm$18.32 &                   82.24$\pm$9.40 &          64.78 \\
      FBCSP+SVM &                   84.44$\pm$16.00 &                   80.39$\pm$16.05 &                   67.91$\pm$15.63 &                   79.65$\pm$18.63 &                   75.07$\pm$16.97 &                   58.45$\pm$13.93 &                   81.44$\pm$17.89 &                   65.63$\pm$21.64 &                   76.81$\pm$18.88 &                   92.64$\pm$5.01 &          76.24 \\
          FgMDM &                   86.53$\pm$12.14 &                   79.28$\pm$15.25 &                   72.90$\pm$12.70 &                   87.02$\pm$13.20 &                   81.34$\pm$13.93 & \textbf{68.46}$\pm$\textbf{19.06} &                   86.71$\pm$13.79 &                   70.86$\pm$23.36 &                   78.41$\pm$14.85 &                   92.54$\pm$6.67 &          80.41 \\
     LogVar+LDA &                   77.96$\pm$15.09 &                   78.51$\pm$15.25 &                   64.49$\pm$10.08 &                   78.71$\pm$11.69 &                   66.21$\pm$12.06 &                   61.94$\pm$14.41 &                   78.44$\pm$13.76 &                   61.78$\pm$22.77 &                   74.13$\pm$10.40 &                   88.39$\pm$8.57 &          73.06 \\
     LogVar+SVM &                   75.86$\pm$16.45 &                   78.30$\pm$15.18 &                   65.46$\pm$11.71 &                   81.73$\pm$12.40 &                   73.83$\pm$13.85 &                   62.35$\pm$16.87 &                   79.42$\pm$13.66 &                   61.38$\pm$22.68 &                   74.85$\pm$11.33 &                   88.47$\pm$8.50 &          74.17 \\
            MDM &                   81.69$\pm$14.94 &                   77.66$\pm$15.78 &                   63.39$\pm$13.69 &                    64.29$\pm$8.04 &                   70.23$\pm$13.87 &                   54.76$\pm$16.79 &                   61.53$\pm$16.41 &                   62.99$\pm$21.25 &                   58.80$\pm$16.13 &                   90.70$\pm$7.11 &           68.6 \\
 ShallowConvNet &                   86.17$\pm$13.74 &                   72.36$\pm$18.05 &                   73.84$\pm$14.95 &                   86.53$\pm$13.00 &                   75.83$\pm$15.04 &                   65.19$\pm$15.80 &                   84.82$\pm$15.29 &                   60.80$\pm$19.27 &                   79.10$\pm$12.63 & \textbf{95.65}$\pm$\textbf{5.55} &          78.03 \\
      TRCSP+LDA &                   79.84$\pm$16.28 &                   79.78$\pm$15.22 &                   71.85$\pm$13.84 &                   78.29$\pm$16.66 &                   76.26$\pm$15.41 &                   67.24$\pm$17.23 &                   79.14$\pm$15.91 &                   67.30$\pm$23.19 &                   79.33$\pm$14.43 &                   93.53$\pm$6.38 &          77.25 \\
          TS+EL &                   86.44$\pm$13.20 &                   79.75$\pm$15.44 & \textbf{76.23}$\pm$\textbf{14.21} & \textbf{89.25}$\pm$\textbf{12.00} & \textbf{84.74}$\pm$\textbf{13.19} &                   67.91$\pm$20.03 & \textbf{88.65}$\pm$\textbf{12.98} &                   68.68$\pm$23.64 & \textbf{85.29}$\pm$\textbf{12.10} &                   94.35$\pm$6.04 & \textbf{82.13} \\
          TS+LR &                   87.41$\pm$12.58 &                   80.09$\pm$15.01 &                   75.01$\pm$13.71 &                   87.60$\pm$13.20 &                   83.09$\pm$13.46 &                   67.28$\pm$19.19 &                   87.22$\pm$13.83 &                   69.31$\pm$23.06 &                   83.62$\pm$13.88 &                   94.16$\pm$6.33 &          81.48 \\
         TS+SVM &                   86.48$\pm$13.58 &                   79.41$\pm$15.26 &                   74.62$\pm$14.19 &                   88.08$\pm$13.58 &                   83.57$\pm$14.08 &                   68.18$\pm$19.92 &                   87.64$\pm$13.48 &                   68.45$\pm$24.25 &                   83.72$\pm$14.28 &                   93.37$\pm$6.30 &          81.35 \\
        \midrule
        Average &                              81.1 &                             76.35 &                             68.47 &                             79.02 &                             73.18 &                             62.15 &                             79.72 &                             63.44 &                             74.74 &                            89.47 &          74.76 \\
\bottomrule
\end{tabular}\end{adjustbox}
\end{table*}

\begin{table*}
\centering
\caption{Summary of performances via average on all the P300 datasets, for classification using right hand vs. feet tasks motor imagery task. Intra-session validation.  Bold numbers represent the best score in each dataset. }
\label{tab:rf_agg_dataset}
\begin{adjustbox}{width=1.15\textwidth}
\begin{tabular}{c|ccccccccc|c}
\toprule
       pipeline &             AlexandreMotorImagery &                     BNCI2014-001 &                      BNCI2014-002 &                     BNCI2015-001 &                      BNCI2015-004 &            PhysionetMotorImagery &                Schirrmeister2017 &                        Weibo2014 &                         Zhou2016 &        Average \\
\midrule
     ACM+TS+SVM & \textbf{86.56}$\pm$\textbf{12.26} & \textbf{97.32}$\pm$\textbf{3.35} & \textbf{88.60}$\pm$\textbf{10.71} & \textbf{93.01}$\pm$\textbf{8.09} & \textbf{62.60}$\pm$\textbf{14.62} &                   93.33$\pm$8.46 &                   98.67$\pm$3.06 & \textbf{93.25}$\pm$\textbf{4.12} & \textbf{97.18}$\pm$\textbf{3.00} & \textbf{90.06} \\
        CSP+LDA &                   77.19$\pm$17.58 &                  91.52$\pm$10.39 &                   80.98$\pm$14.79 &                  88.52$\pm$10.75 &                   54.02$\pm$11.33 &                  86.41$\pm$13.96 &                   97.02$\pm$5.17 &                   88.59$\pm$6.36 &                   95.20$\pm$3.17 &          84.38 \\
        CSP+SVM &                   78.59$\pm$20.14 &                  91.04$\pm$10.35 &                   81.21$\pm$15.30 &                  89.19$\pm$10.08 &                   52.08$\pm$11.05 &                  88.04$\pm$12.57 &                   97.50$\pm$4.90 &                   88.64$\pm$5.90 &                   94.95$\pm$3.53 &          84.58 \\
DLCSPauto+shLDA &                   77.03$\pm$18.93 &                  91.54$\pm$10.37 &                   80.45$\pm$15.52 &                  88.87$\pm$10.42 &                   53.02$\pm$10.75 &                  86.81$\pm$13.34 &                   96.95$\pm$5.22 &                   88.48$\pm$6.53 &                   94.43$\pm$3.41 &          84.18 \\
    DeepConvNet &                   61.88$\pm$19.05 &                  88.27$\pm$12.19 &                   87.56$\pm$11.25 &                  88.12$\pm$13.19 &                   57.08$\pm$12.29 &                  71.49$\pm$15.88 &                   95.90$\pm$7.14 &                  79.29$\pm$12.63 &                   95.92$\pm$3.66 &          80.61 \\
       EEGITNet &                    47.50$\pm$9.46 &                  75.98$\pm$13.09 &                   70.90$\pm$17.50 &                  71.95$\pm$16.76 &                    51.41$\pm$6.40 &                  54.69$\pm$11.97 &                   96.04$\pm$8.62 &                  62.54$\pm$12.32 &                  80.40$\pm$17.12 &          67.93 \\
         EEGNeX &                   52.34$\pm$14.81 &                  64.36$\pm$13.49 &                   69.95$\pm$20.12 &                  72.34$\pm$19.83 &                    53.02$\pm$9.69 &                  51.77$\pm$12.06 &                  89.49$\pm$16.91 &                  60.18$\pm$11.70 &                  64.80$\pm$16.64 &          64.25 \\
     EEGNet-8,2 &                   64.22$\pm$16.01 &                  88.55$\pm$14.92 &                   83.93$\pm$16.31 &                  90.43$\pm$11.75 &                    54.20$\pm$8.20 &                  73.78$\pm$15.59 &                   96.50$\pm$8.07 &                  78.15$\pm$14.46 &                   94.58$\pm$3.21 &          80.48 \\
       EEGTCNet &                   61.09$\pm$22.06 &                  75.21$\pm$18.53 &                   73.92$\pm$19.02 &                  77.21$\pm$18.55 &                    51.22$\pm$5.84 &                  57.03$\pm$13.25 &                   97.15$\pm$7.70 &                  62.37$\pm$12.42 &                  85.46$\pm$16.42 &          71.19 \\
      FBCSP+SVM &                   80.78$\pm$18.86 &                   93.55$\pm$6.29 &                   80.39$\pm$16.83 &                   91.57$\pm$7.66 &                    52.51$\pm$9.82 &                  83.97$\pm$12.43 &                   97.40$\pm$4.18 &                   88.27$\pm$7.91 &                   94.63$\pm$3.94 &          84.78 \\
          FgMDM &                   79.84$\pm$17.80 &                   93.52$\pm$8.18 &                   84.77$\pm$11.26 &                   90.18$\pm$9.77 &                   58.31$\pm$12.63 &                  89.67$\pm$10.65 &                   98.48$\pm$3.45 &                   88.56$\pm$4.63 &                   96.04$\pm$2.67 &           86.6 \\
            MDM &                   74.22$\pm$21.19 &                  89.13$\pm$10.38 &                   77.48$\pm$14.11 &                  86.20$\pm$12.99 &                    48.45$\pm$9.62 &                  81.78$\pm$11.64 &                  84.67$\pm$13.13 &                   65.18$\pm$9.75 &                   92.21$\pm$4.31 &           77.7 \\
 ShallowConvNet &                   64.22$\pm$18.33 &                   93.00$\pm$8.05 &                   87.60$\pm$12.05 &                  91.41$\pm$10.88 &                   57.23$\pm$12.36 &                  74.75$\pm$14.98 &                   98.06$\pm$4.35 &                   88.70$\pm$5.60 &                   97.06$\pm$1.86 &          83.56 \\
          TS+EL &                   81.41$\pm$21.36 &                   94.45$\pm$6.74 &                   85.98$\pm$11.38 &                   91.19$\pm$8.49 &                   58.70$\pm$13.37 &                   94.09$\pm$7.17 &                   98.56$\pm$3.01 &                   92.32$\pm$3.98 &                   96.59$\pm$2.82 &          88.14 \\
          TS+LR &                   83.75$\pm$17.47 &                   94.45$\pm$7.06 &                   85.86$\pm$11.01 &                   91.09$\pm$8.71 &                   61.01$\pm$14.22 &                   93.15$\pm$7.40 &                   98.60$\pm$3.08 &                   91.53$\pm$4.53 &                   96.76$\pm$2.58 &          88.47 \\
         TS+SVM &                   82.66$\pm$18.16 &                   94.01$\pm$7.60 &                   86.19$\pm$11.50 &                   90.81$\pm$8.95 &                   62.55$\pm$15.30 & \textbf{94.27}$\pm$\textbf{7.19} & \textbf{98.72}$\pm$\textbf{2.92} &                   91.84$\pm$4.25 &                   96.11$\pm$2.99 &          88.57 \\
         \midrule
        Average &                             72.08 &                            88.49 &                             81.61 &                            87.01 &                             55.46 &                            79.69 &                            96.23 &                            81.74 &                            92.02 &          81.59 \\
\bottomrule
\end{tabular}\end{adjustbox}
\end{table*}

\begin{table*}
\centering
\caption{Summary of performances via average on all the P300 datasets, for classification using all the labels. Intra-session validation.  Bold numbers represent the best score in each dataset.}
\label{tab:P300_agg_dataset}
\begin{adjustbox}{width=1.15\textwidth}
\begin{tabular}{c|ccccccccccccccc|c}
\toprule
         pipeline &                     BNCI2014-008 &                     BNCI2014-009 &                     BNCI2015-003 &                BrainInvaders2012 &               BrainInvaders2013a &               BrainInvaders2014a &               BrainInvaders2014b &               BrainInvaders2015a &                BrainInvaders2015b &                    Cattan2019-VR &                         EPFLP300 &                      Huebner2017 &                      Huebner2018 &                      Lee2019-ERP &                     Sosulski2019 &        Average \\
\midrule
       ERPCov+MDM &                   74.30$\pm$9.77 &                  81.16$\pm$10.13 &                  76.79$\pm$10.95 &                  78.77$\pm$10.32 &                   80.59$\pm$9.36 &                  71.62$\pm$11.17 &                  78.57$\pm$12.36 &                  80.02$\pm$10.07 &                   75.04$\pm$15.85 &                  80.76$\pm$10.07 &                  71.97$\pm$10.88 &                   94.47$\pm$8.26 &                   95.15$\pm$3.72 &                  74.43$\pm$13.26 &                  68.17$\pm$13.59 &          78.79 \\
ERPCov(svdn4)+MDM &                   75.42$\pm$9.91 &                   84.52$\pm$8.83 &                  76.93$\pm$11.26 &                  79.02$\pm$10.53 &                   82.07$\pm$8.46 &                  72.11$\pm$11.64 &                  76.48$\pm$12.83 &                  77.92$\pm$10.33 &                   77.09$\pm$15.81 &                   80.67$\pm$9.47 &                  71.44$\pm$10.20 &                   96.21$\pm$6.50 &                   96.61$\pm$1.89 &                  82.47$\pm$12.56 &                  70.63$\pm$13.79 &          79.97 \\
        XDAWN+LDA &                   82.24$\pm$5.26 &                   64.03$\pm$3.91 &                   78.62$\pm$7.19 &                   64.41$\pm$4.14 &                   76.74$\pm$7.16 &                   66.60$\pm$7.54 &                  83.73$\pm$10.62 &                  76.02$\pm$10.46 &                   77.22$\pm$13.73 &                   67.16$\pm$6.11 &                   62.98$\pm$5.38 &                   97.74$\pm$2.84 &                   97.54$\pm$1.58 &                   96.45$\pm$3.93 &                   67.49$\pm$7.44 &          77.27 \\
     XDAWNCov+MDM &                   77.62$\pm$9.81 &                   92.04$\pm$5.97 & \textbf{83.08}$\pm$\textbf{7.55} &                   88.22$\pm$5.90 &                   90.97$\pm$5.52 &                  80.88$\pm$11.01 &                  91.58$\pm$10.02 &                   92.57$\pm$5.03 &                   83.48$\pm$12.05 &                   88.53$\pm$7.34 &                   83.20$\pm$9.05 &                   98.07$\pm$2.09 &                   97.78$\pm$1.04 &                   97.70$\pm$2.68 &                   86.07$\pm$7.15 &          88.79 \\
  XDAWNCov+TS+SVM & \textbf{85.61}$\pm$\textbf{4.43} & \textbf{93.43}$\pm$\textbf{5.11} &                   82.95$\pm$8.57 & \textbf{90.99}$\pm$\textbf{4.79} & \textbf{92.71}$\pm$\textbf{4.92} & \textbf{85.77}$\pm$\textbf{9.75} & \textbf{91.88}$\pm$\textbf{9.94} & \textbf{93.05}$\pm$\textbf{4.98} & \textbf{84.56}$\pm$\textbf{12.09} & \textbf{90.68}$\pm$\textbf{6.29} & \textbf{84.29}$\pm$\textbf{8.53} & \textbf{98.69}$\pm$\textbf{1.78} & \textbf{98.47}$\pm$\textbf{0.97} & \textbf{98.41}$\pm$\textbf{2.03} & \textbf{87.28}$\pm$\textbf{6.92} & \textbf{90.58} \\
           \midrule
          Average &                            79.04 &                            83.03 &                            79.67 &                            80.28 &                            84.61 &                             75.4 &                            84.45 &                            83.91 &                             79.48 &                            81.56 &                            74.78 &                            97.04 &                            97.11 &                            89.89 &                            75.93 &          83.08 \\
\bottomrule
\end{tabular}\end{adjustbox}
\end{table*}

\begin{table*}
\centering
\caption{Summary of performances via average on all the P300 datasets, for classification using left vs. right motor imagery task. Intra-session validation.  Bold numbers represent the best score in each dataset.}
\label{tab:SSVEP_agg_dataset}
\begin{adjustbox}{width=1.15\textwidth}
\begin{tabular}{c|ccccccc|c}
\toprule
pipeline &                       Kalunga2016 &                     Lee2019-SSVEP &                            MAMEM1 &                            MAMEM2 &                            MAMEM3 &                     Nakanishi2015 &                          Wang2016 &        Average \\
\midrule
     CCA &                    25.40$\pm$2.51 &                    23.86$\pm$3.72 &                    19.17$\pm$5.01 &                    23.60$\pm$4.10 &                    13.80$\pm$7.47 &                     8.15$\pm$0.74 &                     2.48$\pm$1.01 &          16.64 \\
 MsetCCA &                    22.67$\pm$4.23 &                    25.10$\pm$3.81 &                    20.50$\pm$2.37 &                    22.08$\pm$1.76 &                    27.60$\pm$3.01 &                     7.10$\pm$1.50 &                     4.00$\pm$1.10 &          18.43 \\
     MDM & \textbf{70.89}$\pm$\textbf{13.44} &                   75.38$\pm$18.38 &                   27.31$\pm$11.64 &                    23.12$\pm$6.29 &                    34.40$\pm$9.96 &                   78.77$\pm$19.06 &                   54.77$\pm$21.95 &          52.09 \\
   TS+LR &                   70.86$\pm$11.64 & \textbf{89.44}$\pm$\textbf{13.84} & \textbf{53.71}$\pm$\textbf{24.25} & \textbf{39.36}$\pm$\textbf{12.06} & \textbf{42.10}$\pm$\textbf{14.33} & \textbf{87.22}$\pm$\textbf{15.96} & \textbf{67.52}$\pm$\textbf{20.04} & \textbf{64.32} \\
  TS+SVM &                   68.95$\pm$13.73 &                   88.58$\pm$14.47 &                   50.58$\pm$23.34 &                   34.80$\pm$11.76 &                   40.20$\pm$14.41 &                   86.30$\pm$15.88 &                   59.58$\pm$20.57 &          61.28 \\
    TRCA &                    24.84$\pm$7.24 &                   64.01$\pm$15.27 &                    24.24$\pm$6.65 &                    24.24$\pm$2.93 &                    23.70$\pm$3.49 &                   83.21$\pm$10.80 &                     2.79$\pm$1.03 &          35.29 \\
  \midrule
 Average &                             47.27 &                             61.06 &                             32.58 &                             27.87 &                              30.3 &                             58.46 &                             31.86 &          41.34 \\
\bottomrule
\end{tabular}\end{adjustbox}
\end{table*}

%TODO: no results for all class ?

%\section{Statistical analysis}

%TODO: add figures for significativity.

\end{document}